\documentclass[trackchanges]{aastex7}

\usepackage[table]{xcolor}
\definecolor{lightgray}{gray}{0.9}
\usepackage{graphicx}
\usepackage{amsmath}
\usepackage{booktabs}
\usepackage{float}
\usepackage{makecell}
\usepackage{longtable}
\usepackage{pifont}
\usepackage{xcolor}
\usepackage[normalem]{ulem}

\newcommand{\tbol}{$T_\mathrm{bol}$}
\newcommand{\lbol}{$L_\mathrm{bol}$}

\begin{document}

\title{Nascent Embedded-protostar Survey in Taurus (NEST) I: Protostellar Multiplicity}

\author[orcid=0009-0003-3073-9148, gname=Aislinn, sname=Plante]{Aislinn C. Plante}
\affiliation{National Radio Astronomy Observatory, 520 Edgemont Rd., Charlottesville, VA 22903, USA}
\email[show]{acoleman@nrao.edu}

\author[orcid=0000-0002-6195-0152, gname=John, sname=Tobin]{John J. Tobin} 
\affiliation{National Radio Astronomy Observatory, 520 Edgemont Rd., Charlottesville, VA 22903, USA}
\email{jtobin@nrao.edu}

\author[orcid=0000-0002-9209-8708, gname=Patrick, sname=Sheehan]{Patrick D. Sheehan}
\affiliation{National Radio Astronomy Observatory, 520 Edgemont Rd., Charlottesville, VA 22903, USA}
\email{psheehan@nrao.edu}

\author[orcid=0009-0009-4618-8049, gname=Noshin, sname=Yesmin]{Noshin Yesmin}
\affiliation{Department of Astronomy, University of Virginia, Charlottesville, VA 22903, USA}
\email{pqt7tv@virginia.edu}

\author[orcid=0000-0002-4276-3730, gname=Nicholas, sname=Ballering]{Nicholas P. Ballering}
\affiliation{Space Science Institute, Boulder, CO 80301, USA}
\affiliation{Department of Astronomy, University of Wisconsin-Madison, Madison, WI 53706, USA}
\email{nballering@SpaceScience.org}

\author[orcid=0000-0001-7491-0048, gname=Tyler, sname=Bourke]{Tyler L. Bourke}
\affiliation{SKA Observatory, Jodrell Bank, Macclesfield SK11 9FT, United Kingdom}
\email{Tyler.Bourke@skao.int}

\author[orcid=0000-0002-1031-4199, gname=Josh, sname=Eisner]{Josh Eisner}
\affiliation{University of Arizona Department of Astronomy and Steward Observatory, 933 North Cherry Ave., Tucson, AZ 85721, USA}
\email{jeisner@arizona.edu}

\author[orcid=0000-0002-7402-6487]{Zhi-Yun Li}
\affiliation{Department of Astronomy, University of Virginia, Charlottesville, VA 22903, USA}
\affiliation{Virginia Institute of Theoretical Astronomy, University of Virginia, Charlottesville, VA 22904, USA}
\email[]{zl4h@virginia.edu} 

\begin{abstract}
We present new ALMA 0.9 mm and VLA 9 mm observations in the Taurus Molecular Cloud (TMC) of 25 protostellar systems, containing 40 protostars, observed at 0\farcs3 ($\sim$20 au) resolution. Within separations of 18--10,000 au, the ALMA/VLA-observed Taurus sample has a multiplicity fraction (MF), defined as the fraction of systems with at least one companion, of $0.50 \pm 0.07$, and a companion fraction (CF), defined as the average number of companions per system, of $0.58 \pm 0.20$. To build a more complete census of protostellar multiplicity in this region, we supplement the observed sample with 24 protostars (12 protostellar systems and 5 additional companions associated with systems we observed) previously identified through archival infrared or ALMA observations. 
Together, these 64 individual protostars (37 systems) define our Taurus+ sample, for which we measure higher values of $0.53 \pm 0.06$ and $0.72 \pm 0.19$ for the MF and CF, respectively. These multiplicity statistics in the TMC are notably higher than those reported in the more clustered star-forming regions of Orion and Perseus at the $\sim$3--4$\sigma$ level, suggesting that Taurus may preserve a larger fraction of primordial multiples. The separation distributions in our samples show populations of both close and wide multiples, but a deficit at intermediate separations of 200--300 au. This pattern may suggest two distinct formation pathways: close binaries ($<$200 au) arising primarily from disk fragmentation, and wide multiples ($>$1000 au) from core fragmentation.

\end{abstract}


\section{Introduction} \label{sec:Intro}

Multiplicity is a common outcome of the star formation process, but the likelihood that a star hosts one or more companions depends strongly on its stellar mass. 
The multiplicity fraction increases with stellar mass or spectral type; the most massive stars, O- and B-type, are almost always found in multiple systems, with multiplicity fractions around 90\% \citep{Sana2013,Sana2014,Frost2025}. Solar-type stars (spectral types F, G, and K) have a multiplicity fraction of about 50\% \citep{Raghavan2010}, while M-type (low-mass) stars are found in multiples roughly 30–35\% of the time \citep{Lada2006}. Some of these trends are seen consistently in both field populations and young clusters, at least for close multiples, suggesting that many of these multiple systems are formed early and survive into later evolutionary stages \citep{Bouvier1997,Patience2002,Elliott2014,Deacon2020,Torres2021}. Overall, the prevalence of multiples across stellar masses highlights the importance of studying multiplicity in the context of star formation.

To understand how multiple systems form and evolve, it is essential to observe their initial configurations during the earliest protostellar stages, when multiplicity is closest to its primordial state. Young stellar objects (YSOs) are commonly classified into evolutionary stages using quantities derived from their spectral energy distributions (SEDs), such as bolometric temperature and infrared spectral index. Class~0 sources are the youngest and most deeply embedded protostars, with emission dominated by their cold natal envelopes. They are typically identified by bolometric temperatures below 70 K \citep{Chen1995}, and are thought to be younger than 0.2 Myr years \citep{Dunham2015}. Class~I sources are slightly more evolved ($\sim$0.5 Myr) but remain embedded, with lower envelope densities and reduced infall rates compared to Class 0 sources, as a substantial fraction of the envelope has already been accreted onto the central protostar/disk system. Their SEDs still rise toward longer infrared wavelengths, corresponding to spectral indices $\alpha > 0.3$, because envelope emission and extinction remain significant \citep{Tobin2024}.

During these embedded stages, circumstellar disks form as a natural consequence of angular momentum conservation during collapse \citep{Cassen1981}. Flat Spectrum protostars represent a transitional phase between embedded protostars and more evolved disk sources ($\sim$0.5--1.2 Myr). In these objects, the envelope contribution has decreased while the protostar/disk contribution has become more prominent, producing a comparatively flat infrared SED with $-0.3 < \alpha < 0.3$. Class~II and Class~III sources are more evolved systems that are largely free of envelopes. Class~II sources retain optically thick protoplanetary disks \citep[at $\sim$1-3 Myr]{Kristensen2018}, while Class~III sources have little or no remaining disk material (several Myr). These systems are characterized by declining infrared SEDs, with Class~II sources typically having $\alpha < -0.3$ and Class~III sources having $\alpha < -1.6$ \citep{Greene1994}. By the Class~II phase, the separation distribution has largely already been reshaped and no longer cleanly traces the formation pathways \citep{Offner2023}.

Multiples are believed to form early in the star formation process, especially during the Class~0 stage, when the system still contains most of its mass reservoir. The two primary mechanisms are Disk fragmentation and Core fragmentation. Disk fragmentation involves gravitational rotational instability in a rotationally supported disk, typically forming only closely separated multiples ($\lesssim$300~au; \citet{Kratter2010}) due to the limited size of the disk. In contrast, core fragmentation occurs at the scales of the core, typically forming much wider multiples (from 1,000-10,000 au) via turbulent fragmentation \citep{Offner2010,Lee2019}. It is important to note, the formation of close multiples remains an area of active debate, with some studies suggesting they may primarily originate from core fragmentation followed by substantial migration rather than disk fragmentation \citep{Sadavoy2017,Lee2019, Kuruwita2023}.

Over time, multiplicity can be reshaped after formation through both internal dynamical processes within bound systems and external interactions with the surrounding stellar environment, leading to reduced multiplicity at wide separations in older populations \citep{Offner2023}. Throughout this paper, we use the term "dynamical evolution" to refer broadly to processes that modify the separation distribution of multiple systems after their initial formation. Internal dynamical evolution includes disk-driven migration, inward or outward migration due to angular momentum exchange, and gravitational interactions within higher-order multiple systems that can destabilize the system and eject one or more components \citep[e.g.,][]{Reipurth2012,Tokovinin2020} These internal processes are expected to occur in multiple systems across all star-forming environments, although their efficiency can depend on properties such as mass ratios, component separations, disk structure, and system architecture, which is outside the scope of our analysis.

External dynamical evolution, in contrast, depends more directly on the stellar density of the environment. Encounters between initially unrelated systems can perturb wide companions, leading to inward migration, orbital disruption, or ejection. Dynamical capture of an unrelated source can also occur sometimes \citep{Kuruwita2023, Guszejnov2023}. Because these processes depend on the frequency of close encounters, the stellar surface density of a star forming environment becomes an important variable.

Over time, dynamical interactions and environmental effects reshape multiplicity: wide companions may migrate inward to intermediate or close separations, be ejected, or be disrupted altogether, leading to reduced multiplicity at wide separations in older populations \citep{Reipurth2007, Offner2023}. 

Direct observations of protostellar multiplicity are essential for testing theoretical models and understanding how systems evolve from their initial conditions. However, studying protostars has historically been difficult due to their deeply embedded nature. Fortunately, protostellar envelopes become optically thin at (sub)millimeter and centimeter wavelengths, making high-resolution observations with the Atacama Large Millimeter/submillimeter Array (ALMA) and the NSF’s Karl G. Jansky Very Large Array (VLA) particularly well suited to detecting companions on scales below 1000~au, where mid-infrared instruments typically struggle to resolve them. At these wavelengths, the continuum emission primarily traces warm circumstellar dust associated with embedded protostars, enabling these observations to penetrate the envelope and resolve the dense inner regions where multiple systems are likely to form.

Over the past decade, large interferometric surveys have enabled detailed protostellar multiplicity studies in two major star-forming regions: Perseus and Orion \citep{Tobin2016,Tobin2022}. These surveys revealed a double-peaked separation distribution, with populations of both close and wide multiples--particularly for the youngest sources--broadly supporting the idea that multiple formation pathways (core and disk fragmentation) are at play on distinct scales. In this paper we aim to extend this analysis to the Taurus star-forming region. Among the three regions, Orion is the densest and the only one forming massive (OB) stars, with rich embedded clusters; Perseus is intermediate, combining clustered subregions with a more distributed population; and Taurus is the most diffuse and predominantly distributed, lacking massive stars. Studying multiplicity in Taurus therefore provides an important counterpoint, enabling us to explore how environment influences the formation and survival of multiple systems. A smaller protostellar multiplicity survey has also been carried out in Ophiuchus \citep{Encalada2021}; however, because its limited ALMA field of view and different analysis methodology make direct comparison to the Perseus, Orion, and Taurus surveys difficult, we do not include it in the comparative analysis here. To examine how multiplicity evolves over time, we also compare our findings to a survey of the more evolved Class~II and III population in Taurus \citep{Kraus2011}.

In this study, we will characterize the occurrence rate and separation distribution of multiple protostars in the Taurus Molecular Cloud, capturing multiplicity at an early stage before significant dynamical evolution. We will then evaluate how they compare to those in other star-forming regions and more evolved Taurus populations, to better understand the formation mechanisms of multiple systems. Section~\ref{sec:obs-analys} describes the sample, observations, and data reduction. Section~\ref{sec:results} presents the observational results. Section~\ref{sec:data-analys} details the analysis methodology used to characterize multiplicity, with the resulting multiplicity properties presented in Section~\ref{sec:mult-charac}. We then discuss the implications of these results in Section~\ref{sec:discussion}. Finally, Section~\ref{sec:sumandconclusions} summarizes our conclusions.

\section{Observations \& Sample Selection}\label{sec:obs-analys}

\subsection{The Taurus Sample}\label{subsec:Taurus-samp}

\begin{deluxetable}{lllllllllllll}
\tabletypesize{\scriptsize}
\tablewidth{0pt}
\tablecaption{Taurus Source Catalog: Observed Sample\label{tab:Catalog-obs}}
\tablehead{\colhead{Source} & \colhead{Cl.\textsuperscript{a}} & \colhead{R.A.} & \colhead{Decl.} & \colhead{$L_\mathrm{bol}$} & \colhead{$T_\mathrm{bol}$} & \colhead{$\alpha$\textsuperscript{b}} & \colhead{Dist.\textsuperscript{c}} & \colhead{\#C\textsuperscript{d}} & \multicolumn{2}{c}{\textbf{ALMA (0.9 mm)}} & \multicolumn{2}{c}{\textbf{VLA (9 mm)}} \\[-1.5ex] \colhead{} & \colhead{} & \colhead{(J2000)} & \colhead{(J2000)} & \colhead{($L_\odot$)} & \colhead{(K)} & \colhead{} &  \colhead{(pc)} & \colhead{} & \colhead{\boldmath$I_\nu$, Peak\textsuperscript{e}} & \colhead{rms} & \colhead{\boldmath$I_\nu$, Peak\textsuperscript{e}} & \colhead{rms} \\[-1.5ex] \colhead{} & \colhead{} & \colhead{} & \colhead{} & \colhead{} & \colhead{} & \colhead{} & \colhead{} & \colhead{} & \colhead{(mJy)} & \colhead{(mJy)} & \colhead{(mJy)} & \colhead{(mJy)}}
\startdata
IRAS 04016+2610 & I & 4:04:43.08 & +26:18:56.11 & 3.50 & 222 & 1.10 & 149.0 & 1 & 13.78 & 0.18 & 0.19 & 0.021 \\
IRAS 04108+2803-B & I & 4:13:54.73 & +28:11:32.25 & 0.56 & 210 & 0.74 & 130.0 & 2 & 50.82 & 0.35 & 0.39 & 0.021 \\
IRAS 04158+2805-A & I & 4:18:58.14 & +28:12:22.70 & 0.18 & 423 & 0.35 & 130.0 & 2 & 5.14 & 0.24 & 0.06 & 0.019 \\
IRAS 04158+2805-B & I & 4:18:58.15 & +28:12:22.78 & 0.14 & 427 & 0.33 & 130.0 & 2 & 3.98 & 0.24 & ... & 0.019 \\
IRAS 04166+2706 & 0 & 4:19:42.51 & +27:13:35.79 & 0.37 & 68 & 2.22 & 158.0 & 1 & 107.27 & 0.70 & 1.07 & 0.019 \\
IRAS 04169+2702 & I & 4:19:58.48 & +27:09:56.80 & 1.29 & 165 & 1.37 & 158.0 & 1 & 123.16 & 0.90 & 0.52 & 0.019 \\
IRAS 04181+2654-B & I & 4:21:10.39 & +27:01:37.27 & 0.27 & 278 & 0.40 & 160.0 & 2 & ... & 0.10 & ... & 0.021 \\
IRAS 04181+2654-A & I & 4:21:11.49 & +27:01:08.95 & 0.51 & 294 & 0.42 & 158.0 & 2 & 12.13 & 0.13 & 0.17 & 0.021 \\
IRAM 04191-A & 0 & 4:21:56.90 & +15:29:46.05 & 0.10 & 28 & 1.61 & 144.0 & 4 & 7.21 & 0.11 & ... & 0.020 \\
IRAS 04191+1523-A & I & 4:22:00.43 & +15:30:21.18 & 0.52 & 89 & 1.03 & 144.0 & 4 & 80.95 & 0.75 & 0.16 & 0.022 \\
IRAS 04191+1523-B & I & 4:22:00.09 & +15:30:24.59 & 0.52 & 89 & 1.03 & 144.0 & 4 & 31.39 & 0.75 & ... & 0.022 \\
IRAS 04239+2436-A & I & 4:26:56.26 & +24:43:34.76 & 1.27 & 257 & 0.81 & 128.0 & 2 & 43.86 & 0.37 & 0.64 & 0.040 \\
IRAS 04239+2436-B & I & 4:26:56.28 & +24:43:34.75 & 1.27 & 257 & 0.81 & 128.0 & 2 & 42.23 & 0.37 & 0.55 & 0.040 \\
IRAS 04248+2612-A & I & 4:27:57.34 & +26:19:17.89 & 0.33 & 224 & 0.52 & 128.0 & 3 & 2.57 & 0.10 & ... & 0.022 \\
IRAS 04248+2612-B & I & 4:27:57.32 & +26:19:17.78 & 0.33 & 224 & 0.52 & 128.0 & 3 & 2.30 & 0.10 & ... & 0.022 \\
IRAS 04248+2612-C & I & 4:27:56.37 & +26:19:17.61 & 0.33 & 224 & 0.52 & 128.0 & 3 & 1.15 & 0.10 & ... & 0.022 \\
IRAS 04260+2642 & I & 4:29:05.00 & +26:49:06.79 & 0.08 & 353 & 0.31 & 128.0 & 1 & 30.84 & 0.39 & 0.17 & 0.019 \\
IRAS 04263+2426-A & I & 4:29:23.73 & +24:33:01.01 & 8.62 & 364 & 0.74 & 128.0 & 2 & 86.97 & 0.62 & 0.83 & 0.017 \\
IRAS 04263+2426-B & I & 4:29:23.75 & +24:32:59.66 & 8.62 & 364 & 0.74 & 128.0 & 2 & 65.40 & 0.62 & 1.53 & 0.017 \\
IRAS 04264+2433-A & I & 4:29:30.10 & +24:39:54.54 & 0.39 & 209 & 0.96 & 128.0 & 2 & 16.22 & 0.16 & 0.25 & 0.017 \\
IRAS 04264+2433-B & I & 4:29:30.10 & +24:39:54.93 & 0.39 & 209 & 0.96 & 128.0 & 2 & 4.20 & 0.16 & 0.06 & 0.017 \\
IRAS 04287+1801-A & I & 4:31:34.16 & +18:08:04.68 & 26.79 & 111 & 1.91 & 144.0 & 2 & 322.49 & 2.37 & 3.03 & 0.023 \\
IRAS 04287+1801-B & I & 4:31:34.17 & +18:08:04.26 & 26.79 & 111 & 1.91 & 144.0 & 2 & 191.53 & 2.37 & 2.29 & 0.023 \\
L1551 NE-A & I & 4:31:44.51 & +18:08:31.33 & 3.18 & 101 & 0.90 & 144.0 & 2 & 242.49 & 2.32 & 2.09 & 0.022 \\
L1551 NE-B & I & 4:31:44.48 & +18:08:31.57 & 3.18 & 101 & 0.90 & 144.0 & 2 & 97.87 & 2.32 & 0.87 & 0.022 \\
IRAS 04295+2251 A & I & 4:32:32.08 & +22:57:26.10 & 0.56 & 299 & 0.44 & 161.0 & 2 & 9.49 & 0.42 & 0.56 & 0.016 \\
IRAS 04302+2247 & FS & 4:33:16.50 & +22:53:20.22 & 0.31 & 151 & 0.25 & 161.0 & 1 & 29.15 & 0.70 & 0.15 & 0.016 \\
IRAS 04325+2402-A & I & 4:35:35.42 & +24:08:18.78 & 0.85 & 112 & 0.91 & 128.0 & 2 & 15.65 & 0.27 & 0.20 & 0.019 \\
IRAS 04325+2402-B & I & 4:35:35.32 & +24:08:26.79 & 1.55 & 77 & 1.70 & 128.0 & 2 & 11.18 & 0.27 & 0.09 & 0.019 \\ 
IRAS 04361+2547-A & I & 4:39:13.91 & +25:53:20.33 & 3.30 & 138 & 1.56 & 140.0 & 2 & 71.70 & 0.54 & 1.12 & 0.020 \\
IRAS 04365+2535 & I & 4:39:35.21 & +25:41:44.08 & 2.44 & 180 & 1.23 & 140.0 & 1 & 151.81 & 1.25 & 1.20 & 0.021 \\
IRAS 04368+2557 & 0 & 4:39:53.88 & +26:03:09.40 & 1.50 & 40 & 2.93 & 140.0 & 1 & 100.71 & 0.96 & 1.04 & 0.021 \\
IRAS 04381+2540-A & I & 4:41:12.69 & +25:46:34.63 & 0.62 & 174 & 0.97 & 140.0 & 2 & 25.90 & 0.20 & 0.32 & 0.019 \\
IRAS 04381+2540-B & I & 4:41:12.73 & +25:46:34.65 & 0.62 & 174 & 0.97 & 140.0 & 2 & 13.56 & 0.20 & 0.24 & 0.019 \\
IRAS 04385+2550 & FS & 4:41:38.84 & +25:56:26.30 & 0.45 & 607 & -0.04 & 140.0 & 1 & 42.16 & 0.28 & 0.18 & 0.019 \\
IRAS 04489+3042-A & I & 4:52:06.69 & +30:47:16.93 & 0.39 & 424 & 0.39 & 156.0 & 2 & 22.46 & 0.22 & ... & 0.017 \\
IRAS 04489+3042-B & I & 4:52:06.74 & +30:47:19.83 & 0.39 & 424 & 0.39 & 156.0 & 2 & 0.54 & 0.22 & ... & 0.017 \\
DG Tau-B & I & 4:27:02.58 & +26:05:30.08 & 0.91 & 194 & 1.29 & 128.0 & 2 & 159.96 & 1.71 & 1.40 & 0.017 \\
DG Tau-A & FS & 4:27:04.70 & +26:06:15.66 & 3.58 & 632 & 0.19 & 128.0 & 2 & ... & 1.71 & 0.54 & 0.017 \\ 
HH 30 & II & 4:31:37.49 & +18:12:23.76 & 0.02 & 320 & -0.31 & 144.0 & 1 & 4.79 & 0.15 & ... & 0.020 \\
\enddata
\tablecomments{The entries where the peak flux is absent were the non-detections, with only two such instances in the ALMA data, and ten in the VLA data. All intensities and rms values are reported per beam. Alternate names for each source can be found in Appendix \ref{appendix:obs}. Superscripts indicate: 
\textsuperscript{a}Evolutionary class. 
\textsuperscript{b}Infrared spectral index. 
\textsuperscript{c}Distances to individual sources based on \citet{Luhman2023}.
\textsuperscript{d}Number of components in the system, which include companions from the Taurus+ sample.
\textsuperscript{e}Peak specific intensity.
}
\end{deluxetable}

\begin{deluxetable}{lllllllll}
\tabletypesize{\scriptsize}
\tablewidth{0pt}
\tablecaption{Additional sources that make up the Taurus+ sample}\label{tab:Catalog-ext}
\label{tab:extended-catalog}
\tablehead{\colhead{Source} & \colhead{Class} & \colhead{R.A.} & \colhead{Decl.} & \colhead{$L_\mathrm{bol}$} & \colhead{$T_\mathrm{bol}$} & \colhead{$\alpha$\textsuperscript{a}} & \colhead{Distance\textsuperscript{b}} & \colhead{\# Components\textsuperscript{c}} \\[-1.5ex]
\colhead{} & \colhead{} & \colhead{(J2000)} & \colhead{(J2000)} & \colhead{($L_\odot$)} & \colhead{(K)} & \colhead{} & \colhead{(pc)} & \colhead{}}
\startdata
IRAS 04108+2803-A & FS & 4:13:53.40 & +28:11:22.92 & 0.14 & 731 & -0.26 & 130.0 & 2  \\
IRAS 04154+2823 & FS & 4:18:32.05 & +28:31:14.99 & 0.38 & 517 & 0.27 & 130.0 & 1 \\
IRAS 04166+2708 & I & 4:19:41.53 & +27:16:06.67 & 0.13 & 170 & 0.68 & 158.0 & 1 \\
2MASS J04194657+2712552 & FS & 4:19:46.57 & +27:12:55.22 & 0.05 & 541 & 0.14 & 158.0 & 1  \\
IRAS 04181+2655 & I & 4:21:07.97 & +27:02:20.08 & 0.52 & 378 & 0.51 & 160.0 & 1 \\
IRAS 04181+2654-B & I & 4:21:10.39 & +27:01:37.27 & 0.30 & 306 & 0.33 & 160.0 & 2 \\
IRAM 04191-B & 0 & 4:21:57.67 & +15:29:51.07 & 0.10 & 28 & 1.61 & 144.0 & 4 \\
FS Tau-B & I & 4:22:01.01 & +26:57:35.28 & 0.58 & 174 & 1.18 & 140.0 & 2 \\
FS Tau-A & FS & 4:22:02.20 & +26:57:30.38 & 3.39 & 240 & -0.22 & 140.0 & 2 \\
L1521F & 0 & 4:28:38.96 & +26:51:34.99 & 0.36 & 29 & 1.43 & 140.0 & 1  \\
2MASS J04293209+2430597 & I & 4:29:32.09 & +24:30:59.76 & 0.06 & 360 & 0.83 & 128.0 & 1  \\
LkH$\alpha$358 & FS & 4:31:36.14 & +18:13:43.21 & 0.20 & 875 & -0.16 & 144.0 & 3  \\
HL Tau & I & 4:31:38.51 & +18:13:57.86 & 8.69 & 307 & 0.53 & 147.3 & 3 \\
XZ Tau & II & 4:31:40.08 & +18:13:56.64 & 1.83 & 1362 & -1.52 & 147.3 & 3 \\
Haro 6-13 & FS & 4:32:15.42 & +24:28:59.59 & 1.29 & 654 & -0.03 & 128.6 & 1 \\
IRAS 04295+2251-B & I & 4:32:32.07 & +22:57:26.15 & 0.68 & 337 & 0.44 & 161.0 & 2 \\
HP Tau & FS & 4:35:52.77 & +22:54:23.15 & 1.40 & 912 & -0.24 & 176.3 & 5 \\
HP Tau G3 & II & 4:35:53.38 & +22:54:09.89 & 0.32 & 1285 & -0.91 & 176.0 & 5 \\
HP Tau G2 & III & 4:35:54.00 & +22:54:13.46 & 1.27 & 2233 & -1.69 & 176.0 & 5 \\
Haro 6-28 B & II & 4:35:56.82 & +22:54:35.53 & 0.17 & 1591 & -0.90 & 162.5 & 5 \\
Haro 6-28 A & II & 4:35:56.87 & +22:54:35.82 & 0.17 & 1591 & -0.90 & 162.5 & 5 \\
IRAS S04361+2331 & FS & 4:39:05.26 & +23:37:44.76 & 0.07 & 601 & 0.19 & 161.0 & 1  \\ 
IRAS 04361+2547-B & I & 4:39:13.90 & +25:53:20.13 & 3.30 & 138 & 1.56 & 140.0 & 2 \\
IC 2087 IR & I & 4:39:55.73 & +25:45:01.73 & 3.26 & 573 & 0.84 & 135.0 & 1  \\ 
\enddata
\tablecomments{Alternate names for each source can be found in Appendix \ref{appendix:extended}. 
Superscripts indicate: \textsuperscript{a}Infrared spectral index.
\textsuperscript{b}Distances to individual sources based on \citet{Luhman2023}. \textsuperscript{c}Number of components in the system.}
\end{deluxetable}

The Taurus Molecular Cloud is a nearby star-forming region, at an average distance of $\sim$140 pc \citep{Luhman2023}. 
Unlike more active regions such as Orion or Perseus, Taurus is characterized by a low stellar surface density, measured to be $\sim$5--7 stars pc$^{-2}$ \citep{Kraus2008}. Studies of YSO surface density distributions across star-forming regions show that low-density regions have distributions that peak below $\sim$10 stars pc$^{-2}$, while high-density regions peak above this value \citep{Megeath2016, Kounkel2016}. Taurus falls well within the low-density regime under this definition, as illustrated in Figure~11 of \citet{Megeath2016}.
The low stellar density in Taurus suggests that it provides a laboratory for studying multiplicity in environments with minimal dynamical interactions, where the interaction timescale is expected to be much longer than the age of the region \citep{Kraus2011}.

Taurus also hosts a comparatively more evolved protostellar population. While other multiplicity surveys find that $\sim$29\% of sources in Orion \citep{Tobin2022} and $\sim$50\% in Perseus \citep{Tobin2016} are Class 0, only $\sim$6\% of the Taurus population falls into this youngest stage. This provides an ideal environment in which to test whether multiplicity trends observed in more clustered and younger regions persist in a low-density, more evolved region like Taurus.

The original sample for this study consisted of 3 Class~0, 23 Class~I/FS, and 1 Class~II protostars. These sources were selected based on classifications that are in agreement across multiple studies (e.g., \citet{Motte2001}; \citet{Andrews2005}; \citet{Furlan2008}), which use different datasets and classification methods but yield consistent evolutionary classifications for these sources.

The full source catalog for these observations is presented in Table \ref{tab:Catalog-obs}, which lists the classes, coordinates, distances, bolometric luminosities ($L_\mathrm{bol}$), bolometric temperatures ($T_\mathrm{bol}$), and ALMA and VLA fluxes and rms values for all components in each system.
Initially, Flat Spectrum (FS) sources were excluded from the sample, along with several Class~I sources previously identified as FS, because the original proposal was designed to probe the earliest stages of disk evolution. However, to ensure completeness for this multiplicity survey, the sample was later expanded to include these sources, even where classifications in the literature were inconsistent, which we hereafter refer to as the Taurus+ sample. For clarity, we distinguish between the following samples used throughout this work:

\vspace{2mm}

\noindent\textbf{Taurus Sample:} The original observed sample from our ALMA and VLA program, consisting of 25 protostellar systems, identifying 40 protostars.

\vspace{1mm}

\noindent\textbf{Taurus+ Sample:} An extended sample consists of the 40 protostars in the Taurus Sample, supplemented by 24 additional protostars identified in the literature and 2 previously unreported protostars identified in archival ALMA data, for a total of 64 protostars. These additions comprise 12 new protostellar systems and 5 additional companions associated with systems already included in the Taurus Sample, increasing the sample from 25 to 37 systems. Thus, the Taurus Sample is a subset of the Taurus+ Sample.

\vspace{2mm}

For consistency, we compiled the available archival photometry into SEDs for all sources in both the Taurus and Taurus+ samples and recalculated their evolutionary classifications using bolometric temperature and infrared spectral index. Class~0 sources were classified using $T_\mathrm{bol}$, while Class~I, Flat Spectrum, Class~II, and Class~III sources were classified using the infrared spectral index. This procedure is described in Appendix \ref{appendix:source_seds}.

Information for the additional sources in Taurus+ can be found in Table \ref{tab:Catalog-ext}.
Most of these sources have previously been observed at infrared and/or radio wavelengths with angular resolutions comparable to those of this survey, and published multiplicity information from the literature is incorporated into our analysis where available. For two systems in the Taurus sample, we additionally examined unpublished archival ALMA and VLA data that revealed companions not detected in our observations; these companions are therefore included in the Taurus+ sample. For five systems in the Taurus+ sample, no sub-arcsecond resolution data are currently available. These sources can be associated with other objects at wide separations, but the presence of close companions cannot be assessed.

In total, the Taurus+ extension added 2 Class~0, 9 Class~I, and 8 FS sources. Additionally, we included 4 Class~II sources and 1 Class~III source that are components of systems containing sample members. Table \ref{tab:extended-catalog} provides the full list of these Taurus+ sources, along with their properties, and individual descriptions and literature references for each source are provided in the Appendix \ref{appendix:extended}.

\begin{figure}[ht!]
    \centering
    \includegraphics[width=0.8\textwidth]{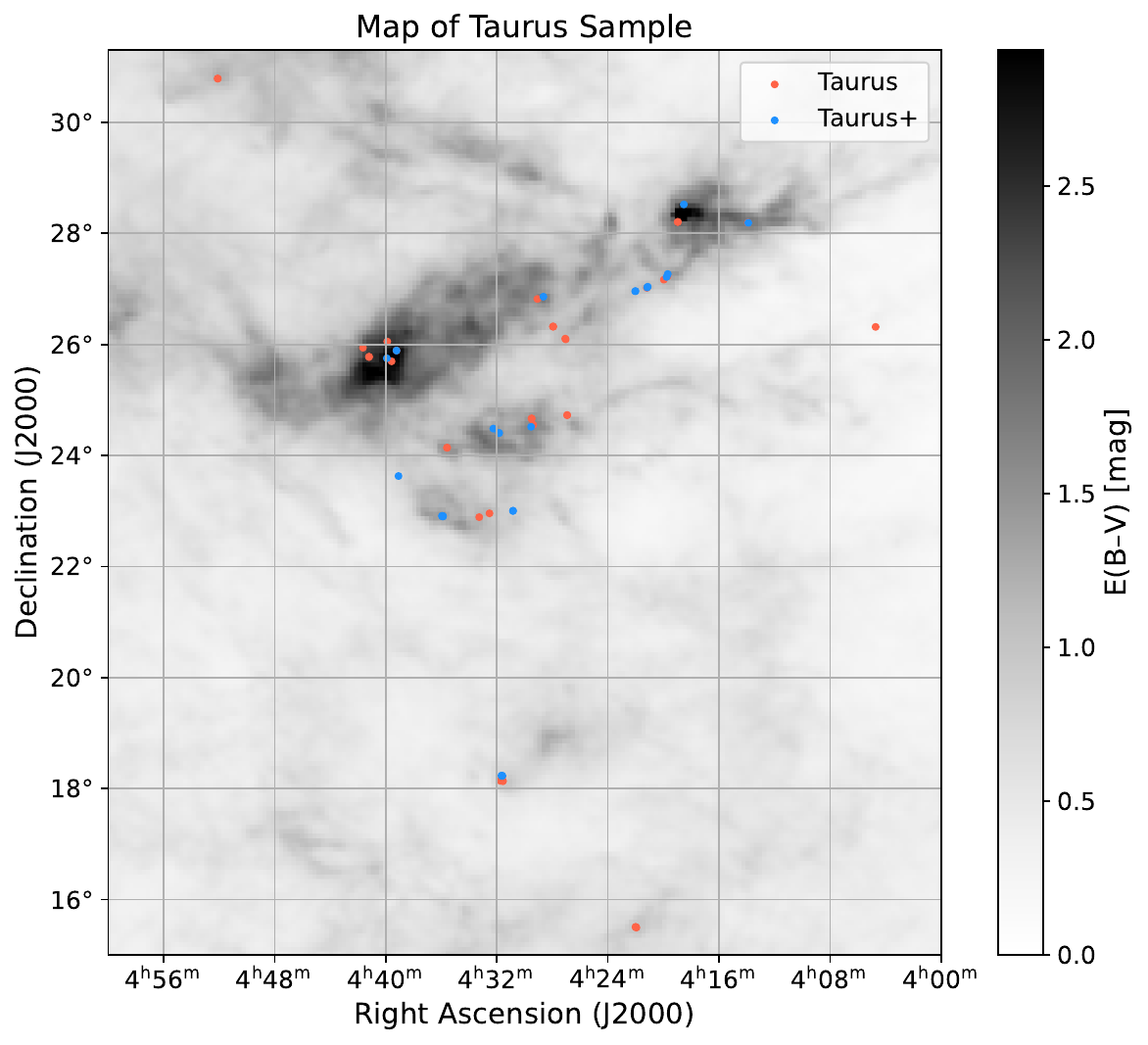}
    \caption{Map of the Taurus Molecular Cloud showing the locations of the sources overlaid on a background extinction map from \citet{Schlegel1998}. Sources from the Taurus sample are shown in red, and the additional sources included in the Taurus+ sample are shown in blue.}
    \label{fig:taurus-map}
\end{figure}

To refine distances to individual sources, we cross-matched our target coordinates with the stellar groups identified by \citet{Luhman2023}, who derived distances to these groups using Gaia data. This approach assigns distances to our sources based on their association with these groups, yielding values spanning 130–160 pc, which are used throughout our analysis.
A map of the Taurus Molecular Cloud and locations of our sources therein can be found in Figure \ref{fig:taurus-map}. 
The background extinction map is based on the dust map from \citet{Schlegel1998}, which provides full-sky estimates of interstellar reddening E(B–V). The map was constructed from a combination of infrared observations by IRAS and COBE/DIRBE, tracing thermal emission from interstellar dust. We use the implementation provided by the dustmaps Python package described in \citet{Green2018}.

\subsection{ALMA Observations} \label{subsec:alma-obs}
We observed 27 pointings in the Taurus star-forming region with ALMA in Cycle 7 (project 2019.1.00847.S) using the 12~m array in Band~7 (340~GHz; 0.9~mm). The observations were conducted in 2019 and 2022 across three scheduling blocks (SBs) grouped by source location, all using the C43-4 configuration, which provided baselines from 15~m to 784 m. The corresponding angular resolution was $\sim$0\farcs3, and the largest recoverable angular scale was $\sim$4\arcsec. The continuum sensitivity was typically $\sim$0.29 mJy/beam. 

Block 1 (2019 October 7) consisted of two executions covering five sources at 0\farcs265 resolution, block 2 (2019 October 7) contained one execution for a single source at 0\farcs309 resolution, and block 3 (2022 August 8) comprised six executions targeting 21 sources at 0\farcs208 resolution. The phase calibrators, number of antennas, and precipitable water vapor (PWV) levels varied between executions and are listed in Table \ref{tab:alma_obs_details}. On-source integration times ranged from 13.3 to 22.7 minutes per field and are provided in Table \ref{tab:obs-per-source}, which summarizes additional observing parameters by source.

The correlator was configured with four basebands. One provided a continuum window centered at 328.750~GHz with 1.875~GHz total bandwidth in 31.25~MHz channels. 
The remaining basebands were configured to target several molecular lines, however, these data are not analyzed in this work. Instead, line-free channels from these spectral windows were identified and combined with the dedicated continuum window, yielding an aggregate continuum bandwidth of $\simeq$3.0~GHz.

The ALMA data taken in 2019 were calibrated with the Common Astronomy Software Applications (CASA) pipeline version 5.6.1-8, while the 2022 data were calibrated with pipeline version 6.2.1-7 \citep{CASA2022, Hunter2023}. The ALMA observations have an absolute flux calibration uncertainty of approximately 10\%, but only statistical uncertainties are considered in our analysis.

\subsection{VLA Observations} \label{subsec:vla-obs}
We observed the same 27 pointings with the VLA under project 18B-179. All sources were observed in B configuration between 2019 June 3 and June 26, with a subset of 14 also observed in BnA configuration between 2019 June 28 and July 17. The B configuration has a maximum baseline of 11.1~km, yielding a resolution of 0\farcs19, while the hybrid BnA configuration extends the maximum baseline along the north arm of the array to 22.8~km, achieving a resolution of 0\farcs15 in that direction.

All observations were conducted in Ka-band (33.3~GHz; 9~mm) using the 3-bit samplers in 8~GHz continuum mode, with two 4~GHz basebands centered at 35.0~GHz and the other at 31.0~GHz. The full 8~GHz bandwidth was divided into 128~MHz spectral windows, each containing 64 channels with 2~MHz resolution. On-source integration times ranged from approximately 7 to 14 minutes and are listed in Table \ref{tab:obs-per-source}. 3C147 served as the flux calibrator, and J0319+4130 as the bandpass calibrator across all observations. The specific phase calibrator varied between scheduling blocks and is also listed in Table \ref{tab:obs-per-source}. We alternated between target and phase calibrator scans, spending roughly 80-100 s on the calibrator after each $\sim$100 s scan of a science target.

The calibration for VLA observations obtained on or before 2019-06-17 used version 5.4.1-32 of the VLA pipeline, while observations taken after that date used version 5.4.2-5.
The only exception was the observation taken on MJD 58662, which failed in the original pipeline and was subsequently reprocessed with the VLA pipeline version 6.5.4-9 \citep{CASA2022}.
The VLA data likewise carry an absolute flux calibration uncertainty of roughly 10\%, though only statistical uncertainties are included in our analysis.

\subsection{Self Calibration \& Imaging} \label{subsec:sc-and-imaging}
We then applied automated self-calibration \citep[v1.2.1]{Sheehan2025} to the pipeline-calibrated ALMA and VLA data, with additional manual flagging applied where necessary. Table \ref{tab:obs-per-source} summarizes which observations were successfully self-calibrated. Imaging was performed with the \texttt{tclean} task in CASA using Briggs weighting. 
For ALMA, we imaged each target using Briggs weighting with robust parameters of $R = -1.0$, $0.0$, $0.5$, and $1.0$. For the VLA, we used $R = 0.5$, $1.0$, and $2.0$. The final image for each target was selected to optimize the trade-off between resolution and sensitivity based on the properties of the source.
All ALMA images cover $31\arcsec\times31\arcsec$. To maintain adequate sampling of the synthesized beam, we adopted a cell size of $0.0155\arcsec\,$pix$^{-1}$ for the higher–resolution weightings ($R=-1.0,\,0.5$; $2000\times2000$ pixels) and $0.031\arcsec\,$pix$^{-1}$ for the lower–resolution weightings ($R=0.0,\,1.0$; $1000\times1000$ pixels). 
VLA images were made with a cell size of $0.020\arcsec\,$pix$^{-1}$, yielding a $122\arcsec\times122\arcsec$ field of view ($6144\times6144$ pixels).

Typical synthesized beams were $0.28\arcsec\times0.20\arcsec$ for ALMA and $0.33\arcsec\times0.18\arcsec$ for the VLA (FWHM; varying with robust and $uv$ coverage), corresponding to $\sim\!39\times28$~au and $\sim\!46\times25$~au at 140 pc. 
The median continuum rms was $\sim\!0.66$ mJy\,beam$^{-1}$ at 0.9~mm (ALMA) and $\sim\!0.020$ mJy\,beam$^{-1}$ at 9~mm (VLA).

\subsection{Archival Data} \label{subsec:additional_vla_data}

We include two previously unpublished archival datasets from the NRAO Science Data Archive, each revealing a companion and adding one protostar to the Taurus+ sample. The observations and data reduction procedures for these datasets are described as follows.

IRAS 04295+2251 was observed by the VLA in Ka-band in B-configuration on 2015 February 19 and during the A-configuration to D-configuration move on 2015 September 25 (Project 15A-381). The correlator was configuration for full 3-bit continuum mode, with one baseband centered at $\sim$29~GHz and the other centered at $\sim$37~GHz, sampling 8~GHz of bandwidth, with a 4 GHz gap between the basebands; C-band data were also observed as part of the same EB, but those data will not be discussed here. The observed basebands were divided into 64, 128 MHz wide spectral windows, each with 64 channels. Approximately 50 minutes were spent on source, after slew overheads in both observations. For both observations, J0440+2728 was observed as the complex gain calibrator, 3C84 was observed as the bandpass calibrator, and 3C147 was observed as the flux density calibrator.
Both the A and B-configuration data were originally calibrated using the scripted VLA pipeline using CASA 4.2.2. The observations conducted during the A to D configuration move had 18 antennas still in the A-configuration positions and the other 9 antennas had already been moved to their D-configuration positions. All antennas were used during calibration of that dataset, but prior to imaging, we flagged all the visibilities with $uv$-distances $<$ 100~k$\lambda$ to avoid creating a beam dominated by short baselines. The data were self-calibrated using the automated self-calibration routines as used on the other ALMA and VLA observations mentioned previously. We first ran the self-calibration on the B-configuration data alone, then we copied the self-calibrated B-configuration data into a new directory and self-calibrated the A-configuration data together with the previously self-calibrated B-configuration data. This yields a superior result because the B-configuration data alone could be self-calibrated to a solution interval that was the length of a single scan, while the self-calibration of the A- and B-configuration together could only be self-calibrated on a solution interval spanning an the entire EB. This is at least in part due to the A-configuration data only have 18 usable antennas for self-calibration.

The proper motion of IRAS 04295+2251 has caused the apparent source position to move relative to the position at the time of the ALMA observations, from the VLA observations being conducted in 2015. Thus, we need to compute the proper motion of the source in order to properly align the coordinates of the source with the ALMA data. To compute the proper motion, we use the VLA data taken in B-configuration on 2015 February 19 (MJD 57072) and the data taken on 2019 June 18 (MJD 58652) to calculate the proper motion of IRAS 04295+2251. The position on MJD 57072 was $\alpha$=04:32:32.0785 $\delta$=22:57:26.204 and on MJD 59652 was $\alpha$=04:32:32.0801 $\delta$22:57:26.158, which were determined using the CASA task imfit to fit Gaussians to the unresolved point source in the images. The proper motion is found to be $\mu_{\alpha}$=5.25 mas yr$^{-1}$ and $\mu_{\delta}$=-10.59 mas yr$^{-1}$. The ALMA observations were taken approximately on 2022 July 2 MJD 59762 and the total shift in position between the VLA data taken in 2015 is 38.67 mas in right ascension and -77.98 mas in declination. We used these values to adjust the reference positions for the image headers of the 2015 VLA data in order to bring their coordinates into agreement with those of the ALMA position from the observations in 2022.

The IRAM 04191 binary system was observed with ALMA Band 6 (project 2016.1.01284.S) using the 12 m array in two executions with different configurations. The first observation was obtained on 2016 October 15 in configuration C40-6, providing baselines from 15 to 1800 m, while the second was taken on 2017 July 21 in configuration C40-5, spanning 17-1100 m. Together, the data yield a synthesized beam of 0\farcs154 and a largest recoverable angular scale (LAS) of 1.95\arcsec, with a continuum sensitivity of 0.017 mJy/beam and a field of view of 25.6\arcsec.
The 2016 execution used J0433+0521 as the phase calibrator, with 41 antennas operational and precipitable water vapor (PWV) of 0.64 mm. The 2017 execution used J0449+1121 as the phase calibrator, with 46 antennas and similar atmospheric conditions (PWV = 0.64 mm). The total on-source integration time for the combined dataset is 3992 s.
The correlator was configured with four basebands. One baseband contained four high-resolution spectral windows centered at 218.436, 218.459, 219.558, and 219.948 GHz. The remaining three basebands provided continuum coverage centered at 221.36, 233.50, and 235.50 GHz, each 2.0 GHz wide with 977 kHz channels, yielding an effective continuum bandwidth of $\sim$6 GHz.

We applied the same self-calibration procedure described in Section \ref{subsec:sc-and-imaging}. No proper-motion correction was performed, as the position of the source identified as IRAM 04191-A (also detected in our original ALMA dataset) showed only a negligible positional shift between observations.

\begin{deluxetable}{llcccc}
\tablecaption{ALMA Observing Details by Execution Block\label{tab:alma_obs_details}}
\tablehead{
\colhead{MOUS\textsuperscript{a}} & \colhead{EB ID\textsuperscript{b}} & \colhead{Date} & \colhead{$N_{\rm ant}$} & \colhead{PWV (mm)} & \colhead{ALMA Phase Calibrator}
}
\startdata
uid://A001/X1467/X260 & uid://A002/Xe1d2cb/Xc90c & 2019-10-07 & 45 & 0.46 & J0431+1731\\
" & uid://A002/Xe1d2cb/Xd227 & 2019-10-07 & 45 & 0.41 & J0431+1731\\
\hline
uid://A001/X1467/X267 & uid://A002/Xe1d2cb/Xda79 & 2019-10-07 & 45 & 0.44 & J0438+3004\\
\hline
uid://A001/X1467/X26e & uid://A002/Xfc69ac/Xa3ed & 2022-08-05 & 46 & 0.40 & J0438+3004\\
" & uid://A002/Xfaf3de/X2fd6 & 2022-08-05 & 45 & 0.33 & J0438+3004\\
" & uid://A002/Xfaf3de/X2af2 & 2022-08-05 & 45 & 0.32 & J0438+3004\\
" & uid://A002/Xfa5545/X397f & 2022-08-05 & 46 & 0.50 & J0438+3004\\
" & uid://A002/Xfa2f45/X18c6d & 2022-08-05 & 41 & 0.46 & J0438+3004\\
" & uid://A002/Xfa0be4/Xfda0 & 2022-08-05 & 41 & 0.37 & J0438+3004\\
\enddata
\tablecomments{\textsuperscript{a}Member Observation Unit Set; \textsuperscript{b}Execution Block. Together, the MOUS and EB ID provide unique identifiers assigned by ALMA.}
\end{deluxetable}
\begin{deluxetable}{llllll}
\tablecaption{Observing Parameters Per Source\label{tab:obs-per-source}}
\tablehead{
\colhead{Source} & \colhead{ALMA TOS} & \colhead{VLA TOS} & \colhead{VLA Phase Calibrator}
}

\tablehead{\colhead{Source} &  \multicolumn{2}{c}{\textbf{ALMA (0.9 mm)}} & \multicolumn{3}{c}{\textbf{VLA (9 mm)}} \\[-1.5ex] \colhead{} & \colhead{T.O.S.\textsuperscript{a}} & \colhead{Self Cal.} & \colhead{T.O.S.\textsuperscript{a}} & \colhead{Self Cal.} & \colhead{Phase Cal.}}
\startdata
04016+2610        & 798s  & \ding{51}  & 717s & \ding{55}  & J0403+2600 \\
04108+2803        & 871s  & \ding{51}  & 789s & \ding{51}  & J0403+2600 \\
04158+2805        & 871s  & \ding{51}  & 813s & \ding{55}  & J0403+2600 \\
04169+2702        & 798s  & \ding{51}  & 688s & \ding{51}  & J0429+2724 \\
04181+2654 A      & 798s  & \ding{51}  & 688s & \ding{55}  & J0429+2724 \\
04166+2706        & 798s  & \ding{51}  & 688s & \ding{51}  & J0429+2724 \\
04381+2540        & 798s  & \ding{51}  & 806s & \ding{55}  & J0429+2724 \\
DG Tau            & 798s  & \ding{51}  & 782s & \ding{51}  & J0429+2724 \\
04260+2642        & 798s  & \ding{51}  & 751s & \ding{55}  & J0429+2724 \\
HH\_30            & 968s  & \ding{51}  & 535s & \ding{55}  & J0431+1731 \\
L1551 NE          & 968s  & \ding{51}  & 419s & \ding{51}  & J0431+1731 \\
04287+1801        & 968s  & \ding{51}  & 502s & \ding{51}  & J0431+1731 \\
04191+1523        & 907s  & \ding{51}  & 551s & \ding{55}  & J0431+1731 \\
IRAM 04191        & 907s  & \ding{51}  & 551s & \ding{55}  & J0431+1731 \\
04264+2433        & 798s  & \ding{51}  & 782s & \ding{51}  & J0426+2327 \\
04302+2247        & 798s  & \ding{51}  & 766s & \ding{55}  & J0426+2327 \\
04295+2251        & 798s  & \ding{51}  & 766s & \ding{51}  & J0426+2327 \\
04263+2426        & 798s  & \ding{51}  & 790s & \ding{51}  & J0426+2327 \\
04489+3042        & 1361s & \ding{51}  & 750s & \ding{55}  & J0438+3004 \\
04385+2550        & 798s  & \ding{51}  & 774s & \ding{55}  & J0438+3004 \\
04325+2402        & 798s  & \ding{51}  & 750s & \ding{55}  & J0438+3004 \\
04365+2535        & 798s  & \ding{51}  & 582s & \ding{51}  & J0429+2724 \\
04368+2557        & 798s  & \ding{51}  & 588s & \ding{51}  & J0429+2724 \\
04361+2547        & 798s  & \ding{51}  & 588s & \ding{51}  & J0429+2724 \\
04181+2654 B      & 798s  & \ding{55}  & 782s & \ding{55}  & J0429+2724 \\
04239+2436        & 798s  & \ding{51}  & 798s & \ding{51}  & J0429+2724 \\
04248+2612        & 798s  & \ding{55}  & 766s & \ding{55}  & J0429+2724 \\
\enddata
\tablecomments{Because of slew overheads, the actual VLA Time On Source\textsuperscript{a} is $\sim$30\% less than the values reported in this table. Check marks (\ding{51}) indicate observations that were successfully self-calibrated; crosses (\ding{55}) indicate those that were not.}
\end{deluxetable}

\subsection{The Orion \& Perseus Sample} \label{subsec:ori-per-samp}

The comparison samples used in this study come from the VLA/ALMA Nascent Disk and Multiplicity (VANDAM) surveys of the Perseus and Orion molecular clouds. 
These regions represent clustered high-density, actively star-forming environments that differ from the more quiescent Taurus region. 
Following the YSO surface density framework of \citet{Megeath2016}, both Orion and Perseus fall within the high-density regime, with YSO surface density distributions that peak above 10 stars pc$^{-2}$. These regions reach significantly higher densities than Taurus, with peak values up to $\sim$800 stars pc$^{-2}$ in Perseus and $\sim10^{4}$ stars pc$^{-2}$ in Orion.

The first of these surveys was conducted by \citet{Tobin2016} towards Perseus which lies at a distance of $\sim$300 pc \citep{Ortiz-Leon2018}.
They used the VLA at 9 mm at a resolution of 0\farcs065 (15 au), and detected companions with separations as small as 18.4 au (0\farcs08). 
In total, the survey observed 104 systems in total: 55  Class~0 protostars, 37  Class~I and FS protostars, and 12  Class~II systems. These sources were primarily selected from \citet{Enoch2009}, and span a bolometric luminosity range of $\sim$0.1 to 120 $L_\odot$.

The second survey was toward Orion at a distance of $\sim$400 pc \citep{Tobin2022}.
The VANDAM Orion sample includes 94  Class~0 protostars, 128  Class~I protostars, and 103 Flat Spectrum (FS) sources, for a total of 325 sources drawn primarily from the Herschel Orion Protostar Survey (HOPS) \citep{Furlan2016}. They observed these sources with ALMA at 0.87~mm at an angular resolution at 0\farcs1 and observed a subset of these sources with the VLA at 9.1~mm and $\sim$0\farcs07 resolution. With these observations, the smallest companion separation they were able to resolve was 21.8~au (0\farcs055). The bolometric luminosities for these sources span a wide range, from $\sim$0.1 to $\sim$ 1400 $L_\odot$. 

Both Orion and Perseus host significantly larger protostellar populations than Taurus, with a much higher proportion of Class~0 sources. Orion additionally forms high-mass stars within dense clusters, whereas Perseus is dominated by low- to intermediate-mass star formation but still exhibits a higher degree of clustering than Taurus. Owing to the high YSO surface densities in these regions, corrections for chance alignments were applied when identifying companions.
This was done by calculating a weighted probability for each candidate companion based on the local surface density of YSOs, the projected separation between sources, and the statistical likelihood of random alignment. We will proceed referencing the contamination corrected results for our comparisons. No such correction was applied to the Taurus sample due to its significantly lower source density ($\sim$5--7 stars pc$^{-2}$), which reduces the chance of spurious associations.
The Orion and Perseus samples provide a benchmark for multiplicity in denser, more clustered star-forming environments and serve as a useful contrast to Taurus.

\subsection{More Evolved Taurus Sample} \label{subsec:Kraus-samp}

While Orion and Perseus allow us to compare Taurus to other star-forming regions, it is also important to examine how multiplicity evolves within the same environment over time. To this end, we compare our results to a sample of more evolved  Class~II and III stars in Taurus from \citet{Kraus2011}. This provides a direct view of how the multiplicity properties of protostars may change as the populations age.

\citet{Kraus2011} conducted a high-resolution multiplicity survey of Class~II/III stars in the Taurus Molecular Cloud using the Keck and Palomar telescopes. Their sample included 128 sources and was sensitive to separations ranging from 3 to 5000~au, meaning that their survey did not cover the widest multiples probed in our study (up to 10,000~au).

A key aspect of their analysis was to investigate how multiplicity varies with stellar mass. Stellar masses were estimated from spectral types, with uncertainties of up to 20\%. Their full sample spanned 0.25–2.5$M_\odot$, and they divided it into subsamples of low-mass (LM; 0.25-0.7 $M_\odot$) and solar-mass (SM; 0.7-2.5 $M_\odot$) primaries. They found that lower-mass stars showed a notable lack of companions at separations above approximately 200~au. Overall, their findings supported the idea that binary formation operates through two distinct mechanisms: large-scale fragmentation of collapsing cores leading to wide multiples, and disk fragmentation producing closer companions.

\section{Continuum Imaging Results}\label{sec:results}

\begin{figure}[ht!]
    \centering
    \includegraphics[width=0.8\textwidth]{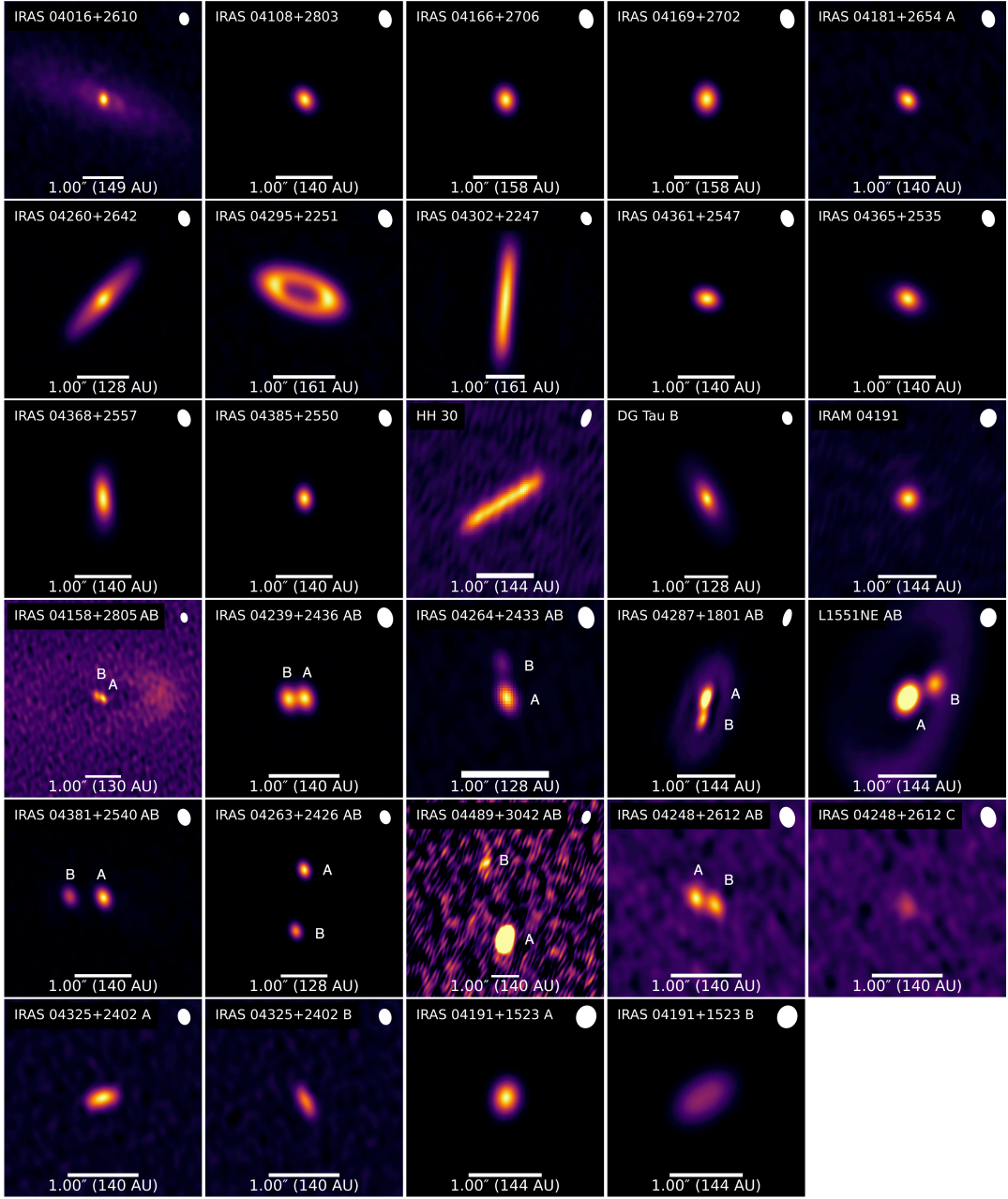}
    \caption{ALMA 0.9 mm continuum images of protostars in Taurus, displayed in order of single protostars, close binary systems, and zoomed-in views of wide multiples. Large scale views of these wide multiples can be found in Figure \ref{fig:wide_multiples}. Synthesized beams in the top-right of each panel give the angular resolution, and 1\arcsec\ scale bars are indicated.}
    \label{fig:alma-observations}
\end{figure}

\begin{figure}[ht!]
    \centering
    \includegraphics[width=0.8\textwidth]{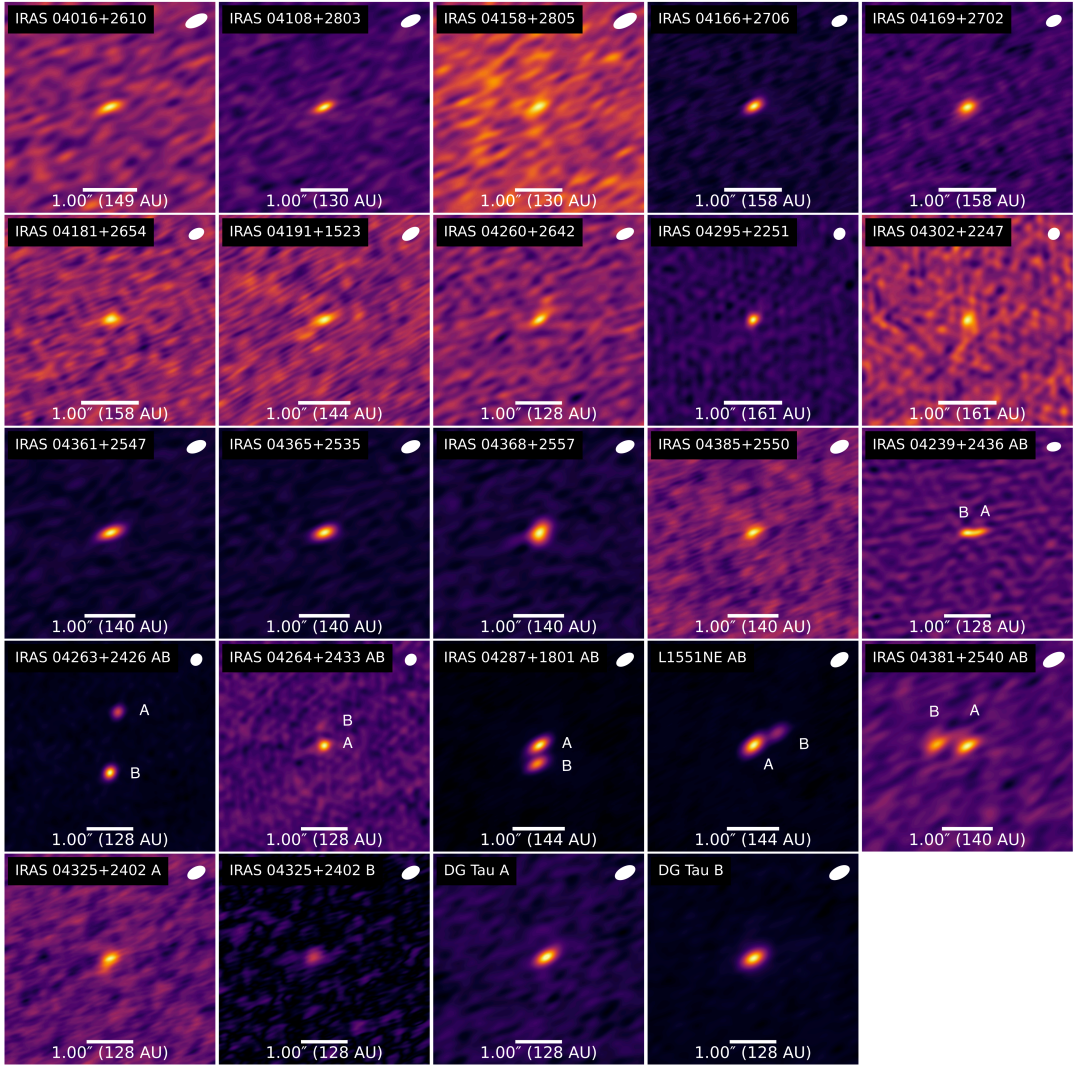}
    \caption{VLA 9 mm continuum images of protostars in Taurus, displayed in order of single protostars, close binary systems, and zoomed-in views of wide multiples.
    Large scale views of these wide multiples can be found in Figure \ref{fig:wide_multiples}. Synthesized beams in the top-right of each panel give the angular resolution, and 1\arcsec\ scale bars are indicated.}
    \label{fig:vla-observations}
\end{figure}

\begin{figure}[ht!]
    \centering
    \includegraphics[width=0.9\textwidth]{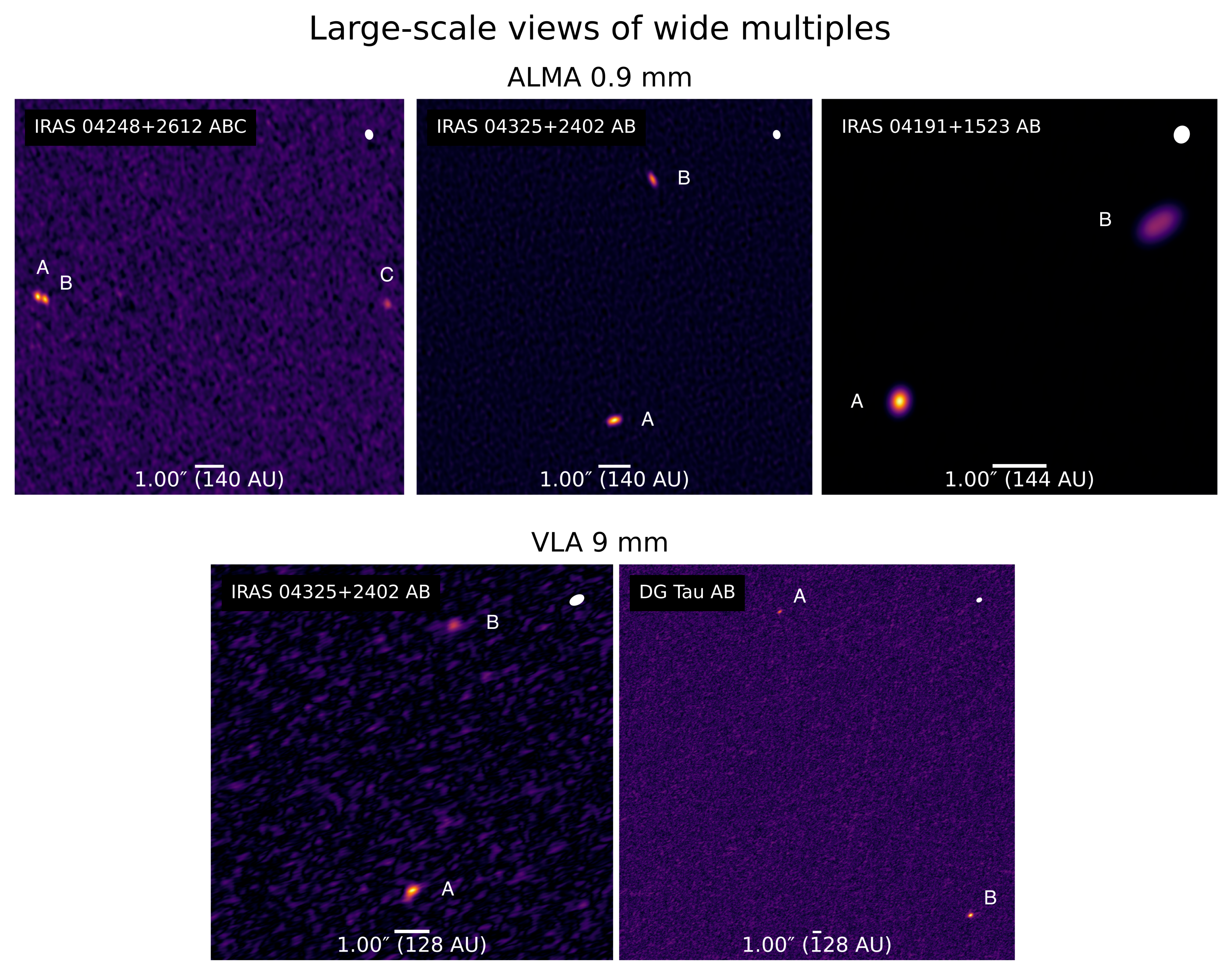}
    \caption{Large-scale views of the wide multiples from the ALMA observations (top) and VLA observations (bottom) to show their full projected separations. Small scale views of the individual protostars can be found in Figures \ref{fig:alma-observations} and \ref{fig:vla-observations}. Synthesized beams in the top-right of each panel give the angular resolution, and 1\arcsec\ scale bars are indicated.}
    \label{fig:wide_multiples}
\end{figure}

The sources of the Taurus sample were observed in continuum with ALMA at 0.9~mm and the VLA at 9~mm.
The observed continuum maps from our primary observations, with close-up views, are shown in Figures~\ref{fig:alma-observations} (ALMA) and~\ref{fig:vla-observations} (VLA), while Figure~\ref{fig:wide_multiples} presents wider-field views of the most widely separated multiple systems.
All of these companions were previously detailed in literature at infrared wavelengths (references in Appendix \ref{appendix:sourceinfo}) except for the close companion to IRAS~04489+3042 which we identified at a separation of 3\arcsec or 467 au.

We present the multiple systems observed in our Taurus sample in Table \ref{tab:companion-table}, including their classifications and projected separations. In systems with more than two components, the listed separation to a third component or sub-system refers to the distance from the geometric center of the first pair. This allows for a consistent measure of separation across multi-source systems. This association process is described in Section \ref{subsec:assoc-mult-sys}.

Within the Taurus sample, we detect 12 single protostars, and 14 multiples (Figure \ref{fig:alma-observations} \& \ref{fig:vla-observations}). Five of these multiples have separations greater than 800 au, and are therefore shown with their full fields of view in \ref{fig:wide_multiples}.
The smallest separation we resolve is 0\farcs21 (27.6~au), while the widest system we observe within a single image in the Taurus sample spans 54\arcsec, corresponding to a projected separation of $\sim$6900~au. When including systems from the Taurus+ sample, we identify an even wider multiple system with five components and projected separations up to 9138~au, which is shown as the last listing in Table \ref{tab:companion-table}. Sources included only in the Taurus+ sample are not shown in the images presented here because they have previously been imaged and reported in the literature; the corresponding references are provided in Appendix \ref{appendix:sourceinfo}.

For some targets, detections were made at only one wavelength. IRAS~04181+2654 B was not detected with either ALMA or the VLA, while HH 30, IRAS~04489+3042, and IRAS~04248+2612 ABC were undetected with the VLA. We retain IRAS~04181+2654 B in the Taurus+ sample given its confirmation in previous high-resolution studies \citep{2008AJ....135.2496C}. In all cases of non-detection, we proceeded with analysis based solely on the available ALMA or VLA data in the Taurus sample.
Some systems also showed differing multiplicity signatures between instruments. For example, IRAS~04158+2805 and IRAS~04191+1523 appeared as single sources with the VLA but were resolved as binaries with ALMA. Conversely, DG Tau A and B were both detected with the VLA, but only DG Tau B was detected with ALMA due to ALMA’s smaller field of view.

We detect dust continuum emission with disk-like morphology around nine sources in our sample. Six of these are consistent with circumstellar disks surrounding single protostars, three of which appear edge-on. The remaining three sources, IRAS 04158+2805, L1551 NE, and IRAS 04287+1801, exhibit circumbinary disks encompassing close binaries, each showing evidence of a large central cavity, supporting the idea that large cavities in the protostellar phase can be indicators of binary formation \citep{Sheehan2020}. Our ALMA observations of IRAS 04295+2610 show a disk with a large central cavity but no detected inner source, whereas our VLA data reveal a single central point source. Based on the disk morphology, we infer the presence of a close companion and examine archival higher-resolution data toward this target to test this possibility.

\begin{deluxetable}{llll}
\tabletypesize{\scriptsize}
\tablewidth{0pt}
\tablecaption{Companion Separations}
\label{tab:companion-table}
\tablehead{\colhead{Source Pair} & \colhead{Separation} & \colhead{Separation} & \colhead{Classes} \\ 
[-2.1ex]
\colhead{} & \colhead{(arcsec)} & \colhead{(au)} & \colhead{}}
\startdata
04295+2251 A $\leftrightarrow$ 04295+2251 B* & 0.116$\pm$0.010 & 18.622$\pm$1.578 & CI--CI \\
04158+2805 A $\leftrightarrow$ 04158+2805 B & 0.212$\pm$0.009 & 27.567$\pm$1.181 & CI--CI \\
04239+2436 A $\leftrightarrow$ 04239+2436 B & 0.240$\pm$0.001 & 30.745$\pm$0.131 & CI--CI \\
04248+2612 A $\leftrightarrow$ 04248+2612 B & 0.297$\pm$0.004 & 38.010$\pm$0.536 & CI--CI \\
04264+2433 A $\leftrightarrow$ 04264+2433 B & 0.394$\pm$0.005 & 50.489$\pm$0.658 & CI--CI \\
04287+1801 A $\leftrightarrow$ 04287+1801 B & 0.422$\pm$0.003 & 60.829$\pm$0.485 & CI--CI \\
L1551NE A $\leftrightarrow$ L1551NE B & 0.549$\pm$0.006 & 79.106$\pm$0.839 & CI--CI \\
04381+2540 A $\leftrightarrow$ 04381+2540 B & 0.609$\pm$0.001 & 85.200$\pm$0.195 & CI-CI \\
Haro 6-28 A* $\leftrightarrow$ Haro 6-28 B* & 0.663 & 107.728 & CII--CII \\
04361+2547 A $\leftrightarrow$ 04361+2547 B* & 0.228$\pm$0.001 & 31.975$\pm$0.190 & CI--CI \\
04263+2426 A $\leftrightarrow$ 04263+2426 B & 1.363$\pm$0.002 & 174.461$\pm$0.213 & CI--CI \\
04489+3042 A $\leftrightarrow$ 04489+3042 B & 2.994$\pm$0.212 & 467.139$\pm$32.997 & CI--CI \\
04191+1523 A $\leftrightarrow$ 04191+1523 B & 6.055$\pm$0.005 & 871.946$\pm$0.663 & CI--CI \\
04325+2402 A $\leftrightarrow$ 04325+2402 B & 8.116$\pm$0.004 & 1038.823$\pm$0.452 & CI--CI \\
HP Tau G3* $\leftrightarrow$ HP Tau G2* & 9.358 & 1647.022 & CII--CII \\
IRAM 04191 A $\leftrightarrow$ IRAM 04191 B* & 12.708$\pm$0.009 & 1829.900$\pm$1.320 & C0--C0 \\
04248+2612 C $\leftrightarrow$ (04248+2612 B -- 04248+2612 A) & 12.869$\pm$0.012 & 1647.211$\pm$1.538 & CI--(CI--CI) \\
FS Tau A* $\leftrightarrow$ FS Tau B* & 16.592 & 2322.900 & FS--CI \\
04108+2803 A* $\leftrightarrow$ 04108+2803 B & 19.874$\pm$0.001 & 2583.610$\pm$0.154 & FS--CI \\
(HP Tau G3 -- HP Tau G2)* $\leftrightarrow$ HP Tau* & 17.111 & 3011.575 & (CII--CII)--FS \\
XZ Tau* $\leftrightarrow$ HL Tau* & 22.397 & 3299.022 & CII--CI \\
04181+2654 A $\leftrightarrow$ 04181+2654 B* & 31.897$\pm$0.002 & 5039.699$\pm$0.285 & CI--CI \\
LkH$\alpha$358* $\leftrightarrow$ (XZ Tau -- HL Tau)* & 47.042 & 6774.019 & FS--(CII--CI) \\
DG Tau A $\leftrightarrow$ DG Tau B & 53.815$\pm$0.004 & 6888.373$\pm$0.499 & FS--CI \\
(04191+1523 A -- 04191+1523 B) $\leftrightarrow$ (IRAM 04191 A -- IRAM 04191 B*) & 54.774$\pm$0.003 & 7887.484$\pm$0.378 & (CI--CI)--(C0--C0) \\
\makecell[tl]{[(HP Tau G3 -- HP Tau G2) -- HP Tau]*\\ $\leftrightarrow$ (Haro 6-28 B -- Haro 6-28 A)*} & 51.922 & 9138.274 & [(CII--CII)--FS]--(CII--CII) \\
\enddata
\tablecomments{This table lists the separations used to characterize the hierarchical architecture of each identified multiple system. In the Source Pair column, the ($\leftrightarrow$) symbol indicates the two components or sub-systems for which the listed separation is reported, while the (--) signs indicate components grouped within a sub-system. Parentheses and square brackets are used to show the hierarchy of higher-order systems. For higher order systems, each sub-system will have its own row with its respective separations. The two separation columns give the angular separation in arcseconds and the projected physical separation in au. The final column gives the evolutionary classes of the components included in each pair or sub-system. To keep the table compact, abbreviated source names are used here for all IRAS sources, but are listed in full in the source catalog. An asterisk (*) denotes a source from the Taurus+ sample. Separation entries without associated errors are from the added Taurus+ sources, for which right ascension and declination values were retrieved from the literature or SIMBAD and positional uncertainties were not provided.}
\end{deluxetable}

\subsection{Archival data results}\label{subsec:additional-results}

\begin{figure}[ht!]
    \centering
    \includegraphics[width=0.8\textwidth]{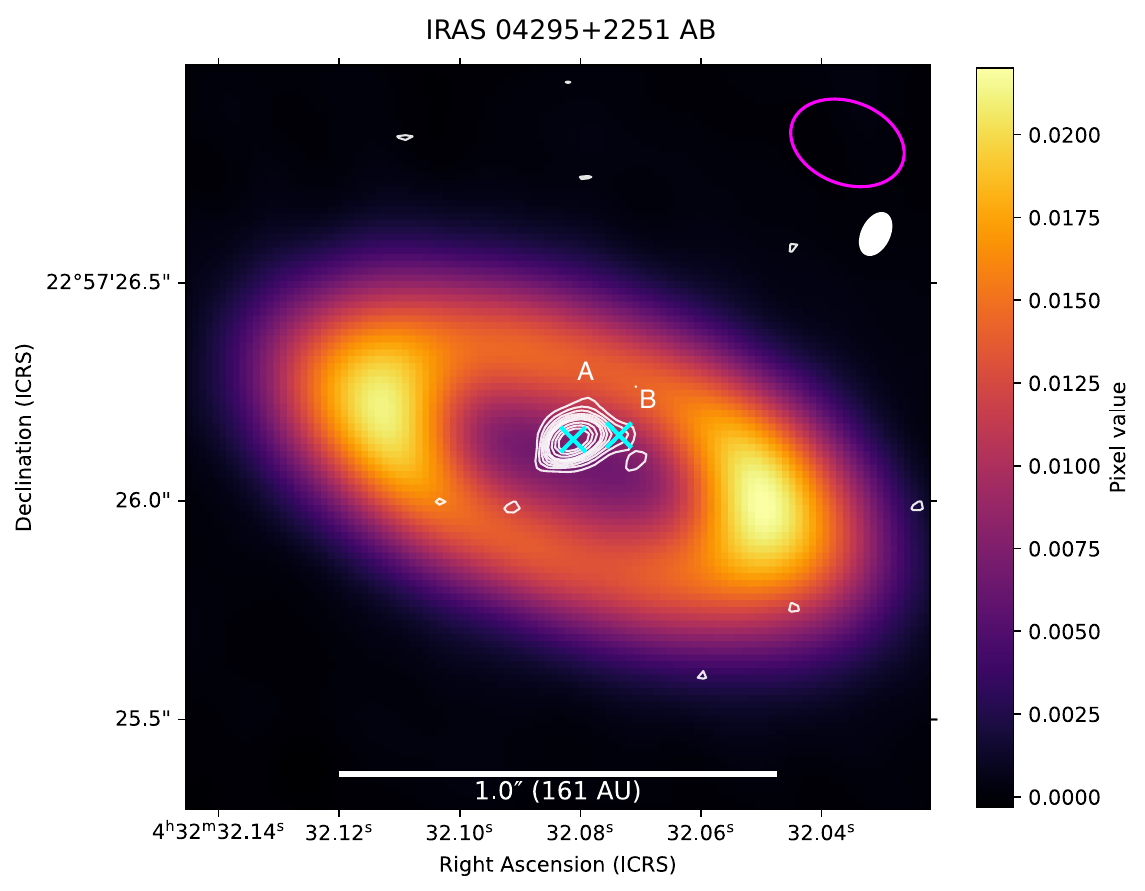}
    \caption{ALMA 0.9 mm continuum image of the disk/circumbinary structure of IRAS 04295+2251 AB, shown with archival VLA 9 mm continuum contours overlaid to search for compact emission inside the central cavity. The ALMA emission shows a ring-like structure with a large central cavity, while the higher-resolution VLA contours reveal compact emission near the cavity center, consistent with the proposed location of an embedded companion. Cyan crosses mark the positions of the two proposed protostars, which are separated by 18~au. The magenta ellipse in the upper-right corner shows the synthesized beam of the ALMA image, and the filled white ellipse below it shows the synthesized beam of the VLA data. VLA contours are drawn from $3\sigma$ to $30\sigma$ in steps of $3\sigma$, where $\sigma = 10.5~\mu\mathrm{Jy~beam}^{-1}$ is the map rms.}
    \label{fig:04295-image}
\end{figure}

The image from the combined VLA B and A-configuration of IRAS 04295+2251 (Figure \ref{fig:04295-image}) shows a prominent point source and a faint extension to the west, this extension could be a secondary source that is becoming marginally-resolved with the higher resolution. The proper motion correction of the source position to match the 2022 July observation puts the source within the observed central cavity of the disk detected with ALMA at 0.9 mm.

We performed the following to assess the robustness of this extension being a secondary source within the central cavity of the disk around IRAS 04295+2251. We first fitted single Gaussian, with a size and orientation identical to the synthesized beam, to the main source. Subtracting this fit leaves a residual that has a peak of 61.2~$\mu$Jy to the west, and this source has a significance of 5.8$\sigma$, relative to the noise level of 10.5~$\mu$Jy. When fitted together as two Gaussians, the primary has a peak of 477$\mu$Jy and the secondary has a peak of 75.8~$\mu$Jy, with uncertainties on these peaks of 12.8~$\mu$Jy; thus the significance of the secondary peak when fitted is 5.9$\sigma$. We note that there are other noise peaks near the source that have peaks of 40.9~$\mu$Jy~beam$^{-1}$ (to the west) and 36.3~$\mu$Jy~beam$^{-1}$ (to the southeast), but both of these are below 4$\sigma$ significance. Thus, while this secondary component might have a significance above 5$\sigma$, because of its blending and the fact that it stands out most after subtracting the main source, we only regard this as a potential companion, and include it in the Taurus+ sample. However, additional observations with longer integration time, better $uv$-coverage, and higher resolution will be able to confirm (or reject) the presence of this putative companion.

The presence of a companion source would offer a natural explanation for the presence of such a large cavity in the disk of IRAS 04295+2251, similar to the large cavity in the disk surrounding the binary source IRAS 04158+2805 and more widely separated sources like L1551NE and IRAS 04287+1801 (L1551 IRS5) that have rings surrounding their companions. The lack of detection by ALMA in the current observations may be due to blending with the surrounding disk and the fact that the circumstellar disks around the protostars may be very small limiting their brightnesses.

\begin{figure}[ht!]
    \centering
    \includegraphics[width=0.8\textwidth]{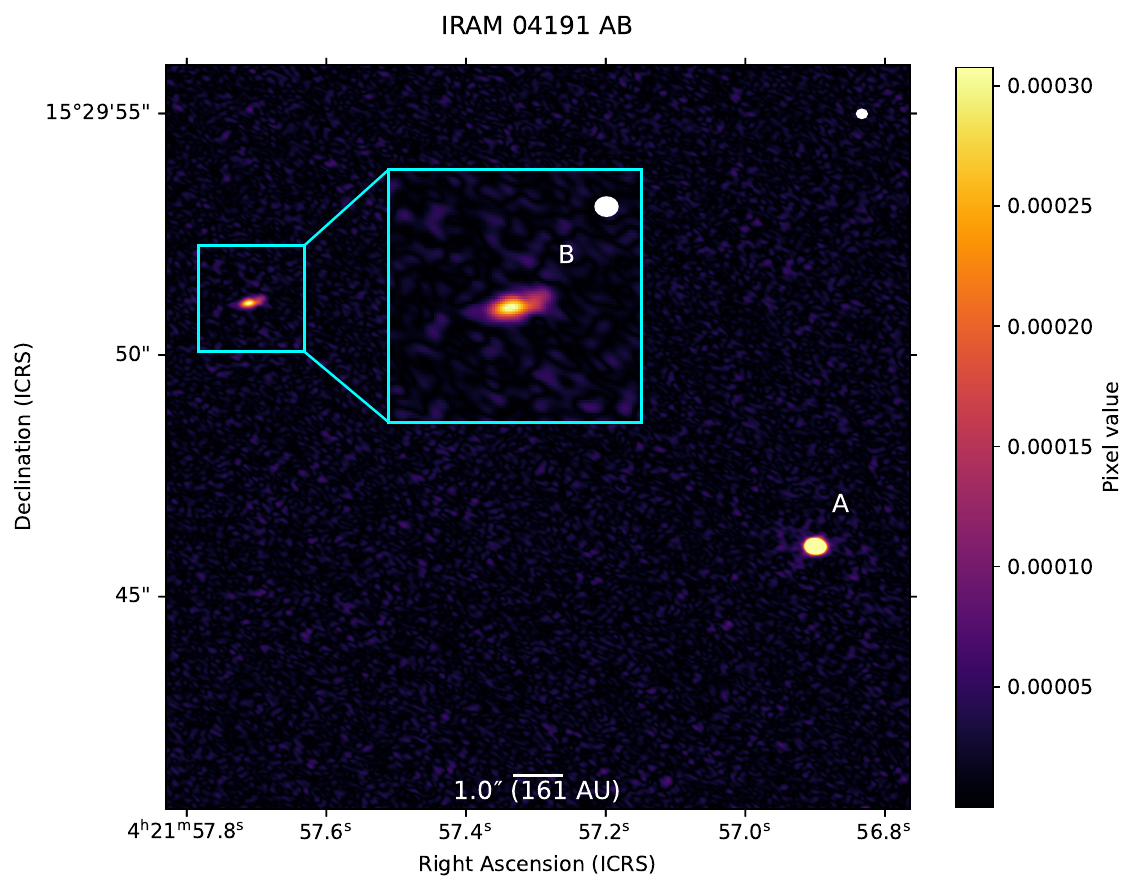}
    \caption{ALMA 1.3 mm continuum image of IRAM 04191 AB with a synthesized beam corresponding to a resolution of 0\farcs154. The source on the lower right is IRAM 04191 A which was detected in our original Band 7 observations (Figure \ref{fig:alma-observations}), while the source in the upper left is a new detection, not previously reported, which we designate as IRAM 04191 B. A zoomed-in view of this new source is shown in the cyan box.}
    \label{fig:04191-image}
\end{figure}

In the archival ALMA Band 6 data, we identify a secondary continuum source, hereafter IRAM 04191 B, located $\sim$12\farcs7 ($\sim$1800 au) northeast of IRAM 04191 A. This source has not been reported in previous studies of the system \citep[e.g.,][]{Andre1999,Chen2012,Kim2016,Kim2019,Luhman2010}. IRAM 04191 B is detected at a peak intensity of $\sim$3.18 mJy beam$^{-1}$ with an rms noise level of $\sim$0.22 mJy beam$^{-1}$. Because no infrared counterpart has been identified for this source, we adopt the Class 0 designation of IRAM 04191 A. As in the Band 7 ALMA observations (Figure \ref{fig:alma-observations}), IRAM 04191 A appears as a compact point source. In the zoomed-in view of IRAM 04191 B, we detect faint extended emission that may trace an edge-on disk around this protostar (Figure \ref{fig:04191-image}).

\section{Analysis Methods} \label{sec:data-analys}
Source identification and flux measurements were performed using the CASA \texttt{imfit} task, which fits two-dimensional elliptical Gaussians to the continuum emission. For some systems with complex or blended structures, multiple Gaussians were fit to better capture the morphology.
To characterize multiplicity across the sample, we created a merged source catalog incorporating detections from both ALMA and VLA. This approach helps account for non-detections from either observatory and ensures a more complete census of companions.
We also assembled photometry and fit SEDs for each target to derive bolometric quantities and evolutionary classes. A full description of this process and the resulting SEDs are provided in Appendix \ref{appendix:source_seds}.

\subsection{Associating Multiple Systems} \label{subsec:assoc-mult-sys}

To study multiplicity, we must first identify which sources are physically associated as part of the same system. We do this using spatial association techniques that group sources based on their projected separations on the sky. 
We adopt the iterative “modified nearest neighbor” method introduced in \citet{Tobin2022}, which builds multiple systems by first identifying the closest pair of sources within a minimum threshold of 15~au, which is smaller than the separation of any source in our sample, chosen conservatively to ensure no close pairs are missed. When a pair is found, the algorithm computes their geometric average position, replaces the pair with that average, and repeats the search for additional companions relative to the new location. The original sources are removed from the catalog after each association to ensure that no source is linked to more than one system. This procedure is performed across 100 logarithmically spaced radial bins, with a maximum separation of 10,000~au.
We adopt 10,000~au as our maximum separation limit because it approximately corresponds to the typical radius of dense cores in which protostars form \citep{Benson1989,Bergin2007,Lane2016,Kirk2017}.

A key advantage of this method over more traditional approaches, which typically define a primary source and apply a fixed radial cutoff, is that it can recover more complex and extended systems. This method also avoids the need to designate one component of a system as the “primary.” Such designations are problematic, as the brightest or most luminous protostar at infrared, millimeter, and/or centimeter wavelengths is not always the most massive. Protostellar luminosity can be dominated by accretion processes rather than intrinsic stellar properties, and dust continuum emission (our main observable) does not directly correlate with stellar mass \citep{Dunham2014,Fischer2017}.

\subsection{Multiplicity Statistics} \label{subsec:mult-stats1}

To characterize the multiplicity of protostars in Taurus, we use two main metrics: the multiplicity fraction (MF) and the companion fraction (CF).

The MF is the fraction of stars that are part of multiple systems (i.e., have at least one companion within 10,000 au), and is calculated by adding up the number of binaries (B), triples (T), quadruples (Q), etc, and dividing by the total number of systems ($N_\mathrm{sys}$), including the single stars as given by
\begin{equation}
MF = \frac{B + T + Q + \dots}{S + B + T + Q + \dots}.
\end{equation}

The CF, which represents the number of companions each source has on average, is given by
\begin{equation}
CF = \frac{B + 2T + 3Q + \dots}{S + B + T + Q + \dots}.
\end{equation}

To estimate the uncertainty in this analysis, we compute Wilson score intervals \citep{Wilson1927}, a binomial statistics method, for all MFs and CFs across different separation ranges, given by

\begin{equation}
\sigma_\mathrm{MF} = \frac{1}{1 + \frac{z^2}{N_\mathrm{sys}}} \left( \mathrm{MF} + \frac{z^2}{2N_\mathrm{sys}} \right) \pm \frac{z}{1 + \frac{z^2}{N_\mathrm{sys}}} \sqrt{\frac{MF \left( 1.0-MF \right)}{N_\mathrm{sys}} + \frac{z^2}{4N_\mathrm{sys}^2}}.
\end{equation}

This shows the formula for calculating the uncertainty in MF and can be applied analogously to CF by replacing MF with CF, where z is the standard normal deviate corresponding to the desired confidence level.
Using z=1.0, which corresponds to a 68\% confidence interval (i.e., 1$\sigma$), yields uncertainties consistent with standard 1-sigma error estimates.
This method provides asymmetric confidence intervals for binomial proportions, which is especially important when the number of systems ($N_{\mathrm{sys}}$) is small or the proportion is near 0 or 1, where normal approximations become unreliable. 

Because the protostellar populations in all regions considered here are well cataloged, our samples represent a substantial fraction of the total number of systems. When this is the case, treating the sample as if it were drawn from an effectively infinite population exaggerates the sampling uncertainty.
We therefore incorporate a finite-population correction (FPC; \citealp{Oneill2021}) into the Wilson interval. The FPC modifies the variance of a binomial proportion by a factor $f$, where

\begin{equation}
    f=\sqrt{\frac{N-n}{N-1}}
\end{equation}

where $n$ is the sample size and $N$ is the estimated total population of protostars in the region. Inside the Wilson formula, this simply replaces the normal deviate $z$ with a smaller effective value $z_\mathrm{eff}= zf$. Thus the  structure of the interval remains the same, but the error bars become narrower in proportion to the sample’s completeness.

We adopt a conservative completeness of 75\% when estimating $N$ for Orion, Perseus, and the Taurus+ sample. This results in smaller uncertainties for MFs and CFs of Orion and Perseus reported here because \citet{Tobin2022} did not use the FPC. For the 
Taurus sample, we use the same $N$ inferred from Taurus+, which implies a completeness of roughly 45\%. This choice yields more realistic—and likely still slightly conservative—uncertainty estimates compared to using the uncorrected Wilson interval.
This completeness accounts for protostars that are undetected in infrared surveys because they are too cool or faint, as well as objects that may be misclassified due to degeneracies inherent in assigning evolutionary classes from broadband SEDs. This also accounts for the 5 systems that have not yet been observed at sub-arcsecond resolution, for which close companions may remain unresolved. 

\subsection{Comparing Multiplicity Properties} \label{subsec:comp-mult-props}

MF and CF offer a general overview of how common multiples are in a given region. These statistics can be calculated for all systems or within specified separation ranges to examine how multiplicity varies with separation.
Since protostellar systems may contain a mix of evolutionary classes, especially at wider separations (see Section~\ref{subsec:evolution}), we include certain Class~II sources when they are apparent companions to  Class~0, I, or Flat Spectrum protostars. 

To explore more detailed trends, we analyze the distribution of companion separations using histograms and cumulative distribution functions (CDFs). These allow us to visually and statistically compare different regions. CDFs are particularly useful for revealing how companions are distributed across a range of separations: steep slopes indicate that companions are clustered at smaller separations, while flatter slopes suggest a more even spread.

In Taurus, we construct standard CDFs since no correction for background YSO density is applied. A CDF is a curve that starts at 0 and rises to 1, where the x-axis is projected separation (in~au), and the y-axis represents the fraction of systems with separations less than or equal to a given value. The empirical CDF is defined as
\begin{equation}
\mathrm{CDF}(d_n) = \frac{n}{N}, \quad \text{for } n = 1, 2, \ldots, N,
\end{equation}

where $N$ is the total number of companion systems and $d_{n}$ is the $n$th smallest projected separation.

To quantitatively compare separation distributions between regions, we use the Kolmogorov–Smirnov (KS) test and the Anderson-Darling (AD) test on the CDFs of companion separations, and we report results from both tests. The KS test measures the maximum difference between two cumulative distributions and is most sensitive around the median, while the AD test gives more weight to the distribution tails.
We compare the Taurus and Taurus+ samples with Orion, Perseus, and the more evolved Taurus protostars (low- and solar-mass) from \citet{Kraus2011}, as well as with benchmark distributions including a log-flat model and the solar-type field stars from \citet{Raghavan2010}.
These statistical tests evaluate the null hypothesis that two samples are drawn from the same parent population. If the resulting p-value is below 0.05, we reject the null hypothesis and conclude that the distributions are statistically distinct.

While MF and CF statistics and CDFs all provide insight into multiplicity, they probe different aspects of the population and can yield different conclusions. For example, two regions might have similar MF and CF values but different shaped separation distributions—or vice versa. Therefore, using both approaches allows for a more complete understanding of how protostellar multiplicity varies across environments.

\section{Multiplicity characterization} \label{sec:mult-charac}

\subsection{Bolometric Luminosities and Temperatures} \label{subsec:lbol-tbol}

\begin{figure*}[ht!]
\centering
\includegraphics[width=0.495\textwidth]{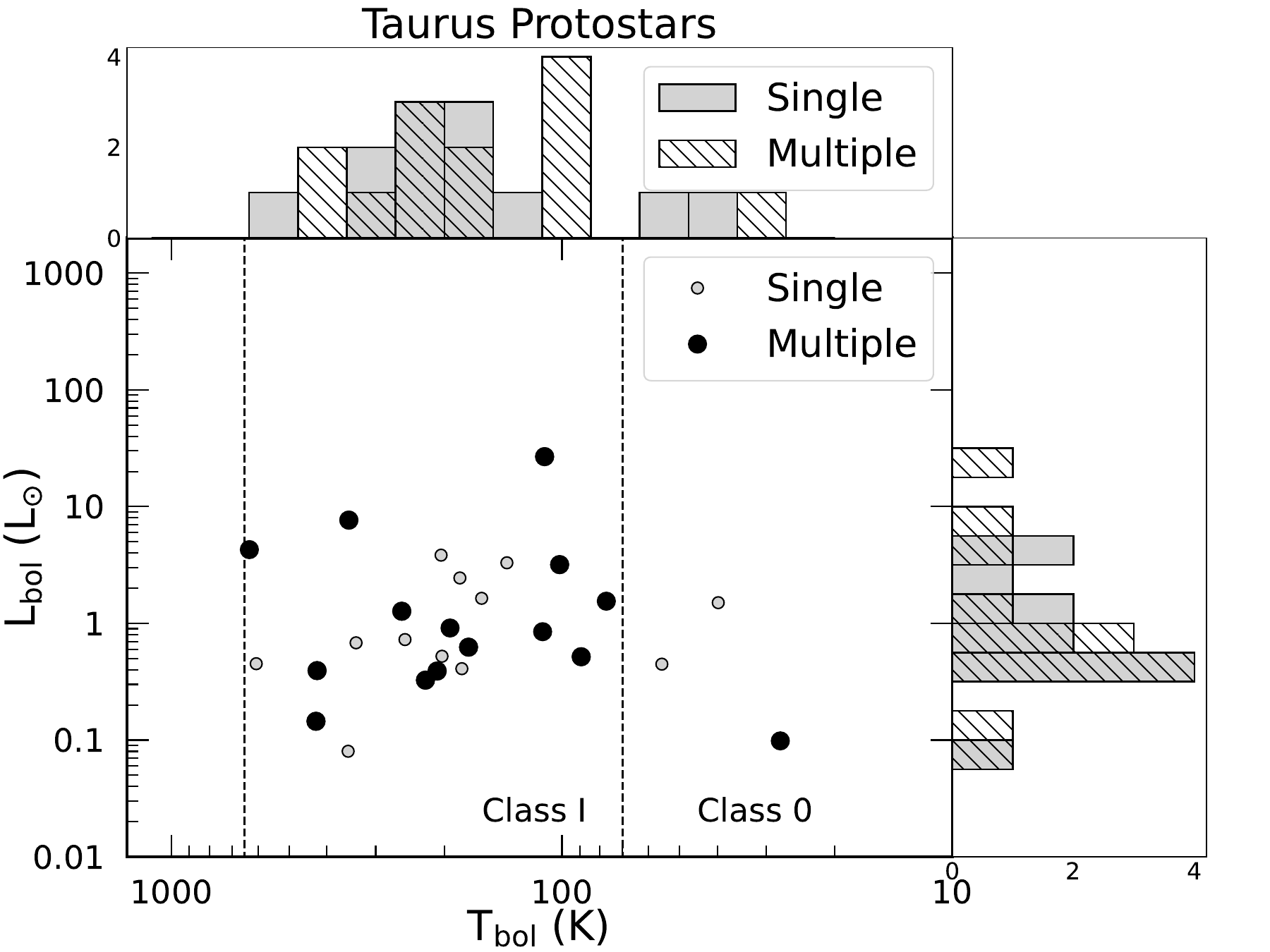}
\includegraphics[width=0.495\textwidth]{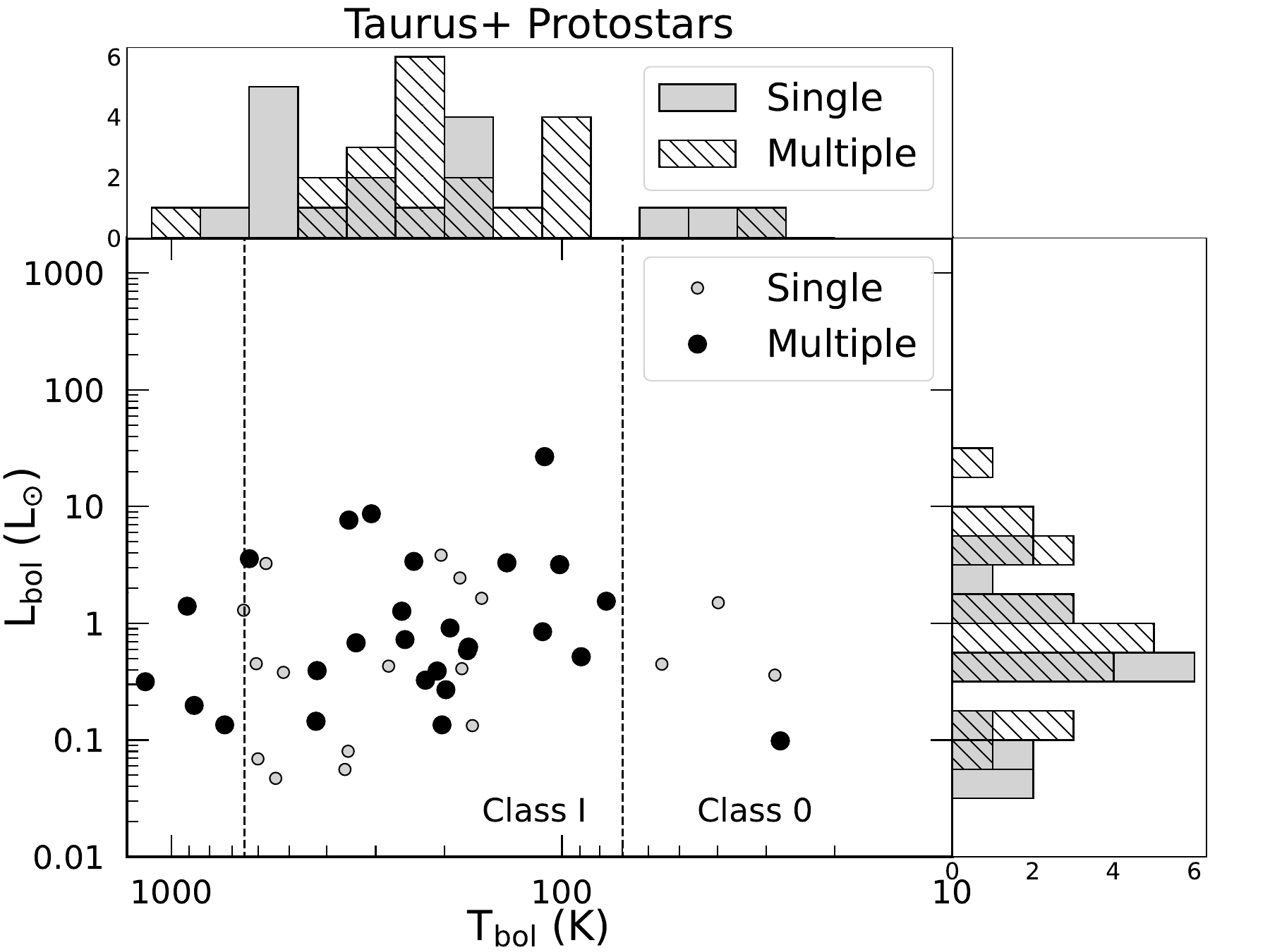}
\caption{Bolometric luminosity (\lbol) versus bolometric temperature (\tbol) for protostars in Taurus. The left panel shows the Taurus sample, and the right panel includes sources from the Taurus+ sample. In both plots, small gray points represent single protostar systems, while larger black points indicate components of multiple systems. Some sources that were classified as single in the Taurus sample are resolved into multiple systems in the Taurus+ sample and therefore appear as larger points in the right panel. Histograms along the top and right axes show the distributions of \tbol{} and \lbol{} for singles (gray fill) and multiples (hatched). Vertical dashed lines mark the boundaries between  Class~0 and  Class~I (\tbol $= 70$ K) and between  Class~I and  Class~II ($T_\mathrm{bol} = 650$ K). Flat Spectrum protostars can appear throughout the  Class~I range and sometimes into  Class~II as that classification is based on the spectral index from 2-24 \micron\ and not \tbol.}
\label{fig:lbol-tbol-plots}
\end{figure*}

Figure \ref{fig:lbol-tbol-plots} shows \lbol{} and \tbol{} of single and multiple protostar systems in Taurus. This plot highlights multiple systems with separations ranging from 18 to 10,000~au. For multiples within $\sim$1000 au separations, where components share an infrared SED and thus are assigned identical \lbol{} and \tbol{} values, only one point is plotted per system.
In contrast, wider multiples and higher-order systems are plotted with one point per component when the sources have distinct infrared fluxes.

Histograms along the top and right axes of the scatter plot show the distributions of \tbol{} and \lbol, respectively.
We compare the \lbol{} distributions of single and multiple systems in our samples. The median \lbol{} for multiples includes both components with individually resolved SEDs and components in unresolved systems that share a blended SED. For the latter, the cataloged system-level \lbol{} is assigned to each component.
In the Taurus sample, single protostars had a median \lbol{} of 0.70 $L_\odot$, while multiples had a median of 0.85 $L_\odot$. In the Taurus+ sample, the medians were 0.43 $L_\odot$ for singles and 0.70 $L_\odot$ for multiples. Thus, multiple systems have somewhat higher median \lbol{} values than single systems in both samples. However, the shapes of the distributions are quite coarse due to the small sample size. The \tbol{} distributions appear less correlated, with no clear similarities in shape between singles and multiples. These distributions show that our sample is dominated by  Class~I protostars and not particularly biased toward very young or very evolved protostars.

To assess the significance of these differences, we performed a KS test comparing the \lbol{} distributions of singles and multiples. We could not reject the null hypothesis that the bolometric luminosity distributions of single and multiple systems are drawn from the same parent population for either sample, with $p = 0.50$ for the Taurus sample and $p = 0.19$ for the Taurus+ sample.

In comparison, Orion and Perseus show stronger trends. In both regions, median luminosities of multiple systems are roughly three times higher than those of single systems ($\sim$3 $L_\odot$ vs. $\sim$1 $L_\odot$), and the KS test for Orion yielded a $p$-value $<$ 0.01, suggesting statistically significant differences. For Perseus, the result is more marginal, with a $p$-value of 0.025, just above the threshold for significance. The \tbol{} distributions for single and multiple protostars are statistically consistent with being drawn from the same parent population in Perseus.

The expectation that multiple protostars should be more luminous arises in part because accretion onto several components releases more total energy than accretion onto a single star at the same rate \citep{Tobin2022}. Multiplicity may also correlate with initial core mass, since more massive cores are more prone to fragmentation and capable of forming brighter systems \citep{Duchene2013}. In addition, higher-mass protostars themselves are more likely to be in multiple systems, which could further augment the luminosities observed in multiples. 
The link between stellar mass and multiplicity is well documented \citep{Duchene2013}. While some of these higher-mass main-sequence stars may have acquired companions via capture, \citet{Beuther2025} describe how disks around massive stars are more prone to instability and fragmentation, contributing to their higher multiplicity rates.
In Taurus, we observe only a slight skew toward higher luminosities in multiples, and that trend is not statistically significant in the Taurus+ sample and not observed at all in the Taurus sample. 

\subsection{Multiplicity Statistics} \label{subsec:mult-stats}

Table~\ref{tab:multiplicity_stats} presents the multiplicity statistics for the Taurus and Taurus+ samples, and compares them to the star-forming regions of Orion and Perseus. Although the Perseus and Orion observations have a larger field of view than our Taurus observations, we adopt a common upper separation limit of 10,000~au for all regions (including Taurus+), since pairs wider than 10,000~au are unlikely to be gravitationally bound and have little impact on the measured multiplicity statistics. The spatial resolution of each survey is comparable and companions are detected down to $\sim18$~au in all regions, resulting in similar sensitivity to close multiples across regions.

Across separations of 18–10,000~au, we measure MF $= 0.50 \pm 0.07$ and CF $= 0.58 \pm 0.20$ for the observed Taurus sample, and MF $= 0.53 \pm 0.06$ and CF $= 0.72 \pm 0.19$ for the Taurus+ sample. The MF in Taurus is significantly higher than in Orion (MF $= 0.29 \pm 0.01$) at the $\sim3\sigma$ level and moderately higher than in Perseus (MF $= 0.36 \pm 0.04$) at the $\sim1.7\sigma$ level, while the MF in Taurus+ exceeds that of Orion and Perseus at the $\sim4\sigma$ and $\sim2.4\sigma$ levels, respectively. In contrast, differences in CF are less statistically significant where Taurus exceeds Orion and Perseus at only the $\sim0.8\sigma$ and $\sim0.3\sigma$ levels, while Taurus+ exceeds Orion and Perseus at the $\sim1.6\sigma$ and $\sim1\sigma$ levels, respectively. These results indicate that a larger fraction of systems in Taurus form and survive as multiple systems relative to more clustered environments, while the total number of companions per system is less significantly differentiated across regions.

However, when we restrict the comparison to wider separations of 500--10,000~au, the MF in Taurus+ decreases substantially and is only marginally higher than in Orion and Perseus. The CF shows a somewhat larger difference in this separation regime, with Taurus+ having CF $=0.29 \pm 0.04$, compared to CF $=0.21 \pm 0.01$ in Orion and CF $=0.21 \pm 0.03$ in Perseus. These correspond to differences of $\sim1.9\sigma$ and $\sim1.6\sigma$, respectively.

This suggests that the elevated multiplicity fraction in Taurus is strongest when close companions are included, rather than being driven solely by an excess of wide systems. Because close companions are probed at comparable physical resolutions in Taurus, Orion, and Perseus, this difference is not due to resolution effects.

The CF in the observed Taurus sample closely tracks the MF across all separation ranges, implying that most multiple systems in the observed sample are simple binaries rather than higher-order multiples. An exception to this trend appears in the Taurus+ sample at wide separations, which includes one triple system and one quintuple system entirely within Taurus+, along with a quadruple system that includes a Taurus+ component.

\begin{deluxetable}{lclll}
\tabletypesize{\scriptsize}
\tablewidth{0pt}
\tablecaption{Multiplicity Statistics}
\label{tab:multiplicity_stats}
\tablehead{\colhead{Region} & \colhead{Separation Range (au)} & \colhead{Counts} & \colhead{MF} & \colhead{CF}}
\startdata
\hline
\rowcolor{lightgray}
Taurus  & 18 - 10,000 & 12:10:2:0:0:0:0:0:0:0 & 0.50$_{-0.07}^{+0.07}$ & 0.58$_{-0.20}^{+0.20}$  \\
\rowcolor{lightgray}
Taurus+ & 18 - 10,000 & 17:15:2:1:1:0:0:0:0:0 & 0.53$_{-0.06}^{+0.06}$ & 0.72$_{-0.19}^{+0.19}$  \\
Orion  & 18 - 10,000 & 217:69:11:2:4:1:0:0:0:1 & 0.29$_{-0.01}^{+0.01}$ & 0.42$_{-0.02}^{+0.02}$  \\
Perseus & 18 - 10,000 & 45:18:4:2:1:0:0:0:0:0 & 0.36$_{-0.04}^{+0.04}$ & 0.51$_{-0.11}^{+0.11}$  \\
\hline
\rowcolor{lightgray}
Taurus  & 18 - 1,000 & 18:10:0:0:0:0:0:0:0:0 & 0.36$_{-0.06}^{+0.07}$ & 0.36$_{-0.06}^{+0.07}$  \\
\rowcolor{lightgray}
Taurus+ & 18 - 1,000 & 36:13:0:0:0:0:0:0:0:0 & 0.27$_{-0.04}^{+0.04}$ & 0.27$_{-0.04}^{+0.04}$  \\
Orion & 18 - 1,000 & 306:57:4:0:0:0:0:0:0:0 & 0.17$_{-0.01}^{+0.01}$ & 0.18$_{-0.01}^{+0.01}$  \\
Perseus & 18 - 1,000 & 63:20:1:0:0:0:0:0:0:0 & 0.25$_{-0.03}^{+0.03}$ & 0.26$_{-0.03}^{+0.03}$  \\
\hline
\rowcolor{lightgray}
Taurus & 18 - 500 & 20:9:0:0:0:0:0:0:0:0 & 0.31$_{-0.06}^{+0.06}$ & 0.31$_{-0.06}^{+0.06}$  \\
\rowcolor{lightgray}
Taurus+ & 18 - 500 & 38:12:0:0:0:0:0:0:0:0 & 0.24$_{-0.04}^{+0.04}$ & 0.24$_{-0.04}^{+0.04}$  \\
Orion & 18 - 500 & 330:48:2:0:0:0:0:0:0:0 & 0.13$_{-0.01}^{+0.01}$ & 0.14$_{-0.01}^{+0.01}$  \\
Perseus & 18 - 500 & 71:16:1:0:0:0:0:0:0:0 & 0.19$_{-0.03}^{+0.03}$ & 0.20$_{-0.03}^{+0.03}$  \\
\hline
\rowcolor{lightgray}
Taurus & 500 - 10,000 & 21:1:2:0:0:0:0:0:0:0 & 0.09$_{-0.04}^{+0.05}$ & 0.15$_{-0.04}^{+0.05}$  \\
\rowcolor{lightgray}
Taurus+ & 500 - 10,000 & 29:3:2:1:1:0:0:0:0:0 & 0.15$_{-0.03}^{+0.04}$ & 0.29$_{-0.04}^{+0.04}$  \\
Orion & 500 - 10,000 & 267:21:9:2:4:1:0:0:0:1 & 0.11$_{-0.01}^{+0.01}$ & 0.21$_{-0.01}^{+0.01}$  \\
Perseus & 500 - 10,000 & 62:2:3:2:1:0:0:0:0:0 & 0.09$_{-0.02}^{+0.02}$ & 0.21$_{-0.03}^{+0.03}$  \\
\hline
\enddata
\tablecomments{This table presents the multiplicity fraction (MF) and companion fraction (CF) for each region and separation range. Grey rows highlight the Taurus and Taurus+ samples analyzed in this work, while the Orion and Perseus values are from \citet{Tobin2022} and \citet{Tobin2016}. Counts are listed by system multiplicity in the order $N_1:N_2:N_3:N_4:N_5:N_6:N_7:N_8:N_9:N_{10}$, where $N_1$ is the number of single systems, $N_2$ is the number of binary systems, and so on.}
\end{deluxetable} 

\subsection{Separation Distributions} \label{subsec:sep-dist}

We also characterize the typical projected separations between protostellar companions and explore how these distributions may reflect underlying star formation processes. To do this, we analyze the full set of projected separations measured for multiple systems (see Table \ref{tab:companion-table}), using the methodology outlined in Section \ref{sec:data-analys}. We visualize the separation distributions using both histograms and cumulative distribution functions (CDFs) to investigate more subtle differences between populations. 
Histograms are binned uniformly in log space, with a bin width of 0.25 dex in projected companion separation.
While in principle the distributions could be broken down by evolutionary class, we include all Class~0, I, Flat Spectrum, and II sources together for this analysis because of the limited sample size in Taurus.

\subsubsection{Taurus Samples} \label{subsubsec:taurus-samp}

\begin{figure*}[ht!]
\centering
\includegraphics[width=0.495\textwidth]{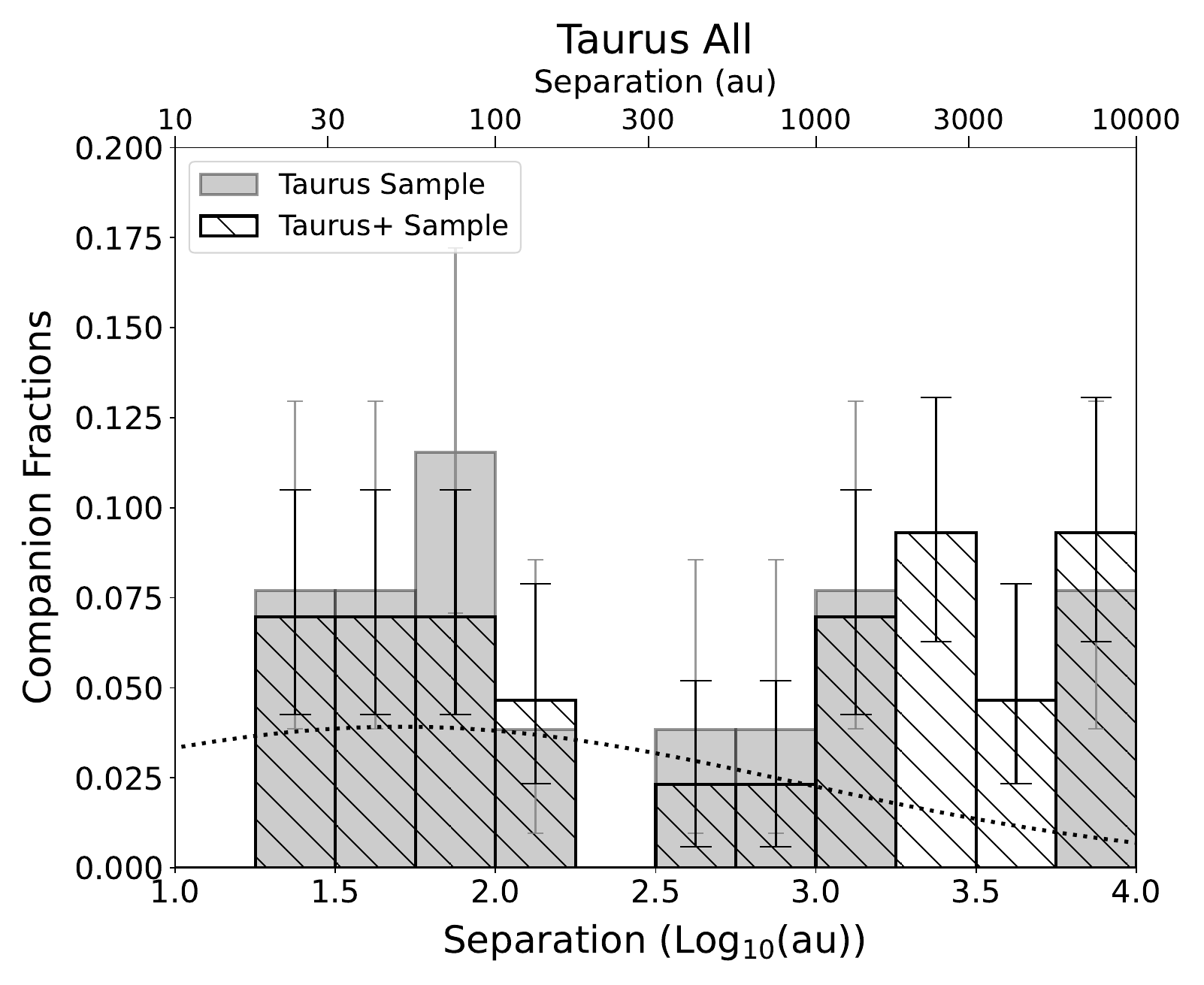}
\includegraphics[width=0.495\textwidth]{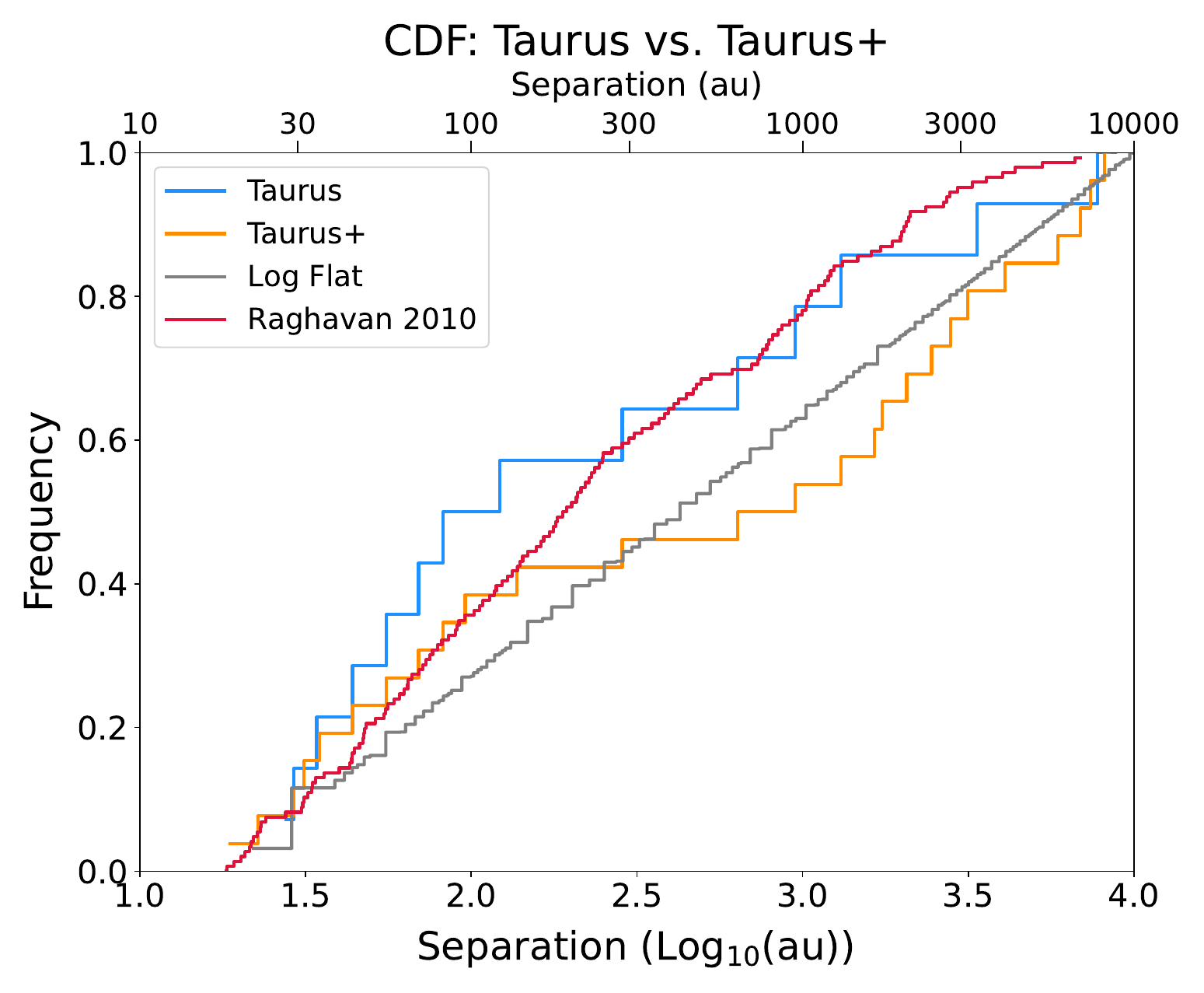}
\caption{The panel on the left shows histograms of companion separation distributions for the observed Taurus sample (gray) and the Taurus+ sample (hatched). The Gaussian fits of the separation distribution of solar-type main sequence field stars from \citet{Raghavan2010} is also included as a dotted line for comparison. Each bin’s error bar is calculated based on binomial statistics, following the method outlined in Section \ref{subsec:mult-stats1}. The right panel displays cumulative distribution functions of separations, plotted in terms of companion frequency, for Class~0/I/FS protostars in the Taurus and Taurus+ samples. For reference, a log-flat distribution is shown in gray, and the field star distribution from \citet{Raghavan2010} is overlaid in red.}
\label{fig:Taurus-histo-cdf}
\end{figure*}

We first compare the separation distributions of the observed Taurus sample and the Taurus+ sample, as shown in Figure \ref{fig:Taurus-histo-cdf}. In the Taurus sample, the distribution peaks at $\sim$75~au. There is a noticeable drop in the number of systems around 200~au, and a broader gap in the distribution around 3000~au.

In the Taurus+ sample, the $\sim$75~au peak persists, though it appears slightly less pronounced. The gap near 200~au also remains, and there are fewer multiples between 200-1000~au.
However, the Taurus+ sample reveals additional structure, with rising multiplicity toward 10,000~au. This rise at the widest separations is likely driven primarily by the Taurus sample, but the inclusion of additional sources in the Taurus+ sample helps fill in the gap around 3000~au noted in the Taurus distribution. The result is a broader and more continuous distribution for Taurus+, suggesting a population of wide companions that is similar in number to the $<$300~au multiples.

The cumulative distribution functions in Figure \ref{fig:Taurus-histo-cdf} highlight these differences. The observed Taurus sample has a steeper rise at separations $<$300~au and then becomes more shallow at wider separations. In other words, the Taurus sample is more heavily weighted toward close multiples. The flattening of both CDFs near 200–300~au reflects the lack of systems at intermediate separations, while steeper slopes below and above that range indicate some clustering at small and wide separations, respectively. These trends are similar in the Taurus+ sample, but the CDF only reaches 0.5 at 1000~au reflecting the greater contribution to wide multiples in the Taurus+ sample.

\subsubsection{Orion \& Perseus Comparisons} \label{subsubsec:ori-per-comp}

\begin{figure*}[ht!]
\centering
\includegraphics[width=0.95\textwidth]{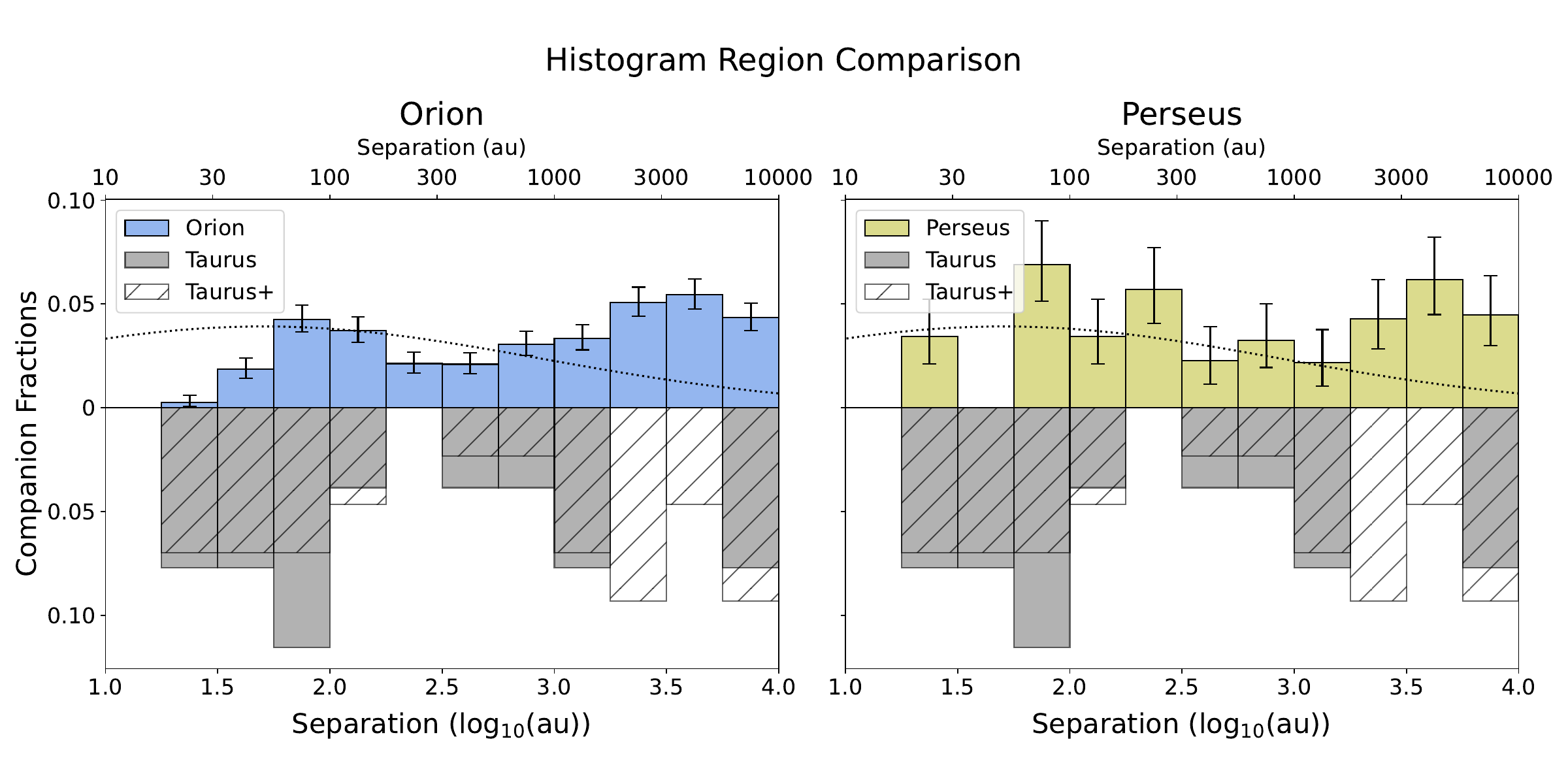}\\[3pt] 
\includegraphics[width=0.5\textwidth]{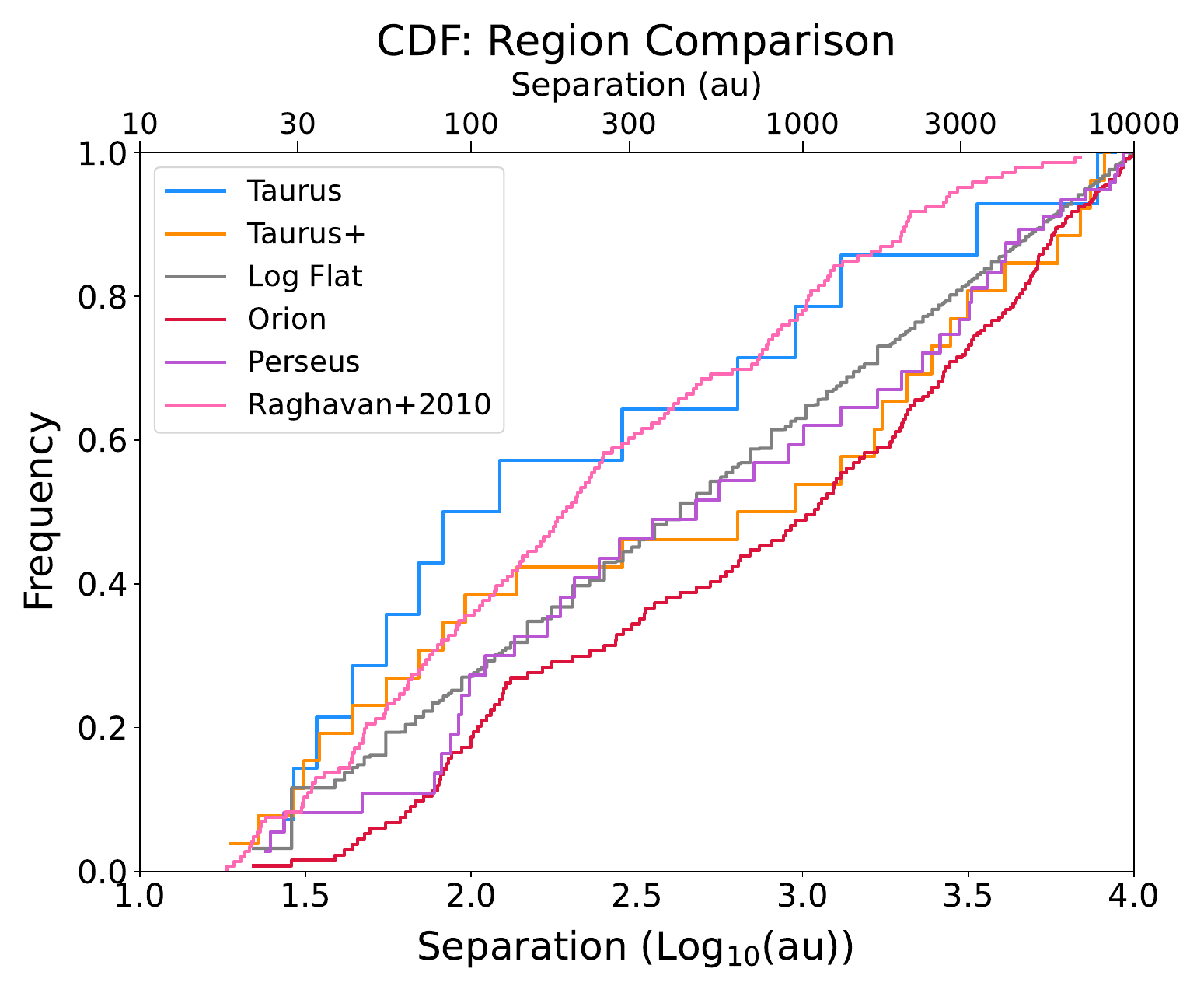} 
\caption{Mirrored histograms (top panels) show the companion separation distributions for Orion (left, blue) and Perseus (right, green). The corresponding Taurus and Taurus+ distributions are mirrored below each. For Orion and Perseus, we adopt the contamination-corrected samples. The dotted line indicates the distribution of solar-type main-sequence field stars from \citet{Raghavan2010}. The bottom panel presents the CDFs of companion separations, expressed in terms of companion frequency, for protostars in Taurus, Orion, and Perseus. The red curve represents the CDF for solar-type field stars from \citet{Raghavan2010}, while the gray curve shows a reference log-flat distribution.}
\label{fig:ori-per_histo_cdf}
\end{figure*}

We compare the Taurus separation distributions to those of Orion and Perseus in Figure \ref{fig:ori-per_histo_cdf} (histograms and CDFs). Both Orion and Perseus were corrected for YSO density, and the multiple systems were weighted by the probability that each companion was real. This correction primarily affected systems at large separations but had little impact on close companions. For comparison with Taurus, we used the contamination corrected probabilities documented for Orion and Perseus in \citet{Tobin2022}.

Both Orion and Perseus exhibit a characteristic double-peaked structure in their histograms (top panels, Figure \ref{fig:ori-per_histo_cdf}), with a peak near 75~au and a second broader rise near 4000~au. These peaks are separated by a local minimum around 300~au, giving the appearance of a bimodal separation distribution. This bimodality is more pronounced in Orion, but still apparent in Perseus. In contrast, the Taurus sample is dominated by a single peak at close separations ($\sim$75~au) and sparsely populates the wide-separations, making it difficult to identify any significant structure beyond a few hundred~au. However, the Taurus+ sample fills in the gap near 3000~au, potentially revealing a broader bimodal distribution more reminiscent of Orion and Perseus.

\citet{Tobin2022} showed that in Orion and Perseus, this bimodality is largely driven by  Class~0 systems, whereas more evolved  Class~I and Flat Spectrum sources exhibit flatter separation distributions. Taurus, which includes only five Class~0 sources (just two of which comprise a single binary), may therefore lack the strong bimodal signal seen in Orion. 

The CDFs displayed in the bottom panel of Figure \ref{fig:ori-per_histo_cdf} offer additional insight by showing the companion frequency per system as a function of separation. At separations $\leq 300 \mathrm{au}$, the Taurus sample has the steepest rise—about 60\% of its companions lie within this range—indicating a strong concentration of close binaries. Taurus+ and Perseus both reach $\sim$40\% by 300~au, while Orion remains the shallowest, with only $\sim$30\% of companions at close separations. Notably, Orion is the only region in which wide companions outnumber close ones, explaining why its CDF remains lowest across the entire separation range.

\subsubsection{Comparison with older Taurus Population} \label{subsubsec:kraus-comp}

\begin{figure}[H]
    \centering
    \includegraphics[width=1.0\textwidth]{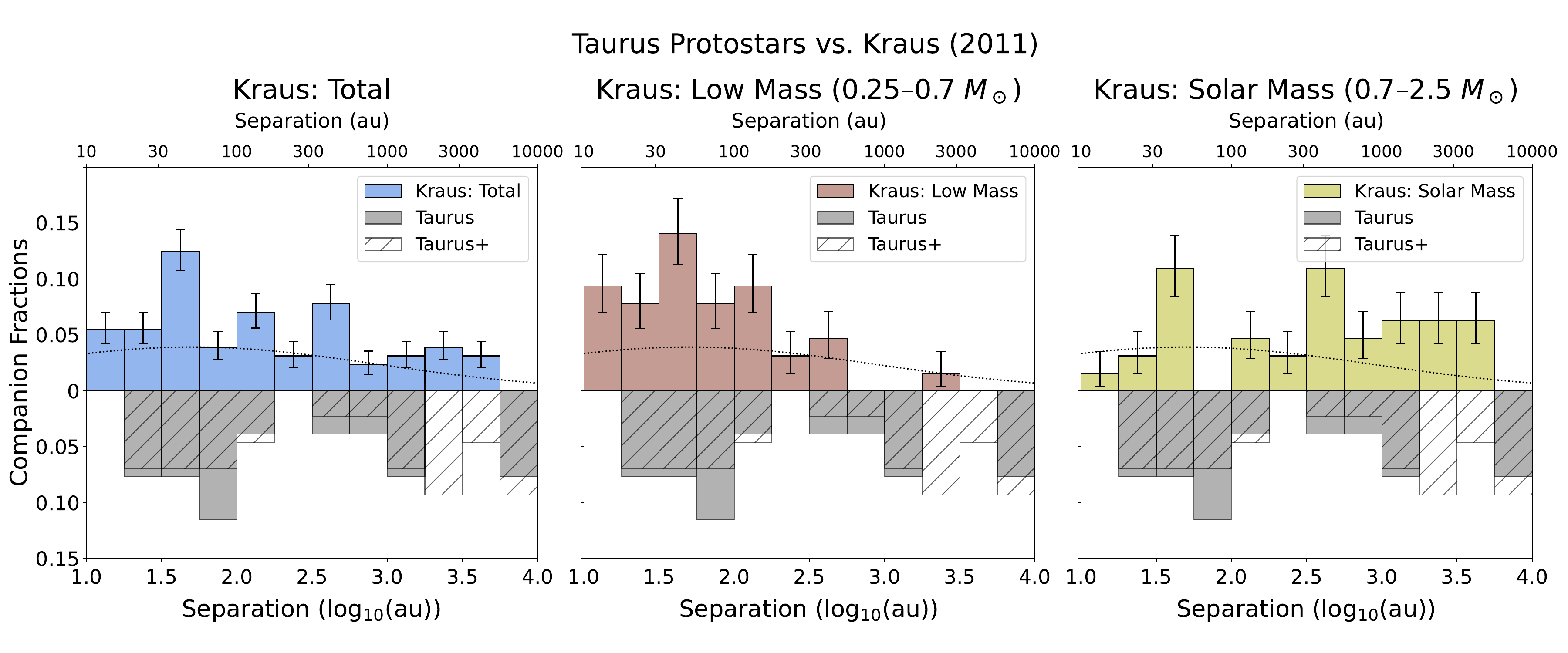}
    \caption{Mirrored histograms comparing separation distributions for the Taurus and Taurus+ protostellar samples with the more evolved Taurus sample from \citet{Kraus2011}. The \citet{Kraus2011} samples are shown above the horizontal axis, while the Taurus and Taurus+ protostellar samples are mirrored below. From left to right, the panels show the full \citet{Kraus2011} sample, the lower-mass sample, and the solar-mass sample, as defined in their study. Since the total number of systems used to calculate the CF for the low- and solar-mass samples was not specified in \citet{Kraus2011}, we assume 64 systems in each bin, corresponding to half of the full sample, based on the roughly equal numbers of low- and solar-mass systems.}
    \label{fig:kraus-histo}
\end{figure}

We compare our Taurus separation distributions to the more evolved  Class~II and III stars in Taurus from \citet{Kraus2011}, shown in Figure \ref{fig:kraus-histo}. The histogram displays the full sample, which includes stars across a range of masses (0.25-2.5 M$_\odot$), along with separate histograms for the low-mass ($<$0.7~M$_\odot$) and solar-mass ($>$0.7~M$_\odot$) primary subsamples. 
The full \citet{Kraus2011} sample displays a peak around 40~au, visible in both the low-mass and solar-mass subsamples. At wider separations, the distribution flattens out, balancing the lack of wide companions in the low-mass sample with an excess in the solar-mass sample. The low-mass systems are strongly concentrated at close separations and show a scarcity of wide multiples.
Due to resolution limits, we cannot assess the innermost separation bins (3-18~au) probed by \citet{Kraus2011}. 

\begin{figure}[H]
    \centering
    \includegraphics[width=0.6\textwidth]{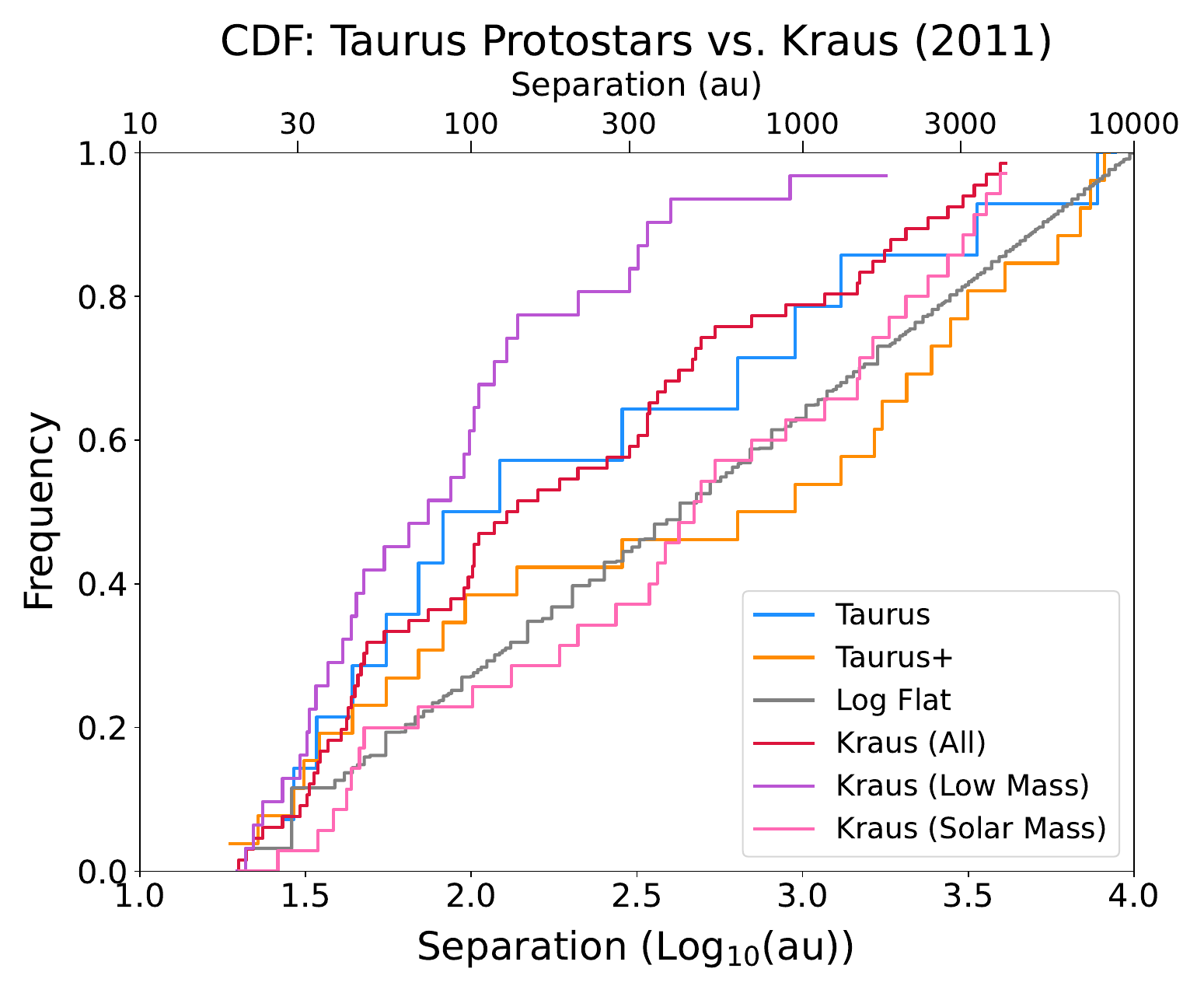}
    \caption{Cumulative distribution functions of companion separations in Taurus. The dark blue and orange curves show  Class~0/I protostars from our Taurus and Taurus+ samples, respectively. The more evolved  Class~II/III systems from \citet{Kraus2011} are also shown: the full Kraus sample in pink, low-mass primaries (0.25–0.7 $M_\odot$) in light blue, and solar-mass primaries (0.7–2.5 $M_\odot$) in red. The gray curve represents a reference log-flat distribution.}
    \label{fig:kraus-cdf}
\end{figure}

The Taurus+ sample resembles the \citet{Kraus2011} solar-mass subsample more closely than the low-mass one, with both distributions showing a lack of intermediate separation companions and possible peaks at both close and wide separations. However, the lull in intermediate Taurus+ companions occurs at slightly wider separations ($\sim$250~au) compared to the $\sim$75~au turnover in the \citet{Kraus2011} solar-mass sample. These similarities are also apparent in the CDFs shown in Figure \ref{fig:kraus-cdf}, where both Taurus+ and \citet{Kraus2011} solar-mass share a similar overall shape and curvature. The \citet{Kraus2011} low-mass sample, however, has the steepest CDF rise and the shortest overall separation range, further reinforcing the dominance of close companions. Since we cannot compare the closest separations probed by \citet{Kraus2011}, we omit companions within $\sim$30~au from the CDF.

\subsection{Quantitative Separation Distribution Comparisons} \label{subsec:quantitative-sep-dist}

To assess whether different protostellar separation distributions are statistically consistent with one another, we performed KS tests and AD tests on the CDFs of companion separations. 
The results are presented in Table \ref{tab:ks-test-results}.
We adopt a significance p-value threshold of $p<$0.05 to indicate that two samples are unlikely to be drawn from the same parent distribution.
Among the comparisons tested, most p-values exceed this threshold, suggesting broad consistency between distributions. However, three cases show statistically significant differences.

Hereafter, we denote the KS and AD test p-values as $p_\mathrm{KS}$ and $p_\mathrm{AD}$, respectively.
First, the Taurus+ sample differs significantly from the field solar-type binary population of \citet{Raghavan2010} ($p_\mathrm{KS}=0.014$, $p_\mathrm{AD}=0.006$), allowing us to reject the null hypothesis that these distributions are drawn from the same parent population. This likely reflects the excess of wide multiples in Taurus+ compared to the narrower separation range characteristic of field multiples. Similarly, the Taurus+ sample differs from both the full \citet{Kraus2011} sample of Class~II/III stars ($p_\mathrm{KS}=0.028$, $p_\mathrm{AD}=0.016$) and its low-mass subsample ($p_\mathrm{KS}=0.001$, $p_\mathrm{AD}=0.001$). These differences are again likely driven by the relative abundance of wide systems in Taurus+ and the lack of wide companions among more evolved, low-mass stars in the \citet{Kraus2011} sample. The Taurus sample is unlikely to be drawn from the same parent distribution as Orion, based on the AD test ($p_\mathrm{AD}=0.012$). The Taurus+ vs. Orion comparison does not meet our adopted significance threshold of $p<0.05$, although the p-values remain relatively small.

In contrast, there is no evidence that the Taurus or Taurus+ distributions differ significantly from those Perseus or the solar-mass component of the \citet{Kraus2011} sample. For example, Taurus+ is statistically consistent with the \citet{Kraus2011} solar-mass subsample ($p_\mathrm{KS}=0.486$, $p_\mathrm{AD}=0.213$)
The observed Taurus sample also matches well with the full \citet{Kraus2011} sample ($p_\mathrm{KS}=0.879$, $p_\mathrm{AD}=0.650$), though the small sample size (N = 14) limits the statistical power of this comparison.

We also tested both Taurus samples against a log-flat separation distribution, often referred to as Öpik's law \citep{Opik1924}. Such a distribution has been shown to describe companion separations in some populations, like the intermediate-mass binaries in the Scorpius OB2 association \citep{Kouwenhoven2007}.
In both cases, the results were inconclusive (p$>$ 0.05), and we cannot reject the hypothesis that the data are consistent with a log-flat distribution. This suggests that while hints of bimodality are present in the histograms, they are not statistically significant enough to definitively rule out a flat distribution in log space.

Among all comparisons, the null hypothesis can only be rejected (p $<$ 0.05) for the Taurus+ vs. \citet{Kraus2011} total and low-mass samples, and the Taurus+ vs. Raghavan 2010 samples suggesting they are not drawn from the same parent distributions. These differences indicate that the Taurus+ separation distribution contains a larger contribution from wide companions than either the more evolved low-mass Taurus Class~II/III population or the field solar-type population.

\begin{deluxetable}{llll}
\tabletypesize{\scriptsize}
\tablewidth{0pt}
\tablecaption{Separation Distributions Statistical Comparisons}
\label{tab:ks-test-results}
\tablehead{\colhead{Comparison} & \colhead{ Samples} & \colhead{ KS Test} & \colhead{ AD Test} \\[-2.5ex]
\colhead{} & \colhead{} & \colhead{ p-value} & \colhead{ p-value}
}S
\startdata
\textbf{Taurus vs. Orion}  & 14, 128 & 0.057 & $\mathbf{0.012}$\\
Taurus vs. Perseus         & 14, 38  & 0.143 & 0.087\\
Taurus vs. Log-flat        & 14      & 0.290 & 0.136\\ 
Taurus vs. Raghavan+2010   & 14, 135 & 0.464 & 0.427\\
Taurus vs. Kraus+2011      & $14, 62$ & $0.879$ & 0.650\\ 
Taurus vs. Kraus+2011 LM   & $14, 28$ & $0.173$ & 0.087\\ 
Taurus vs. Kraus+2011 SM   & $14, 34$ & $0.294$ & 0.390\\ 
\hline
Taurus+ vs. Orion          & $26, 128$ & $0.261$ & 0.090\\ 
Taurus+ vs. Perseus        & $26, 38$ & $0.478$ & 0.464\\ 
Taurus+ vs. Log-flat       & $26$ & $0.176$ & 0.403\\ 
\textbf{Taurus+ vs. Raghavan+2010}  & $26, 135$ & $\mathbf{0.014}$ & $\mathbf{0.006}$\\ 
\textbf{Taurus+ vs. Kraus+2011}     & $26, 62$ & $\mathbf{0.028}$ & $\mathbf{0.016}$\\ 
\textbf{Taurus+ vs. Kraus+2011 LM}  & $26, 28$ & $\mathbf{0.001}$ & $\mathbf{0.001}$\\ 
Taurus+ vs. Kraus+2011 SM  & $26, 34$ & $0.486$ & 0.213\\
\enddata
\tablecomments{Results of the Kolmogorov–Smirnov and Anderson–Darling tests applied to the companion separation distributions for all evolutionary classes over the separation range 18--10,000~au. Bolded rows indicate comparisons with statistically significant differences ($p<0.05$).} 

\end{deluxetable}

\section{Discussion} \label{sec:discussion}
In this section, we discuss what the Taurus multiplicity measurements imply for the formation and early evolution of multiple systems. We use the projected companion separation distribution as a diagnostic of possible formation pathways and subsequent dynamical evolution. While our continuum observations do not measure orbital evolution directly, population-level comparisons provide a way to place the Taurus separation distribution in an evolutionary context. We compare the Taurus protostellar population to protostellar populations in Orion and Perseus, more evolved Class II/III systems in Taurus, and field main-sequence populations. These comparisons allow us to examine how companion separations and multiplicity fractions vary across both environment and evolutionary stage. We first interpret the Taurus separation distribution in this framework, then discuss the elevated MF and CF in Taurus relative to Orion and Perseus, and finally compare the protostellar population to more evolved Taurus and field populations.

\subsection{Implications of the Observed Separation Distributions}

Multiple star formation is currently thought to proceed through two dominant physical mechanisms, each operating at different spatial scales. Core fragmentation typically occurs on scales of thousands of~au, resulting from instabilities within a collapsing molecular core \citep{Offner2010, Lee2019}. In contrast, disk fragmentation takes place within the rotationally supported circumstellar disk and is expected to form companions at separations of hundreds of au or less \citep{Kratter2010}. As a result, if both mechanisms are at play, we expect the resulting multiplicity to span from tens of au to thousands of au, which is consistent with the full range of separations observed in our Taurus samples.

While these mechanisms act during the formation stage, subsequent dynamical evolution can significantly reshape the observed distribution. Simulations by \citet{Offner2010}, \citet{Lee2019} and \citet{Kuruwita2023} demonstrate that wide companions, initially formed at separations beyond $\sim$1000~au, can migrate inward to separations below 100~au on timescales of less than a few 100,000 years. 
Because such evolution happens rapidly, the study of young protostellar populations offers a unique opportunity to observe multiplicity properties closer to their primordial state, providing critical insight into the conditions and mechanisms that govern multiple star formation.
Observations of protostellar multiplicity in other regions support the presence of these dual formation pathways. In Perseus, the survey by \citet{Tobin2016} revealed a bimodal separation distribution, with peaks near 75~au and 3000~au. They suggested that the wide and close peaks reflect the roles of core fragmentation and disk fragmentation, respectively. Additionally, \citet{Tobin2016,Tobin2018} and \citet{Reynolds2021,Reynolds2024} found that many close multiples reside within rotating disk-like structures, and that at least one disk is likely currently gravitationally unstable. Moreover, disk alignments at close separations suggest that these systems may have formed through disk fragmentation.

The Orion survey by \citet{Tobin2022} strengthened this idea. By sampling a large population of  Class~0,  Class~I, and Flat Spectrum sources, they found that the bimodal structure is primarily driven by the  Class~0 systems, while more evolved sources displayed flatter or more uniform distributions. They interpreted the changing separation distributions with evolutionary class as evidence that the primordial multiplicity distribution may begin with distinct peaks associated with both fragmentation modes, which are then gradually smoothed out by migration and dynamical evolution.
Supporting evidence for this bimodal morphology is also found in earlier work by \citet{2008AJ....135.2496C}, who identified a similar valley between close and wide companions in samples of  Class~I sources across multiple star-forming regions. 
Its repeated appearance in different star-forming regions suggests that two distinct mechanisms may be responsible for this recurrent feature, rather than it being a short-lived or transient pattern.

Figure \ref{fig:Taurus-histo-cdf} suggests that Taurus may also exhibit a bimodal separation distribution, with the observed sample showing a peak at close separations and a local minimum at intermediate scales, while the Taurus+ sample reveals an increase in multiplicity toward 2000–10,000~au. However, statistical tests indicate that we cannot definitively rule out a log-flat distribution for either the Taurus or Taurus+ samples.

The statistical comparisons do not show a robust difference between Taurus or Taurus+ and the Orion or Perseus separation distributions. The only comparison below our adopted $p=0.05$ threshold is the AD test for the Taurus--Orion sample ($p_\mathrm{AD}=0.012$); the corresponding KS test is still low but non-rejecting ($p_\mathrm{KS}=0.057$), and the Taurus--Perseus comparison remains above the threshold for both tests ($p_\mathrm{KS}=0.143$, $p_\mathrm{AD}=0.087$). For the Taurus+ sample, all comparisons are above the rejection threshold, with $p_\mathrm{KS}=0.261$ and $p_\mathrm{AD}=0.090$ relative to Orion, and $p_\mathrm{KS}=0.478$ and $p_\mathrm{AD}=0.464$ relative to Perseus. Thus, the low AD probability in the observed Taurus--Orion comparison may reflect incompleteness in the observed sample rather than a real regional difference. While these results do not allow us to claim a statistically significant similarity or difference, the qualitatively similar appearance of the distributions is compelling.

We suggest that the Taurus separation distribution may reflect a similar bimodal structure to that seen in Orion and Perseus. The primary limitation here is likely small number statistics: although our sample is nearly complete, the absolute number of protostellar systems in Taurus is much lower than in Orion or Perseus. With fewer sources, statistical fluctuations become more prominent, and clear bimodal features are more difficult to resolve.

The Taurus separation distribution therefore provides a useful test of whether core fragmentation alone can explain the observed companion population. Radiation-hydrodynamic simulations by \citet{Offner2010} suggest that turbulent core fragmentation can reproduce wide low-mass multiple systems, but that radiative feedback suppresses disk fragmentation. Similarly, gravo-magnetohydrodynamic simulations by \citet{Lee2019} reproduce observed separation distributions over $\sim$100--3000~au, but underpredict companions at separations $<$100~au. These results suggest that core fragmentation alone may not fully explain the close-separation population, and that disk fragmentation and/or subsequent orbital evolution may also be required.

\citet{Kuruwita2023} explored whether dynamical evolution following core fragmentation could produce close multiples by simulating orbital evolution down to $\sim$20~au. Their results show that dynamical interactions, including ejections, captures, and the initial kinematics of fragments, can transform initially wide companions into close multiples and produce a bimodal separation distribution similar to those observed in Perseus and Orion. However, because externally driven dynamical evolution should be less efficient in low-density environments like Taurus, where the interaction timescale is longer than the age of the region \citep{Kraus2011}, the close companion population in Taurus may indicate that external dynamical processing alone cannot fully account for the observed distribution.

A related question is to what degree has dynamical evolution affected the observed separation distribution in Taurus? Ideally, the primordial separation distribution would be characterized using only the youngest protostars, before substantial dynamical evolution has occurred. However, Taurus contains only five Class~0 systems in our sample, and only one of these is in a binary system. As a result, the Taurus separation distribution is dominated by Class~I and Flat Spectrum sources, making it difficult to isolate the earliest multiplicity distribution directly.

Nevertheless, if the bimodal distribution seen among the youngest Orion protostars is representative of an early multiplicity configuration, as suggested by \citet{Tobin2022}, then the abundance of both close and wide companions in Taurus may indicate that some aspects of this early separation structure persist to later protostellar stages in Taurus. In this picture, the low-density environment may help preserve wide systems that would be more efficiently disrupted or reconfigured in denser regions. However, this remains speculative without additional knowledge of the amount of internal dynamical evolution and its diagnostics such as component masses, mass ratios, disk truncation, and orbital information.

This interpretation is broadly consistent with simulations in which local stellar density shapes the survival and evolution of multiple systems. Using radiation-magnetohydrodynamic simulations from the STARFORGE project, \citet{Guszejnov2023} investigated how different environments influence multiplicity and found that stellar density is a particularly important factor. In high-density regions like Orion and Perseus, dynamical interactions between young stars are frequent, leading to disrupted or reconfigured systems and increasing the likelihood of orbital migration that populates the close-binary regime. 
Given the much lower stellar density of Taurus, such interactions are expected to be less frequent, allowing multiple systems to remain more stable and retain their initial separations for longer, potentially leading to wider multiples on average.

This environmental difference could explain why the bimodal distribution in Orion is driven by Class~0 sources. These systems are actively being reshaped by their environment, while the  Class~I and Flat Spectrum sources in Taurus retain a distribution that may still reflect their formation. 

\subsection{Elevated Multiplicity in Taurus}

A key distinction between the Taurus Molecular Cloud and the more clustered regions of Orion and Perseus is its overall elevated multiplicity fraction and companion fraction. The \citet{Guszejnov2023} simulations described in the preceding section, when looking at the overall multiplicity statistics found a strong anti-correlation between stellar density and multiplicity, as lower density regions more effectively retained their wide companions due to reduced dynamical interactions. This behavior is qualitatively consistent with the low-density environment of Taurus and with the elevated multiplicity observed in our survey.

Across separations of 18--10,000~au, both the observed Taurus sample and the more complete Taurus+ sample show elevated multiplicity relative to Perseus and Orion (Table \ref{tab:multiplicity_stats}). In the Taurus+ sample, the MF is higher than in Orion by a factor of $\sim$1.8 and higher than in Perseus by a factor of $\sim$1.5. The corresponding CF is also elevated, by factors of $\sim$1.7 relative to Orion and $\sim$1.4 relative to Perseus. Thus, Taurus+ contains both a larger fraction of multiple systems and a larger number of companions per system than the more clustered regions. The MF enhancement is statistically significant at the $\sim$4$\sigma$ level relative to Orion and at the $\sim$2$\sigma$ level relative to Perseus, while the CF differences are less significant, at the $\lesssim$2$\sigma$ level.

Although the uncertainties on the Taurus samples are larger due to the smaller protostellar population, the near completeness of the Taurus+ sample supports the interpretation that Taurus intrinsically retains a higher fraction of multiple systems. Nevertheless, these results should be interpreted with some caution given the effects of small-number statistics and the fact that a subset of Taurus+ protostars have not yet been observed at sub-arcsecond resolution, which may bias against the detection of close companions. 

The elevated multiplicity observed in Taurus suggests that environmental density may play a role, but this parameter alone cannot fully explain the differences. If low stellar density were sufficient to preserve a higher fraction of multiple systems, then the lower-density subregions of Orion, which are more comparable to Taurus, would also be expected to show elevated multiplicity. However, when \citet{Tobin2022} divided Orion using a YSO surface density cutoff of 30~stars~pc$^{-2}$, they found that the Class~0 MF and CF are comparable between the high- and low-density subregions, while the Class~I and Flat Spectrum populations in the lower-density subregions have consistently lower MF and CF values than those in the higher-density subregions (see Figure 17 in \citet{Tobin2022}). Thus, the relationship between YSO surface density and multiplicity is not monotonic.

The high multiplicity in Taurus may therefore reflect differences in the initial conditions of star formation, rather than dynamical evolution alone. These differences could be related to the physical conditions of the parent molecular clouds and the environments in which the systems formed. A detailed comparison of these cloud-scale properties is beyond the scope of this work, but the contrast between Taurus and Orion suggests that present-day YSO surface density is not by itself sufficient to predict protostellar multiplicity.

The interpretation that Taurus is intrinsically multiplicity-rich is further supported by multiplicity measurements of the more evolved Taurus population. The Class~II and III survey by \citet{Kraus2011} reported a high MF of 0.73. Approximately 10\% of those multiples have separations below 18~au, below the resolution limits of this study; when adjusted to match our separation range, the MF becomes 0.56, comparable to the Taurus MF of 0.50 and Taurus+ MF of 0.53. This suggests that elevated multiplicity is a persistent feature of Taurus, extending from the protostellar phase into the T~Tauri population.

Taken together, these results suggest that the high multiplicity in Taurus is unlikely to be explained by dynamical processing alone. The low-density environment of Taurus may help reduce the loss of companions, particularly at wide separations where systems are most vulnerable to disruption. However, the persistence of high multiplicity into the Class~II/III population, combined with the lack of similarly elevated multiplicity in the low-density subregions of Orion, suggests that Taurus may also form a larger fraction of multiple systems. In this sense, the present-day multiplicity of Taurus is likely a reflection of both reduced dynamical disruption and an environment more likely to initially form multiples.

\subsection{Evolution Toward More Mature Populations} \label{subsec:evolution}

While the elevated multiplicity and possible prolonged bimodal appearance of the separation distribution in Taurus may suggest that its low-density environment helps preserve companions during the embedded phase, these systems are not expected to remain fixed in their current configurations. The Taurus protostars in our sample are likely younger than $\sim$1~Myr \citep{Dunham2015}, whereas the Class~II/III population probed by \citet{Kraus2011} is several Myr old \citep{Tobin2024}. Over this interval, internal interactions, orbital evolution, gas-driven migration, mass loss during envelope dispersal, and the loss of weakly bound wide companions can still reshape multiple systems, even if external dynamical processing proceeds more slowly in Taurus than in denser regions such as Orion or Perseus. Having compared the Taurus protostellar population to other embedded surveys, we now consider how these systems may evolve within Taurus itself by comparing our results to the more evolved Class~II/III population from \citet{Kraus2011}.

We expect most dynamical evolution of multiple systems to occur before the end of the Class~III phase. Studies of Class~II/III stars in Orion show separation distributions that are already similar to the field population \citep{Reipurth2007}. In older and more spatially dispersed regions such as Taurus, where cluster dispersal is ongoing, most dynamical processing is therefore expected to be complete at this stage. This expectation is supported by the low-mass sample of \citet{Kraus2011}, whose separation distribution follows that of field stars (Figure~\ref{fig:kraus-histo}).

For the Taurus Class~II/III comparison, \citet{Kraus2011} divided their sample into low-mass (0.25--0.7~M$_\odot$) and solar-mass (0.7--2.5~M$_\odot$) subsamples. They reported an overall MF of 0.73, with values of 0.77 for the low-mass group and 0.62 for the solar-mass group. After correcting these values to match the resolution limits of our survey (18-10000~au), we obtain an overall MF of 0.56, with the low- and solar-mass subsamples yielding 0.59 and 0.49, respectively, which remain in close agreement with our findings. Although multiplicity is generally expected to decline over time due to the dynamical loss of companions, the \citet{Kraus2011} results suggest that, within Taurus, the overall MF remains relatively stable across these early stages of evolution.

Our separation distribution analysis offers further insight into this evolution. As seen in Figure \ref{fig:kraus-histo}, the Taurus+ sample more closely resembles the solar-mass subsample from \citet{Kraus2011} particularly due to the lack of wide companions in the Kraus low-mass group. This is supported by our statistical analysis, which shows that Taurus+ and the Kraus low-mass sample are unlikely to be drawn from the same parent distribution ($p_\mathrm{KS}=0.001$, $p_\mathrm{AD}=0.001$), while we cannot reject the null hypothesis for the solar-mass comparison ($p_\mathrm{KS}=0.486$, $p_\mathrm{AD}=0.213$). 

However, our observed sample is dominated by sources in the low-mass regime (0.25–0.7~M$_\odot$; A. C. Plante et al., in prep), suggesting that these systems may still be subject to evolutionary processes that shift their distributions toward the unimodal, close-separation peak seen in the Kraus low-mass sample. This evolution would be expected to be driven by two main processes. First, systems that remain bound may undergo inward migration, primarily driven by dynamical friction with the surrounding envelope gas. Second, wide companions, particularly those at separations greater than 1000~au, may become unbound as the system loses mass during early evolution, such as through envelope-clearing outflows or external perturbations (e.g., \citet{Offner2014, Sadavoy2017}). 
While \citet{Guszejnov2023} suggest that low-mass systems are especially vulnerable to losing wide companions in dense environments due to frequent dynamical interactions, this would be less effective in a low-density region like Taurus. 
Another possibility for the loss of wide companions is that some may form as unbound systems within the same core. If their relative velocities are sufficiently high and their masses too low, they may never become gravitationally bound.

We also compared our results to the field solar-type binary population from \citet{Raghavan2010}, which also exhibits a unimodal separation distribution peaking between 30-80~au. Our statistical comparison shows that the Taurus+ sample is inconsistent with this distribution ($p_\mathrm{KS}=0.014$, $p_\mathrm{AD}=0.006$), likely due to the excess of wide multiples in Taurus. Given the mass distribution of our sample and the fact that M-type stars are the most common outcome of star formation, many of our sources are likely to evolve into M stars. These stars are observed to have even smaller mean separations \citep[$\sim$20~au;][]{Winters2019} further supporting the idea that dynamical evolution will ultimately reduce the wide companion population in Taurus and shift the distribution toward tighter separations.
However, it is also possible that low-density regions such as Taurus contribute less to the field population than more clustered environments (e.g., Orion-like regions; \citealp{Lada2003}), in which case wide multiples in Taurus may remain largely intact and need not evolve to match the field separation distribution. In this scenario, the field population would be dominated by systems that formed and evolved in denser environments, where dynamical processing is more efficient.

The number of mixed-class systems is difficult to quantify, since evolutionary classes cannot be independently assigned to companions at separations smaller than 1000~au in our data, however, we identify eight systems that appear to contain components with mixed evolutionary classifications. These wide-separation mixed-class systems are unlikely to be chance line-of-sight alignments given the low stellar surface density and large spatial extent of Taurus, and are therefore likely physically associated. Such mixed classifications could reflect later dynamical capture, but can also be naturally explained if fragments within the same dense core collapse at slightly different times \citep{Murillo2016,Luo2022}. Because the free-fall times of dense cores are comparable to protostellar class lifetimes, small offsets in collapse time can produce apparently non-coeval systems \citep{Tobin2022}.

\subsection{Limitations} \label{subsec:limitations}

Despite the strengths of our study, several limitations must be acknowledged. First, while our sample is nearly complete for known protostars in Taurus, the region itself contains a relatively small number of young stellar objects, which limits the statistical power of our analysis. Additionally, our reliance on dust continuum and free–free emission to identify companions introduces a selection bias: systems without sufficient circumstellar material or jet activity may go undetected, potentially underestimating the true MF \citep{Tobin2016}. This bias is mitigated to some extent by the Taurus+ sample, which also includes companions detected in the infrared. We account for the finite size and differing completeness of the Taurus and Taurus+ samples in our uncertainty estimates using the Wilson score interval with a finite-population correction. This treatment reflects that the Taurus Sample does not include all known companions later incorporated into the Taurus+ sample, resulting in larger uncertainties for the Taurus Sample. For the Taurus+ sample, which is complete to the best of our knowledge, we nevertheless adopt a conservative completeness of 75\% to avoid underestimating the uncertainties.

A key limitation of evolutionary classification schemes based on infrared spectral index or bolometric temperature is that both quantities can be affected by viewing geometry \citep{Whitney2003, Crapsi2008}. For example, high-inclination or edge-on systems may appear more embedded because the central protostar and disk are obscured, which can lead to classifications that are earlier than the true evolutionary stage. Some Taurus sources previously classified as Class~I may instead be more evolved systems viewed at high inclination, such as IRAS~04260+2642 \citep{Luhman2010}.
Although these classification methods are widely used, there is not currently a straightforward way to fully correct for inclination effects using the available SED-based diagnostics alone. These limitations should therefore be kept in mind when interpreting observational class as evolutionary stage. However, they are unlikely to strongly affect the main results of this work. While we report the evolutionary classifications of sources in our sample, our primary multiplicity analysis is performed on the sample as a whole rather than separately by class. Similarly, in close multiple systems where the infrared SED blends the emission from multiple components, the resulting system-level classification may not reflect the evolutionary stage of each individual component, but this does not significantly affect our overall multiplicity statistics. These effects are most important for studies that compare multiplicity directly across evolutionary classes; in this work, they primarily affect the interpretation of individual source classifications rather than the global multiplicity results.
 
Furthermore, we lack high-resolution radio data for some sources in the Taurus+ sample, leaving open the possibility that unresolved close companions remain undetected. Our angular resolution also limits our ability to detect companions with projected separations smaller than $\sim$18~au. We therefore do not attempt to characterize multiplicity below this resolution limit, and the multiplicity fractions and companion fractions reported here should be considered lower limits.
Finally, since we observe each system at a single moment in time and only in two dimensions, the projected separations we report are lower limits; true separations may be larger depending on orbital phase, inclination, and eccentricity \citep{Kuiper1935}. However, because eccentric systems spend the most time near apastron, we are likely to observe them near their widest separations. These factors should be considered when interpreting the observed multiplicity statistics and separation distributions.

\section{Conclusions} \label{sec:sumandconclusions}

We conducted a high-resolution, multi-wavelength survey using ALMA 0.9 mm and VLA 9 mm continuum observations of 40 protostars in the Taurus molecular cloud to analyze multiplicity, measuring the frequency and separation distribution of multiple systems in this nearby, low-density star-forming region.
We extended the sample (Taurus+) by including 24 additional Taurus protostars from the literature not covered in our observations. Using these datasets, we compared multiplicity statistics in Taurus to those in the more clustered Orion and Perseus regions, as well as to a survey of more evolved Class~II and III Taurus stars, in order to examine both environmental and evolutionary trends. Our main results are as follows:

\begin{enumerate}
\item Within 18--10,000~au separations, the Taurus sample has an MF of $0.50 \pm 0.0.07$ and a CF of $0.58 \pm 0.20$, while the Taurus+ sample yields MF = $0.53 \pm 0.06$ and CF = $0.72 \pm 0.19$. These values are higher than those measured in Orion (MF = $0.29 \pm 0.01$, CF = $0.42 \pm 0.02$) and Perseus (MF = $0.36 \pm 0.04$, CF = $0.51 \pm 0.11$). Differences in MF are statistically significant at the $\sim2$--$4\sigma$ level (with the strongest significance for Taurus+ relative to Orion), whereas differences in CF are less significant ($\lesssim 2\sigma$). Given the near completeness of the Taurus+ sample, these results suggest that Taurus may have an intrinsically higher fraction of multiple systems compared to more clustered star-forming regions. 

\item The separation distributions in Taurus and Taurus+ (18--10,000~au) both peak near 75~au and show a deficit of companions around 2000~au. The Taurus sample is sparsely populated beyond 2000~au, whereas Taurus+ fills in some of these wider separations and shows an elevated wide-companion population. Although the small number of protostars in Taurus limits the clarity of these trends, the distributions resemble the more clearly bimodal shapes seen in Orion and Perseus. However, statistical tests indicate that both Taurus ($p_\mathrm{KS}=0.290$, $p_\mathrm{AD}=0.136$) and Taurus+ ($p_\mathrm{KS}=0.176$, $p_\mathrm{AD}=0.403$) cannot be ruled out from having been drawn from a log-flat distribution, so bimodality cannot be confirmed.

\item The presence of both close ($<$200~au) and wide ($>$1000~au) multiples, together with the pronounced local minimum around 200–300~au in each sample, suggests that both disk and core fragmentation are needed to reproduce the observed separation distributions, with disk fragmentation primarily forming close multiples and core fragmentation producing the wide multiples.

\item The elevated multiplicity in Taurus compared to Orion and Perseus is broadly consistent with simulations predicting that low-density environments experience fewer dynamical interactions and can therefore retain a higher fraction of primordial multiples \citep{Guszejnov2023}. However, the low-density subregions of Orion, which are more comparable to Taurus in present-day YSO surface density, show the opposite trend, with consistently lower MFs than the higher-density Orion subregions. This suggests that stellar density alone does not fully determine protostellar multiplicity. While the low stellar density of Taurus likely helps preserve some multiple systems against external dynamical disruption, additional environmental or initial-condition differences may also contribute to Taurus forming an intrinsically higher fraction of multiples.

\item After restricting the Class~II/III sample of \citet{Kraus2011} to the same separation range as ours, their MF (0.56) matches closely with our results ($\sim$0.5), further suggesting that multiplicity in Taurus may be comparatively stable due to the region’s more quiescent environment. Within this comparison, we found the Taurus separation distributions more closely resemble their solar-mass subsample, as the low-mass systems show a pronounced lack of wide companions. But, because most protostars in our sample are likely to evolve into low-mass stars (A. C. Plante et al., in prep.), we expect the separation distribution to change over time, and potentially evolve toward the unimodal distributions observed for the Kraus low-mass sample (peak at 40~au) and the Raghavan field-star population (peak at 30–80~au).

\end{enumerate}


\begin{acknowledgments}
The authors thank the anonymous referee for a constructive report that helped improve the quality of the manuscript.
This paper makes use of the following ALMA data: ADS/JAO.ALMA\#2019.1.00847.S ALMA is a partnership of ESO (representing its member states), NSF (USA) and NINS (Japan), together with NRC (Canada), MOST and ASIAA (Taiwan), and KASI (Republic of Korea), in cooperation with the Republic of Chile. The Joint ALMA Observatory is operated by ESO, AUI/NRAO and NAOJ. 

The National Radio Astronomy Observatory and Green Bank Observatory are facilities of the U.S. National Science Foundation operated under cooperative agreement by Associated Universities, Inc.

This research has made use of the VizieR catalog access tool, CDS, Strasbourg, France (DOI: 10.26093/cds/vizier). The original description of the VizieR service was published in \citet{Vizier2000}.

This research has made use of NASA’s
Astrophysics Data System Bibliographic Services.

This research has made use of the SIMBAD database,
operated at CDS, Strasbourg, France \citep{Simbad2000}.

N.P.B. acknowledges support from NSF grant no. AST-2205698 and NASA/Space Telescope Science Institute grants JWST-GO-03271 and JWST-GO-05460.

ZYL is supported in part by NASA 80NSSC20K0533 and NSF AST-2307199.
\end{acknowledgments}
\facilities{VLA, ALMA}



\software{Astropy \citep[http://www.astropy.org; ][]{astropy2013,astropy2018,astropy2022}, CASA \citep[http://casa.nrao.edu; ][]{CASA2022}, dustmaps \citep{Green2018}}

\appendix

\section{Modeled Source SEDs}\label{appendix:source_seds}

To establish evolutionary classes and calculate bolometric quantities for each target, we assembled an SED based on archival infrared photometry using the VizieR SED tool \citep{Vizier2000}. We used query radii ranging from 4\arcsec\ to 20\arcsec, depending on nearby sources and the available photometry, centered on the coordinates listed in Tables~\ref{tab:Catalog-obs} and \ref{tab:Catalog-ext}. We removed the $12.663 \times 10^{3}$ GHz point from the Spitzer/MIPS 24 $\mu\mathrm{m}$ instrument because it appeared anomalously high across most source SEDs. 
For nine sources, we instead adopted photometric measurements compiled for \citet{Sheehan2017}, which provided more complete wavelength coverage than the corresponding VizieR SEDs, largely because they included Spitzer IRS measurements. Because these measurements were not documented in full in \citet{Sheehan2017}, we include them in our photometry table along with the photometry compiled from VizieR.
We then binned the remaining photometry by wavelength to construct an SED spanning 1.25--1300~$\mu$m. We took the median flux within each bin and used the sample standard deviation as an uncertainty.
To accurately show uncertainties on logarithmic axes, we used log-symmetric error bars rather than linear symmetric ones. These are derived from the fractional (or relative) flux uncertainties given by $r = \sigma_F / F$. Table~\ref{tab:SED-photometry} lists the photometry used to construct the SEDs, including VizieR photometry and additional literature photometry, when applicable. The table also gives the VizieR query radius adopted for each source and identifies the references contributing to the binned photometry.

We fit the SED as the sum of (i) a smooth continuum represented by a 6th-order polynomial in log space over the observed range and (ii) a modified blackbody (isothermal $T=30$ K, $\kappa_\lambda \propto \lambda^{-1.8}$) to extend the SED from the longest observed wavelength to 1000 $\mu$m when necessary; the modified blackbody was scaled to match the observed SED. Bolometric luminosities were computed from
\begin{equation}
L_{\mathrm{bol}} = 4\pi D^2 \int F_\nu \, d\nu,
\end{equation}
and bolometric temperatures from
\begin{equation}
T_{\mathrm{bol}} = 1.25\times10^{-11}\,\langle \nu \rangle \ \mathrm{K},
\end{equation}
where 
\begin{equation}
\langle \nu \rangle = \frac{\int \nu F_\nu \, d\nu}{\int F_\nu \, d\nu}.
\end{equation}

We calculated the infrared spectral index $\alpha$ between 2 and 24 $\mu$m from 
\begin{equation}
\alpha = \frac{\Delta \log (\nu F_\nu)}{\Delta \log \nu},
\end{equation}
where we used the binned SED points closest to those wavelengths, and adopted the standard class boundaries: $\alpha \geq 0.3$ (Class~I), $-0.3 \leq \alpha < 0.3$ (Flat), $-1.6 \leq \alpha < -0.3$ (Class~II), and $\alpha < -1.6$ (Class~III). 
We use both infrared spectral index and bolometric temperature to assign evolutionary classes because neither quantity alone provides a complete classification scheme for our sample. The infrared spectral index distinguishes between Class~I, Flat Spectrum, Class~II, and Class~III sources, but it does not identify Class~0 protostars. Conversely, bolometric temperature is used to identify Class~0 sources, but cannot define the Flat Spectrum class. Therefore, we use $T_\mathrm{bol}$ to classify the four Class~0 sources in our sample, adopting the standard criterion $T_\mathrm{bol} \leq 70$~K \citep{Chen1995}, and use the infrared spectral index for the remaining sources. In most cases, the classifications of Class I and Class II inferred from $\alpha$ and $T_\mathrm{bol}$ agree.
The resulting values and classifications are listed in Tables \ref{tab:Catalog-obs} and \ref{tab:Catalog-ext}.

The photometry is primarily based on mid- to far-infrared observations from space-based infrared telescopes such as Spitzer, WISE, and Herschel, which have angular resolutions $\gtrsim$1\arcsec. At the distance of Taurus (140 pc), these telescopes cannot resolve close multiples with separations $\sim$200-1000~au, meaning many of our high-resolution ALMA and VLA detections appear as unresolved single sources in the mid- to far-infrared.

As a result, all close binary systems in our sample, and even one wide multiple system inherit their \lbol{}, \tbol{}, and classification from the blended infrared source (13 systems total with blended SEDs). In most cases, we were able to resolve individual SEDs for companions with separations greater than 1000~au to get reliable individual $L_{\mathrm{bol}}$ and $T_{\mathrm{bol}}$ estimates, with the exception of IRAS~04248+2612 ABC at separation of 1647~au ($\sim$12\arcsec).

\begin{figure}[H]
    \centering
    \includegraphics[width=0.6\textwidth]{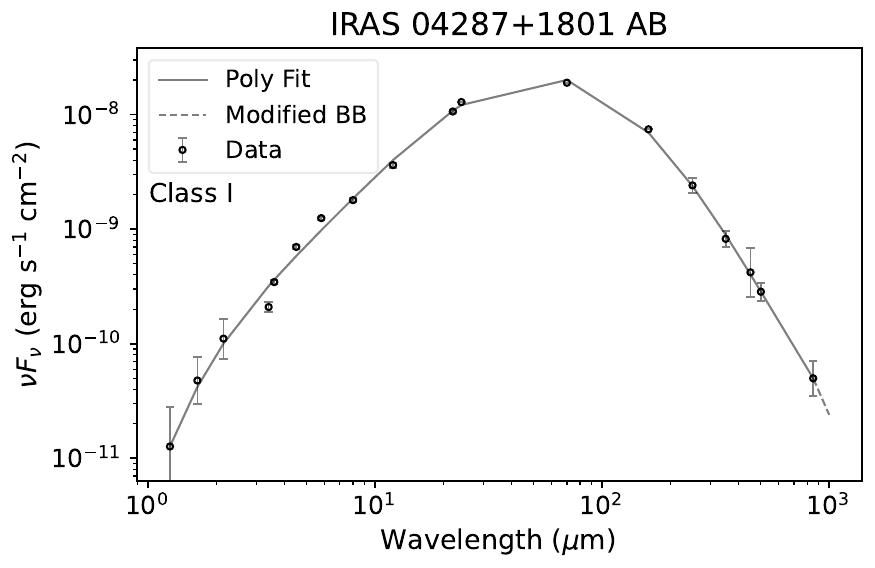}
    \digitalasset
    \caption{Spectral energy distribution and best-fit model for IRAS 04287+1801 AB. The complete figure set (48 images) is available in the online journal.}
    \label{fig:example-sed}
\end{figure}

The complete online figure set presents the SEDs for all sources in our sample, with one example shown in Figure~\ref{fig:example-sed}. Each SED includes the measured photometry (points) and the corresponding polynomial fit and modified blackbody model (when necessary). The large error bars associated with some photometry reflect a large scatter in the measurements available in the literature that can result from differing sensitivity, aperture sizes, and methods.

\figsetstart
\figsetnum{12}
\figsettitle{SEDs for All Sources}

\figsetgrpstart
\figsetgrpnum{12.33}
\figsetgrptitle{IRAS04287+1801AB}
\figsetplot{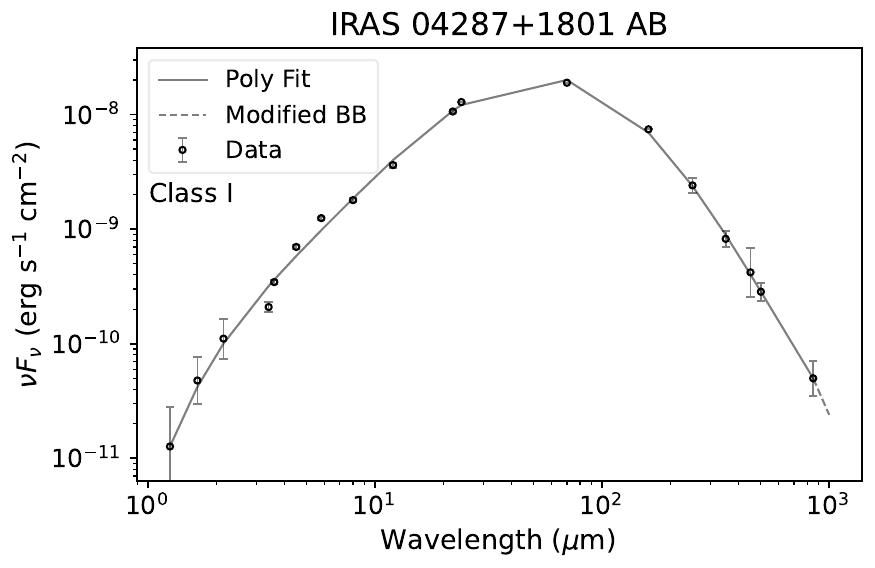}
\figsetgrpnote{Spectral energy distribution and best-fit model for IRAS04287+1801AB.}
\figsetgrpend

\figsetend

\begin{deluxetable}{ccccccc}
\tablecaption{SED Photometry\label{tab:SED-photometry}}
\tablehead{
   \colhead{Source} &
   \colhead{SIMBAD Name} &
   \colhead{Query Radius (\arcsec)} &
   \colhead{$F_{\nu}(1.25\,\mu$m) [Jy]} &
   \colhead{$F_{\nu}(1.65\,\mu$m) [Jy]} &
   \colhead{$F_{\nu}(2.15\,\mu$m) [Jy]} &
   \colhead{References}
}
\digitalasset
\startdata
IRAS 04295+2251 & IRAS 04295+2251 & 10.0 & 0.001750 & 0.016900 & 0.059300 & 2,11,12,\ldots\\
IRAS 04263+2426 & IRAS 04263+2426 & 10.0 & 0.001750 & 0.016900 & 0.059300 & 2,11,12,\ldots\\
IRAS 04191+1523 AB & IRAS 04191+1523 & 10.0 & 0.000316 & 0.001770 & 0.008340 & 2,11,15,\ldots\\
\ldots & \ldots & \ldots & \ldots & \ldots & \ldots & \ldots\\
IRAS 04365+2535 & IRAS 04365+2535 & NaN & 0.000273 & 0.003310 & 0.031200 & 115,116,120,\ldots\\
\enddata
\tablecomments{Wavelengths are given in $\mu$m and flux densities are given in Jy. For sources populated through VizieR, the listed photometry corresponds to the binned values used in the SED construction, and the query radius gives the search radius adopted for that source. Sources with NaN listed for the query radius were not populated using VizieR; for these sources, the photometry was compiled directly from the literature. The reference numbers identify the original catalogs or literature measurements contributing to the photometry for each source and correspond to the citations listed below.} This table is published in its entirety in the machine-readable format. A portion is shown here for guidance regarding its form and content.
\tablerefs{1. \citet{2001KFNT...17..409K} 2. \citet{2004AAS...205.4815Z} 3. \citet{2006AandA...448.1235D} 4. \citet{2006yCat.1304....0C} 5. \citet{2010AJ....139.2440R} 6. \citet{2011MNRAS.416..403F} 7. \citet{2013AJ....145...44Z} 8. \citet{2014yCat.1327....0N} 9. \citet{2017AandA...600L...4A} 10. \citet{2017AJ....153..166Z} 11. \citet{2008AJ....136..735L} 12. \citet{1988SSSC..C......0H} 13. \citet{1990IRASF.C......0M} 14. \citet{1993cio..book.....G} 15. \citet{2003yCat.2246....0C} 16. \citet{2010yCat.2298....0Y} 17. \citet{2012wise.rept....1C} 18. \citet{2008MNRAS.391..136L} 19. \citet{2007MNRAS.379.1599L} 20. \citet{2014PASJ...66...17T} 21. \citet{2014yCat.2328....0C} 22. \citet{2003PASP..115..965E} 23. \citet{2015AandC....10...99A} 24. \citet{2014ASPC..485..223B} 25. \citet{2019MNRAS.487.2522M} 26. \citet{2019MNRAS.490.3158C} 27. \citet{2019ApJS..240...30S} 28. \citet{2021ApJS..253....8M} 29. \citet{2018MNRAS.477.3145J} 30. \citet{2021yCat.2368....0S} 31. \citet{2020AJ....160..120J} 32. \citet{2022ApJS..259...35A} 33. \citet{2014PASP..126..398H} 34. \citet{2019AJ....158..138S} 35. \citet{2021arXiv210804778P} 36. \citet{1999AandA...349..389V} 37. \citet{2016yCat.9048....0M} 38. \citet{2005AandA...438..769D} 39. \citet{2008AandA...487..993K} 40. \citet{2010AandA...514A...2I} 41. \citet{2010AandA...519A..83T} 42. \citet{2014AandA...564A..79D} 43. \citet{2014AandA...570A..29B} 44. \citet{2015AandA...578A..42R} 45. \citet{2016AandA...594A..26P} 46. \citet{2019AandA...628A..66L} 47. \citet{2022AandA...663A..98T} 48. \citet{2024AandA...688A.203M} 49. \citet{2025AandA...699A.145D} 50. \citet{2025AandA...700A.235H} 51. \citet{2026AandA...706A.284T} 52. \citet{1997AandAS..126..479G} 53. \citet{2007AJ....133.1528C} 54. \citet{2007AJ....133.1560W} 55. \citet{2008AJ....135.2496C} 56. \citet{2010AJ....140.1214C} 57. \citet{2012AJ....144...31K} 58. \citet{2019AJ....158...54E} 59. \citet{2020AJ....159..273R} 60. \citet{2021AJ....162..191M} 61. \citet{2024AJ....168..215S} 62. \citet{2006ApJ...645..676L} 63. \citet{2006ApJ...647.1180L} 64. \citet{2009ApJ...696L..84C} 65. \citet{2009ApJ...704..531K} 66. \citet{2010ApJ...709L.114D} 67. \citet{2010ApJ...710.1247S} 68. \citet{Kraus2011} 69. \citet{2012ApJ...751..115H} 70. \citet{2012ApJ...751...52E} 71. \citet{2012ApJ...756...27L} 72. \citet{2013ApJ...769...21S} 73. \citet{2013ApJ...771..129A} 74. \citet{2014ApJ...784..126E} 75. \citet{2015ApJ...799..155D} 76. \citet{2017ApJ...838..150K} 77. \citet{2017ApJ...848...97R} 78. \citet{2017ApJ...849...63R} 79. \citet{2018ApJ...858...41Z} 80. \citet{2018ApJ...867..105T} 81. \citet{2019ApJ...872..158A} 82. \citet{2021ApJ...921...53L} 83. \citet{2022ApJ...928..134S} 84. \citet{2022ApJ...941..104Y} 85. \citet{1997ApJS..112..557K} 86. \citet{1999ApJS..123..233L} 87. \citet{2008ApJS..175..277D} 88. \citet{2008ApJS..176..457S} 89. \citet{2009ApJS..184..138H} 90. \citet{2009ApJS..184...18G} 91. \citet{2010ApJS..186..259R} 92. \citet{2010ApJS..186..406D} 93. \citet{2018ApJS..236...37P} 94. \citet{2020ApJS..247...28H} 95. \citet{1991MNRAS.253..485R} 96. \citet{1999MNRAS.308..897L} 97. \citet{2016MNRAS.458.3479M} 98. \citet{2016MNRAS.459..342M} 99. \citet{2017MNRAS.471..770M} 100. \citet{2018MNRAS.473.4937S} 101. \citet{2019MNRAS.482.5167S} 102. \citet{2024MNRAS.532.4661K} 103. \citet{2025MNRAS.537..931D} 104. \citet{2008PASP..120.1128O} 105. \citet{2015RAA....15.1154Z} 106. \citet{2003yCat.5114....0E} 107. \citet{1990ApJ...361...49S} 108. \citet{2000MNRAS.317...55S} 109. \citet{2006AJ....131.1163S} 110. \citet{2016yCat.7275....0D} 111. \citet{2018MNRAS.479.2374D} 112. \citet{2024yCat.8112....0H} 113. \citet{2011AandA...536A...7P} 114. \citet{2014AandA...571A..29P} 115. \citet{Andrews2005} 116. \citet{Motte2001} 117. \citet{Ladd1991} 118. \citet{Barsony1992} 119. \citet{Young2003} 120. \citet{Moriarty-Schieven1994} 121. \citet{Eisner2012} 122. \citet{Chandler2000} 123. \citet{Ohashi1996} 124. \citet{2015ApJS..218...21L}}
\end{deluxetable}

\section{Source information}\label{appendix:sourceinfo}
\subsection{Observed Sources}\label{appendix:obs}

\begin{description}
  \item[IRAS 04016+2610 (L1489 IRS)] We classify this source as a Class~I protostar, consistent with previous designations in the literature \citep{Kenyon1995, White2004, Luhman2010}. Our ALMA observations clearly detect a circumstellar disk but reveal no companions. The higher resolution eDisk survey likewise resolved the disk and did not identify any additional sources \citep{Ohashi2023}. We supplemented the VizieR SED with additional literature photometry \citep{Andrews2005, Young2003, Moriarty-Schieven1994, Motte2001, Eisner2012, Barsony1992, Ladd1991} and incorporated archival \textit{Spitzer}/IRS and \textit{Herschel}/PACS spectra.

  \item[IRAS 04108+2803 B] Classified as Class I in both \citet{Luhman2010} and \citet{Motte2001}, which our analysis supports. This source has a wide companion observed in previous high-resolution multiplicity studies and is included in the Taurus+ sample \citep[$\sim$0\farcs33;][]{2008AJ....135.2496C}. In our observations, IRAS 04108+2803-B appears as a single point source.
  
  \item[IRAS 04158+2805 AB] \citet{White2004} originally classified this source as a proto-brown dwarf, but \citet{Luhman2010} later argued that it exceeds the hydrogen burning limit and is thus a low-mass source. Studies have found this source consistent with a  Class~I protostar \citep{White2004, Furlan2008, Luhman2010}, which our analysis supports. \citet{Ragusa2021} reported the first detection of this object as a close binary with a circumbinary disk in ALMA data. Our observations confirm the presence of these close companions and the circumbinary structure. 

  \item[IRAS 04166+2706] This source has been classified as Class~I by \citet{White2004, Luhman2010, Motte2001}, but both our analysis and the eDisk study place it in the Class~0 category \citep{Ohashi2023, Phuong2025}. In our observations, IRAS~04166+2706 appears as a single point source with no evidence of companions, consistent with the eDisk results, which resolved the disk but did not identify any additional nearby sources. 

  \item[IRAS 04169+2702] This source has been classified as Class~I by \citet{White2004}, \citet{Luhman2010}, and \citet{Ohashi2023}, consistent with our calculations. In our observations, IRAS~04169+2702 appears as a single point source, with no evidence of companions. \citet{2008AJ....135.2496C} reported a close binary with a separation of 28~au; however, this companion is not confirmed by the higher-resolution eDisk study \citep{Ohashi2023}. The eDisk team also noted the presence of a cavity, which can be indicative of a close binary. They suggested that such a feature could be caused by a companion even closer than 28~au, though no such companion has yet been directly detected. 
  
  \item[IRAS 04181+2654 AB] We detect component A as a point source in both the ALMA and VLA observations, but component B is a non-detection in our data. However, IRAS 04181+2654 B has been reported in other high-resolution surveys \citep{2008AJ....135.2496C} and is therefore included in our Taurus+ sample. Both components are consistently classified as  Class~I sources in the literature \citep{Luhman2010,Fiorellino2023,White2004}, in agreement with our analysis.

  \item[IRAM 04191 (IRAM 04191+1522) AB] First identified by \citet{Andre1999}, IRAM 04191 A was later confirmed as a protostar based on its accretion properties \citep{Kim2016,Kim2019}. \citet{Luhman2010} classified it as a clear  Class~0, consistent with our calculations. We detect it as a point source in our Band 7 ALMA data; however, in archival Band 6 ALMA observations we identify a previously unreported companion at a projected separation of $\sim$1800 au (included in the Taurus+ sample). 
  We associate the pair with the IRAS 04191+1523 AB system. The companion reported by \citet{Chen2012} is not detected in our data. A close companion to IRAM 04191 A was recently reported at $\sim$54 mas ($\sim$8 au at 140 pc) by \citet{Huelamo2026}; however, this separation is below the adopted lower limit of our survey and is therefore not included in our multiplicity analysis.
  
  \item[IRAS 04191+1523 AB] This protostellar binary has been previously reported in the literature \citep{Duchene2004, 2008AJ....135.2496C}. Both components are classified as Class~I by \citet{Luhman2010}, consistent with our own calculations. IRAS 04191+1523 AB forms an intermediate-separation binary ($\sim$900~au), which we further associate with the wider companions IRAM~04191 AB.

  \item[IRAS 04239+2436 AB] Both classified as Class~I in both this work and previous studies \citep{White2004, Luhman2010}. We observe two point source objects at a 30~au separation. Previously identified as a close embedded protostellar binary with HST \citep{Reipurth2000}, more recent SO line emission observations reveal prominent triple spiral arms surrounding the system, with \citet{Lee2023} suggesting it may host three protostars.

  \item[IRAS 04248+2612 ABC] A Class~I triple system classified as a proto–brown dwarf by \citep{White2004} based on its near-IR and optical properties. The system consists of a close binary and a third component at a separation of $\sim$1500~au previously reported by \citet{Duchene2007} but consistent with our results.

  \item[IRAS 04260+2642] This source was initially classified as Class~I by \citet{White2004}, but its evolutionary stage remains uncertain because the edge-on disk obscures the stellar light. Our ALMA observations clearly resolve this disk. Some studies have argued that the system may be more evolved than a typical Class~I; for example, \citet{Luhman2010} reclassified it as Class~II, noting that while its spectral index falls within the Class~I range, the lack of envelope signatures supports a Class~II designation. For consistency with our calculated evolutionary classifications, however, we retain a Class~I designation. No companions have been reported.

  \item[IRAS 04263+2426 (Haro 6-10, GV Tau) AB] This system is a known binary at $\sim$200~au separation \citep{Simon1995, Gibb2007}. We observe two compact sources separated by about 175~au. It has been classified as Class~I by \citet{Luhman2010} and \citet{White2004}, which is in agreement with our calculations. 

  \item[IRAS 04264+2433 AB] This close binary system was first reported by \citet{Duchene2004} in their deep infrared survey, which revealed a faint companion, consistent with our detection. It was previously classified as a  Class~I source by \citet{Luhman2010} and \citet{White2004}, consistent with our results.

  \item[IRAS 04287+1801 (L1551 IRS5) AB] This is a well-known close binary, previously reported in multiple studies \citep{Bieging1985,Rodriguez1995,2008AJ....135.2496C}. With ALMA, we observe this 60~au separation binary with a surrounding circumbinary disk. Both companions are classified as Class~I in our calculations, which is consistent with previous studies \citep{White2004,Luhman2010}.

  \item[L1551NE (IRAS 04288+1802) AB] This source is classified as  Class~I in our calculations and in \citet{Luhman2010}. Our observations resolve it as a binary with a separation of $\sim$80~au, with a faint circumbinary disk, consistent with previous work by \citet{Takakuwa2014}. 
  
  \item[IRAS 04295+2251 (L1536 IRS) AB] Classified as a  Class~I object by \citet{Luhman2010} and \citet{White2004}, from our calculations we likewise adopt a Class~I designation. Our ALMA observations show a clear circumstellar disk with a large central cavity, and we present sub-arcsecond resolution 9 mm archival VLA observations that show a potential close companion (18 au separation) that we include in the Taurus+ sample.
  
  \item[IRAS 04302+2247] Classified as a  Class~I source by \citet{Luhman2010}, we also adopt this Class~I designation. We observe an edge on disk around this single source, which could serve to obscure the central star and make its evolutionary stage uncertain. This edge on disk was also observed in the eDisk survey \citep{Ohashi2023} and by \cite{Wolf2008} with no additional companions reported. 

  \item[IRAS 04325+2402 AB] This source is an intermediate binary ($\sim$1000~au) with two components. It was initially suggested to be a triple system, consisting of a close binary with a wide companion \citep{Hartmann1999}, but later work showed that the proposed close binary is a single protostar with an absorption lane along the line of sight \citep{Scholz2010}. This is a  Class~I system according to our analysis, which is consistent with previous classifications from the literature \citep{White2004,Luhman2010}. With ALMA we observe two resolved disks with orthogonal orientations.

  \item[IRAS 04361+2547 (TMR-1) AB] Classified as  Class~I in both this study and \citet{Luhman2010}. Hubble Space Telescope (HST) imaging by \citet{Terebey1998} revealed a binary with a separation of $\sim$30~au. In our data, we detect only a single source, but include IRAS 04361+2547 B in the Taurus+ sample given its confirmation in high-resolution observations.
  
  \item[IRAS 04365+2535 (TMC 1A)] Classified as  Class~I by \citet{White2004} and \citet{Luhman2010}; our analysis supports the same classification. We detect it as a single point source, in line with previous high-resolution surveys \citep{Duchene2007, 2008AJ....135.2496C}. Later, \citet{Bjerkeli2016} used ALMA to resolve structure down to 6 au and also found no companions. 

  \item[IRAS 04368+2557 (L1527 IRS)] The source is widely recognized as a  Class~0 object, consistent with its low bolometric temperature ($T_\mathrm{bol} < 70$ K, \citet{Chen1995}; \citet{Motte2001}; \citet{Luhman2010}). \citet{Loinard2002} previously reported it as a binary based on VLA 7~mm observations with a separation of 25~au, but \citet{Nakatani2020} and \citet{Sheehan2022} later confirmed it as a single star and noted potential substructure in the disk. In our data, we also detect it as single source and observe a resolved edge on disk. The eDisk study resolved a similar edge on disk and no companions for this source \citep{Ohashi2023}.

  \item[IRAS 04381+2540 (TMC-1, L1534) AB] This source was first published as a binary after being detected with HST by \citet{Apai2005}. With ALMA and VLA we observe two compact sources at a separation of 85~au. We designate it as a Class~I source, which is consistent with the classification in the literature \citep{White2004, Luhman2010}.
  
  \item[IRAS 04385+2550 (Haro 6-33)] Initially classified as a  Class~I source by \citet{White2004}, \citet{Luhman2010} reclassified it as  Class~II, noting that although its spectral index falls within the  Class~I range, the absence of envelope signatures supports a  Class~II designation. For consistency with our calculated evolutionary classifications, we still consider this a  Class~I source. We observe a single point like source and no companions have been previously reported.
  
  \item[IRAS 04489+3042 AB] Initially classified as a proto–brown dwarf by \citep{White2004}, later reinterpreted by \citet{2006ApJ...647.1180L} as a low-mass star. Our observations identify it as a Class~I binary with a faint companion.
  
  \item[DG Tau (IRAS 04240+2559) AB] This is a well-known wide binary system. \citep{White2004} classified both components as  Class~II, while \citet{Luhman2010} classified DG Tau B as  Class~I and DG Tau A as  Class~II. In our analysis, we calculate DG Tau B as Class~I and DG Tau A as FS. With both ALMA and VLA observations centered on DG Tau B, DG Tau A is only detected in the VLA data, owing to its wider field of view. DG Tau B shows a resolved disk in our ALMA observations, and both components appear as compact point sources in the VLA data. previous high-resolution ALMA observations reveal a circumstellar disk around DG Tau A as well \citep{Yamaguchi2024}.

  \item[HH 30] Classified as  Class~I by \citet{Luhman2010}, this source hosts a nearly edge-on disk that obscures the central star, and creates ambiguity for its evolutionary stage. Other studies suggest it may be more evolved \citep{White2004}. We adopt a  Class~II designation based on our own calculations. \citet{Joncour2017} associated this object with LkH$\alpha$358, but their reported separation of $\sim$11,350~au is beyond the separation limit considered in our study. From our data, we measure this source as a single disk.

\end{description}

\subsection{Taurus+ Sources}\label{appendix:extended}

\begin{description}
  \item[IRAS 04108+2803-A] A wide companion ($\sim$2500~au) to 04108+2803 B, though undetected in our observations of that source. We include it in the Taurus+ sample as it has been reported as a companion in previous high-resolution multiplicity studies \citep[$\sim$0\farcs33;][]{2008AJ....135.2496C}.
  This source is classified as Class~I in \citet{Luhman2010} and as Class~II in \citet{Motte2001}, but we designate it a FS source. 

  \item[IRAS 04154+2823] This source was originally classified as Class~II by \citet{Kenyon1995}, but later studies by \citet{Furlan2008} and \citet{Andrews2005} reclassified it as a FS source, noting that its SED appears transitional between Class~I and Class~II. We also have designated it a FS source. No sub-arcsecond IR or radio observations have identified any close companions.

  \item[IRAS 04166+2708] \citet{Luhman2010} and \citet{White2004} classified this source as  Class~I, which is consistent with our calculations. This source has no known companions but has not yet been observed at sub-arcsecond resolution at IR or radio wavelengths.

  \item[2MASS J04194657+2712552 ([GKH94] 41)] \citet{Luhman2010} classified this source as “Class~I?”, while \citet{Furlan2011} found it to be disk-dominated and likely more evolved. Our analysis yields a FS classification. It has no known companions but has never been observed at sub-arcsecond resolution at IR or radio wavelengths.

  \item[IRAS 04181+2655] Located near the IRAS~04181+2654 system but outside the 10,000~au limit to be considered a companion. 
  Classified as Class~I by \citet{White2004} and as  Class~II by \citet{Fiorellino2023} and \citet{Luhman2010}, though the latter note that while its spectral index is consistent with a Class~I source, it lacks evidence for an envelope. For consistency, we adopt the Class~I classification. The IRAS designation refers to a composite position that encompasses multiple nearby sources, and as a result the SIMBAD entry for this source is offset, the correct counterpart is 2MASS J04210795+2702204. Even so, we adopt the IRAS name for consistency with the literature.

  \item[FS Tau (Haro 6-5, IRAS 04189+2650) AB] Identified as a wide binary with a circumbinary disk in the ALMA Band~6 multiplicity survey of \citet{2019ApJ...872..158A} (25–30\arcsec resolution). FS Tau A is a spectroscopic binary \citep{Hartigan2003}, but we do not include this in our sample to stay consistent with our resolution limits. While classified as Class~I by \citet{Fiorellino2023} and \citet{Luhman2010}, and as Class~II by \citet{2019ApJ...872..158A}, our analysis designates FS Tau A as FS and FS Tau B as Class~I.

  \item[L1521F] This source is deeply embedded within one of the densest cores in Taurus. High-resolution ALMA Band~7 (0.87~mm) continuum observations at 25~au resolution revealed compact emission \citep{Tokuda2017}. More recent 1.3~mm observations at 4~au resolution detected spike-like structures extending form the source, but were interpreted as signatures of magnetic flux transport rather than fragmentation \citet{Tokuda2024}. To date, there is no evidence for true protostar companions, and we classify this source as Class~0.

  \item[2MASS J04293209+2430597] This source was listed as “Class~I?” by \citet{Luhman2010}, which our analysis likewise supports, so we adopt a Class~I designation. No companions are currently known, though it has not been observed at sub-arcsecond resolution at IR or radio wavelengths.

  \item[LkH$\alpha$358] While a previous high resolution study by \citet{Wallace2020} ($\sim$0\farcs2) classifies this as a single source, our analysis associates it with HL Tau and XZ Tau at a wide separation of $\sim$6700~au. \citet{Luhman2010} designates LkH$\alpha$358 as a Class~I source, but our calculations place it in the Flat Spectrum category.
  
  \item[HL Tau (Haro 6-14)] A  Class~I protostar with a well-studied protoplanetary disk exhibiting extensive substructure \citep{ALMA2015}. It is a wide companion to XZ Tau (Haro 6-15), a  Class~II spectroscopic binary \citep{Hartigan2003}. The evolutionary classifications for both systems were determined by \citet{Luhman2010} and are confirmed by our analysis.

  \item[Haro 6-13] Listed as Class~II by \citet{Luhman2010} and \citet{White2004}, but reported in the Class~I/FS sample of \citet{Flores2024}. Our analysis classifies it as a FS source. High-resolution studies \citep{Leinert1993, 2008AJ....135.2496C} confirmed it as single, and sub-arcsecond resolution ALMA Band~6 data from \citet{Long2019} likewise showed no evidence of companions.

  \item[HP Tau (IRAS 04328+2248)] A protostar that is widely separated from three known companions: HP Tau G2, HP Tau G3-A, and HP Tau G3 B \citep{Simon1995, 2012ApJ...751..115H, Rizzuto2020}. HP Tau G3 A and B form a spectroscopic binary and are treated as a single source in our analysis to remain consistent with our resolution limits. High-resolution ALMA Band~6 data from \citet{Long2019} showed no companions within their FOV, as expected. We classify HP Tau as Flat Spectrum, while assigning Class~II to HP Tau G2 and G3. Although \citet{Andrews2005} also reported HP Tau as Flat Spectrum, \citet{Luhman2010} classified it as  Class~II and its companions as Class~III.

  \item[Haro 6-28 (V1026 Tau) AB] This close pair ($\sim$100~au separation) was first published by \citet{Leinert1993} and then by \citet{Simon1995}. In our analysis it is part of a larger multiple system with five components. \citet{Flores2024} classified it as  Class~I/FS, whereas \citet{Luhman2010} classified it as  Class~II. From our calculations we designate it a  Class~II source.

  \item[IRAS S04361+2331 (2MASS J04390525+2337450)] Previously classified as “Class~I?” by \citet{Luhman2010}, but our analysis classifies it as a FS source. No companions are currently known, though it has not been observed at sub-arcsecond resolution at IR or radio wavelengths.
  
  \item[IC 2087 IR (IRAS 0437+257P08)] This source was classified as  Class~II by \citet{White2004} but as  Class~I by \citet{Luhman2010}. From our calculations we assign it a  Class~I designation. High-resolution observations ($\sim$0\farcs33) by \citet{2008AJ....135.2496C} indicate that it is a single source. 

\end{description}




\bibliography{bibtry}{}
\bibliographystyle{aasjournalv7}

\end{document}